\documentclass[twocolumn]{emulateapj}
\usepackage{natbib}
\shorttitle{Polarization of protoplanetary disks}
\shortauthors{Kataoka et al.}
\usepackage{graphics,graphicx,subfigure}

\begin{document}

\title{Millimeter-wave polarization of protoplanetary disks due to dust scattering}

\author{Akimasa Kataoka \altaffilmark{1,2,3}, Takayuki Muto\altaffilmark{4}, Munetake Momose \altaffilmark{5}, Takashi Tsukagoshi \altaffilmark{5}, Misato Fukagawa\altaffilmark{2,6}, Hiroshi Shibai\altaffilmark{6}, Tomoyuki Hanawa\altaffilmark{7}, Koji Murakawa\altaffilmark{8}, Cornelis P Dullemond \altaffilmark{1}}
\affil{\altaffilmark{1}Institute for Theoretical Astrophysics, Heidelberg University, Albert-Ueberle-Strasse 2, 69120 Heidelberg, Germany}
\email{kataoka@uni-heidelberg.de}
\affil{\altaffilmark{2}National Astronomical Observatory of Japan, Mitaka, Tokyo 181-8588, Japan}
\affil{\altaffilmark{3}Department of Earth and Planetary Sciences, Tokyo Institute of Technology, Ookayama, Meguro, Tokyo 152-8550, Japan}
\affil{\altaffilmark{4}Division of Liberal Arts, Kogakuin University, 1-24-2 Nishi-Shinjuku, Shinjuku, Tokyo 163-8677, Japan}
\affil{\altaffilmark{5}College of Science, Ibaraki University, 2-1-1 Bunkyo, Mito, Ibaraki 310-8512, Japan}
\affil{\altaffilmark{6}Graduate School of Science, Osaka University, 1-1 Machikaneyama, Toyonaka, Osaka 560-0043, Japan}
\affil{\altaffilmark{7}Center for Frontier Science, Chiba University, 1-33 Yayoi-cho, Inage, Chiba 263-8522, Japan}
\affil{\altaffilmark{8}College of General Education, Osaka Sangyo University, 3-1-1, Nakagaito, Daito, Osaka 574-8530, Japan}

\begin{abstract}
We present a new method to constrain the grain size in protoplanetary disks with polarization observations at millimeter wavelengths.
If dust grains are grown to the size comparable to the wavelengths, the dust grains are expected to have a large scattering opacity and thus the continuum emission is expected to be polarized due to self-scattering.
We perform 3D radiative transfer calculations to estimate the polarization degree for the protoplanetary disks having radial Gaussian-like dust surface density distributions, which have been recently discovered.
The maximum grain size is set to be $100 {\rm~\mu m}$ and the observing wavelength to be 870 ${\rm \mu m}$. 
We find that the polarization degree is as high as 2.5\% with a subarcsec spatial resolution, which is likely to be detected with near-future ALMA observations.
The emission is polarized due to scattering of anisotropic continuum emission.
The map of the polarization degree shows a double peaked distribution and the polarization vectors are in the radial direction in the inner ring and in the azimuthal direction in the outer ring.
We also find the wavelength dependence of the polarization degree: the polarization degree is the highest if dust grains have a maximum size of $a_{\rm max}\sim\lambda/2\pi$, where $\lambda$ is the observing wavelength.
Hence, multi-wave and spatially resolved polarization observations toward protoplanetary disks enable us to put a constraint on the grain size.
The constraint on the grain size from polarization observations is independent of or may be even stronger than that from the opacity index.
\end{abstract}


\section{Introduction}

Dust coagulation is the first step toward planet formation.
Observational constraints on grain size is important in understanding the first stage of planet formation.
The grain growth in protoplanetary disks has been evidenced by a low opacity index of protoplanetary disks at millimeter wavelengths, $\beta\approx0-1$, where $\kappa_{\rm dust}\propto \lambda^{-\beta}$ \citep[e.g.,][]{BeckwithSargent91, Wilner00, Kitamura02, AndrewsWilliams05, Ricci10a, Ricci10b, Guilloteau11}, which suggests that dust grains are grown to millimeter in size \citep[e.g.,][]{MiyakeNakagawa93, DAlessio01, Draine06}.

However, the constraints on the grain size with opacity index still have uncertainties.
The low opacity index at millimeter wavelengths can be reproduced not only by changing grain size, but also by changing chemical composition \citep[e.g.,][]{Pollack94} or shape \citep[e.g.,][]{Min05}.
In addition, optically thick disks may account for the low spectral index of protoplanetary disks \citep[e.g.,][]{Ricci12b}.

Recently, millimeter-wave observations have shown that some protoplanetary disks have a strong asymmetry of the surface brightness distribution although the gas is more broadly and smoothly distributed \citep{Casassus13, vanderMarel13, Fukagawa13, Isella13, Perez14}.
The lopsided continuum emission with the smooth gas distribution may be a signature of dust accumulation by gas density enhancement created by such as vortex \citep[e.g.,][]{Lyra09}.
The dust accumulation accelerates the grain growth and thus it may be a hot spot of planet formation.
Therefore, the dust coagulation at the hot spot should be observationally investigated. 

In this paper, we propose a new method to constrain the grain size in protoplanetary disks, which make use of polarization due to scattering by dust grains.
If grains are sufficiently large, the scattering opacity is large and therefore there is a significant amount of scattered light component in dust continuum emission.
In such cases, in tandem with the anisotropy of the continuum emission, we shall show that the dust continuum emission is polarized.

Polarization of early-stage protoplanetary disks has been detected recently \citep{Rao14, Stephens14, Segura-Cox15}.
It is widely believed that the polarization is due to the grain alignment with magnetic field, as analogous to the interpretations of star-forming regions \citep[e.g.,][]{Girart06, Girart09, Hull13, Hull14}.
We propose another possibility of producing polarized dust continuum emission.
The proposed mechanism will be tested with ALMA, which has a high sensitivity and a high spatial resolution.

This paper is organized as follows.
In Section \ref{sec:grain}, we present the theoretical background of the scattering and the polarization properties of dust grains.
We also present a toy model to illustrate that the anisotropic distribution of background emission produces the polarization.
In Section \ref{sec:radmc}, we perform radiative transfer calculations of protoplanetary disk models.
In Section \ref{sec:discussion}, we discuss detectability of the polarization due to dust scattering.
In Section \ref{sec:conclusions}, we conclude this paper.

\section{Theoretical understanding of scattering of anisotropic radiation field}
\label{sec:grain}
In this section, we describe the theoretical backgrounds for polarization due to dust scattering.
To obtain the polarization, two conditions are required.
One condition is concerning dust grains: the dust size should be appropriate in a way that sufficient polarization degree is produced when incident radiation is scattered.
The other is concerning the radiation field: the incident light should have anisotropic distribution.
After presenting our dust model in Section \ref{sec:dustmodel}, the first condition is discussed in Sections \ref{sec:albedo}-\ref{sec:detectable} and the second condition is discussed in Section \ref{sec:toymodel}.

\subsection{Dust model}
\label{sec:dustmodel}

Opacity of protoplanetary disks is dominated by dust grains.
The opacity depends on dust properties such as their composition, shape, and size.
In this paper, we focus on the dependence on the grain size, which dramatically changes the optical properties.

Dust grains are assumed to be spherical with radius $a$, and their composition is fixed.
The composition is taken to be the same as \citet{Kataoka14}: the mixture of silicate, water ice, and organics, where the mass fractional abundance is set to be consistent with \citet{Pollack94}: $\zeta_{\rm silicate}=2.64 \times 10^{-3}$, $\zeta_{\rm organics}=3.53 \times 10^{-3}$, and $\zeta_{\rm ice}=5.55 \times 10^{-3}$.
The refractive index of astronomical silicate is taken from \citet{WeingartnerDraine01}, organics from \citet{Pollack94}, and water ice from \citet{Warren84}.
We use Mie theory to calculate the absorption and scattering mass opacities and the scattering matrix $Z_{ij}$ (see Appendix for more details).
Dust grains are assumed to have a size distribution of $n(a)\propto a^{-3.5}$.
The minimum size is taken to be $a_{\rm min}=0.01{\rm~\mu m}$ and we change the maximum size $a_{\rm max}$ in the following discussion.
Note that the minimum grain size is small enough not to affect the results.

\subsection{Scattering opacity of large dust grains}
\label{sec:albedo}
\begin{figure}[htbp]
 \begin{center}
  \includegraphics[width=80mm]{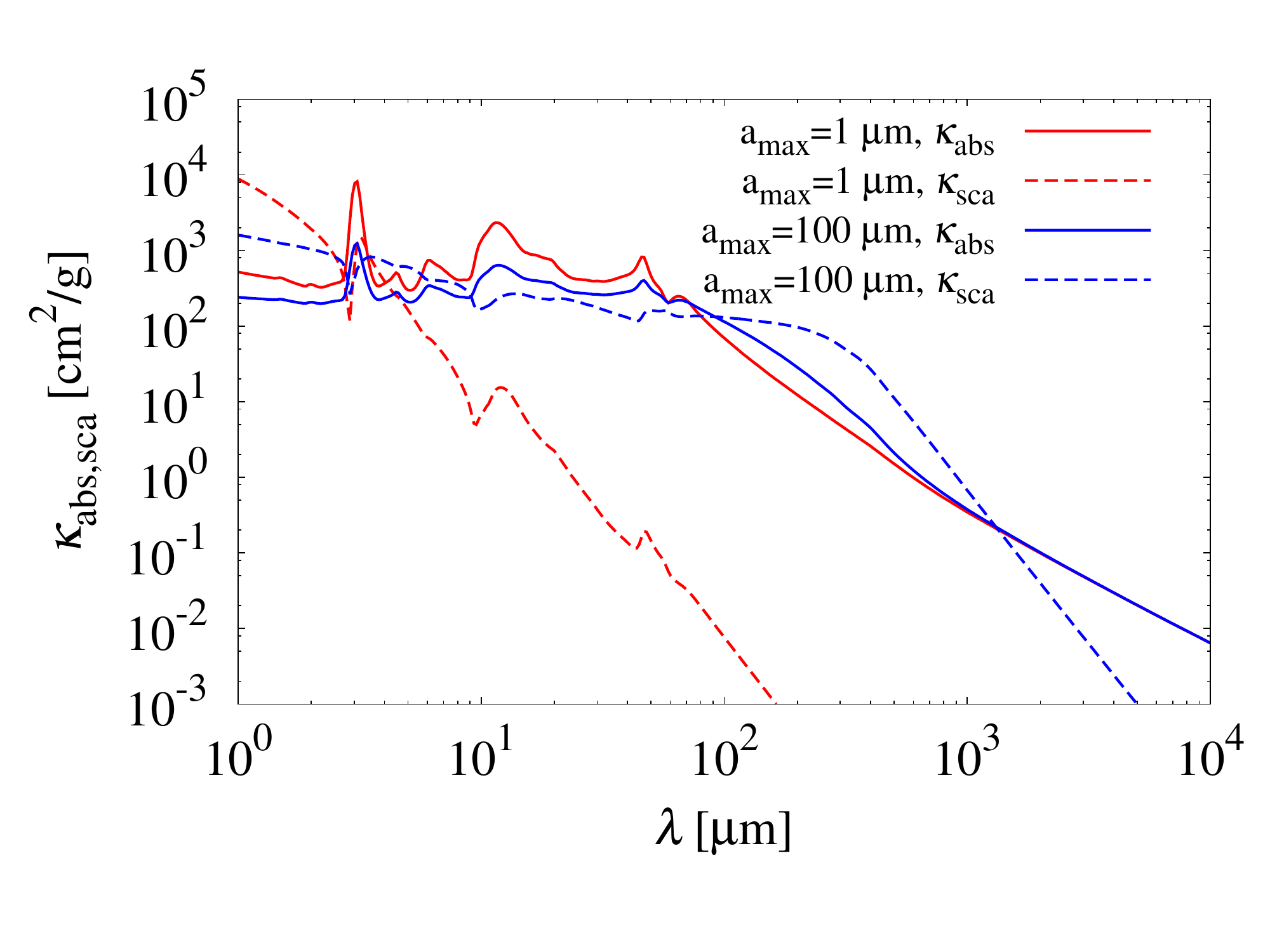}
 \end{center}
 \caption{
 The absorption and scattering mass opacities of dust grains in the cases of $a_{\rm max}=1{\rm ~\mu m}$ and $a_{\rm max}=100{\rm ~\mu m}$.
 The size distribution is $n(a)\propto a^{-3.5}$.
 }
 \label{fig:dustkappa}
\end{figure}

Here, we illustrate the dependence of the scattering efficiency on the grain size.
Figure \ref{fig:dustkappa} shows the absorption and scattering mass opacities of dust grains which have the maximum sizes of 1 ${\rm \mu m}$ and 100 ${\rm \mu m}$.
The mass opacity is written in unit of ${\rm cm}^2$ per gram of dust grains.
In the case of $a_{\rm max}=1{\rm~\mu m}$, the scattering opacity is much less than the absorption at $\sim$ mm wavelengths.
In the case of $a_{\rm max}=100{\rm~\mu m}$, by contrast, the scattering opacity is larger than the absorption.
This implies that the scattering at mm wavelengths is efficient if the dust grains are as large as 100 ${\rm \mu m}$.
This is realized in protoplanetary disks because dust grains are believed to have grown to millimeter size \citep[e.g.,][]{BeckwithSargent91}.

\subsection{Polarization dependence on the scattering angle}
\label{sec:polarization}
\begin{figure*}[htbp]
\centering
\subfigure{
  \includegraphics[width=75mm]{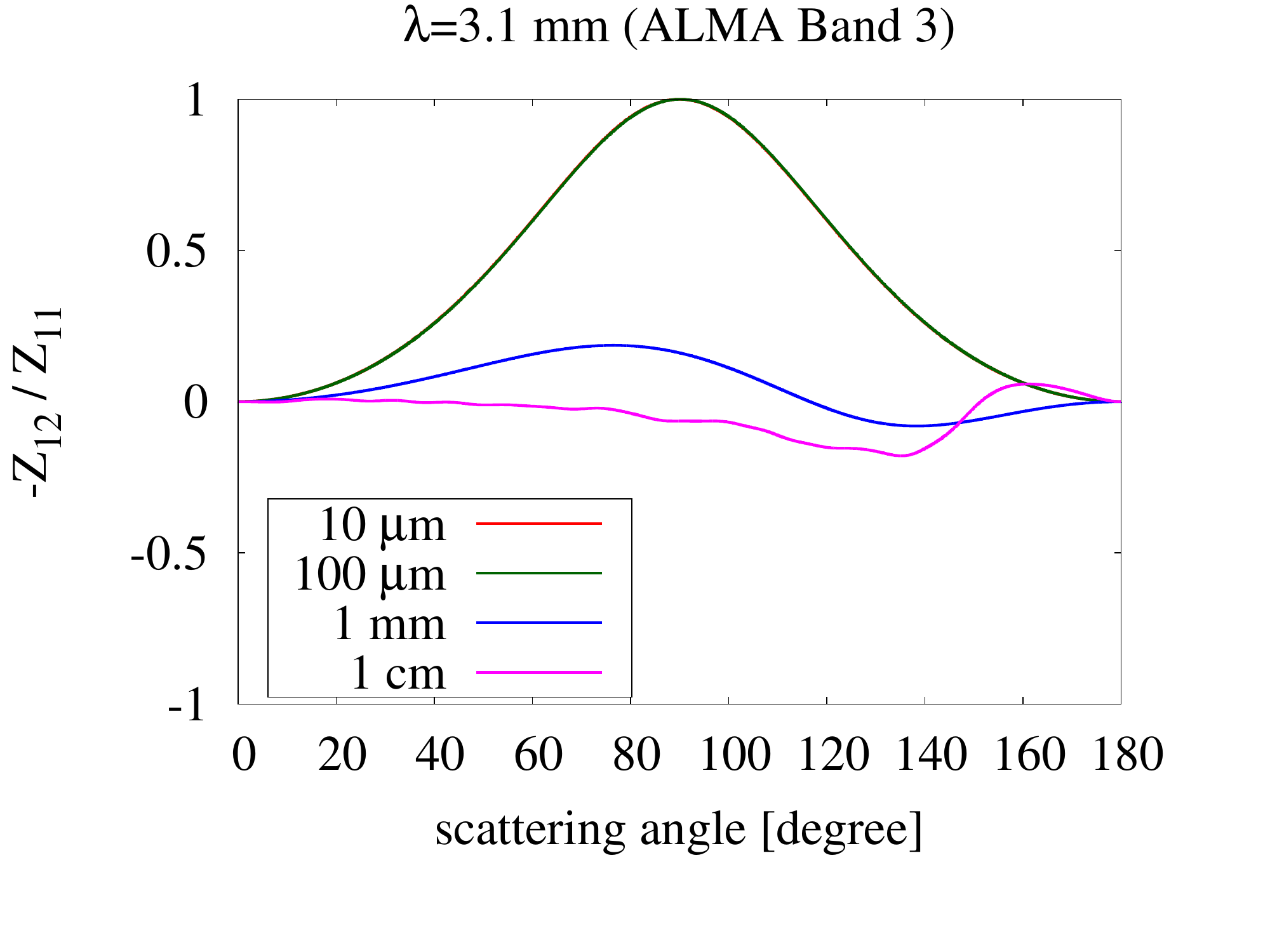}
  }
 \subfigure{
    \includegraphics[width=75mm]{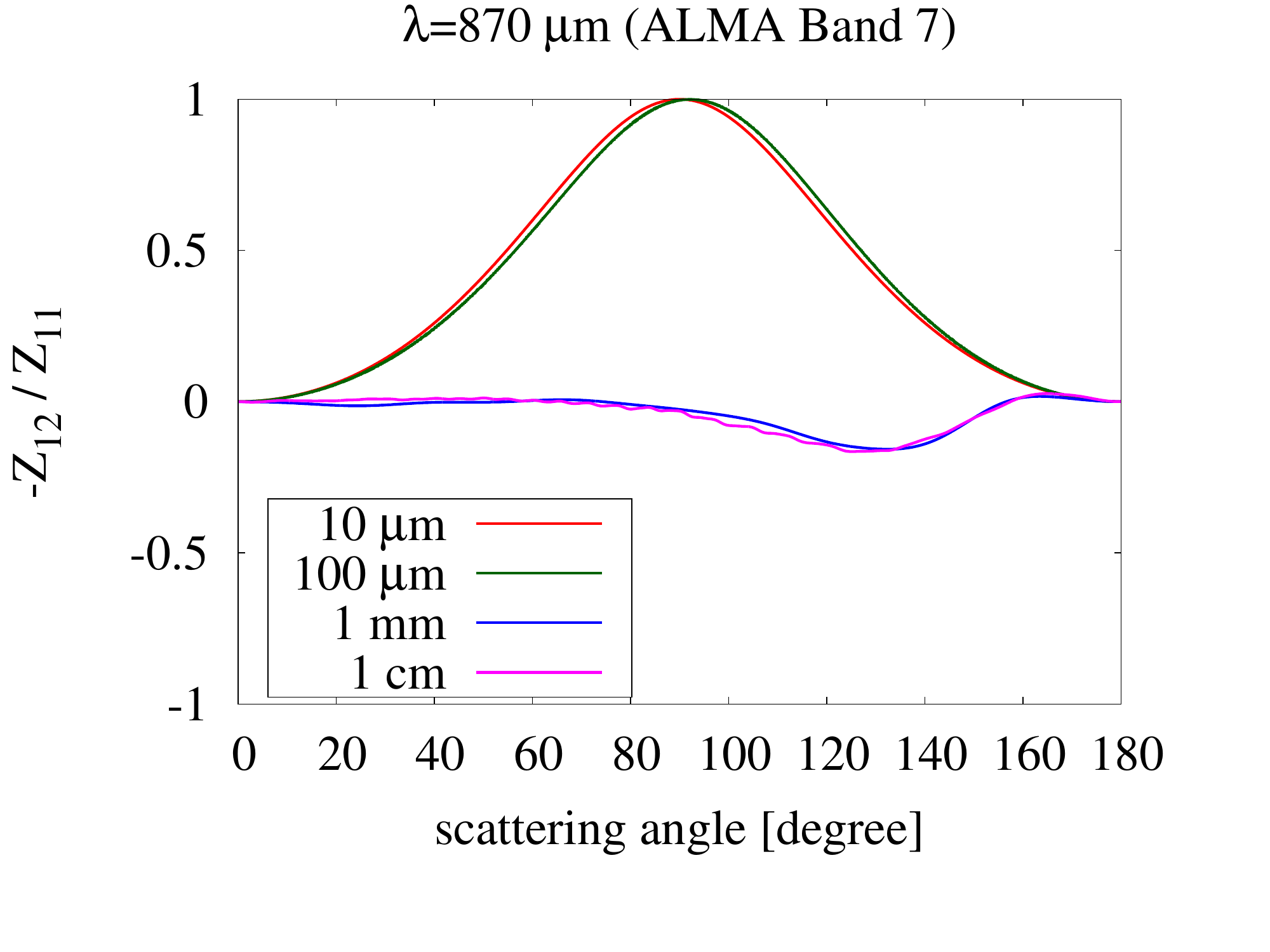}
  }
 \caption{
 The degree of polarization $-Z_{12}/Z_{11}$ of dust grains against scattering angles.
 The left panel is for the case of $\lambda=3.1$ mm and the right panel for $\lambda=870{\rm~\mu m}$.
 The size distribution is assumed to have power law of $n(a)\propto a^{-3.5}$.
 The maximum grain sizes are $10{\rm~\mu m}$, $100{\rm~\mu m}$, $1{\rm~mm}$, and $1{\rm~cm}$.
 }
 \label{fig:polari}
\end{figure*}

\begin{figure*}[htbp]
\centering
\subfigure{
  \includegraphics[width=75mm]{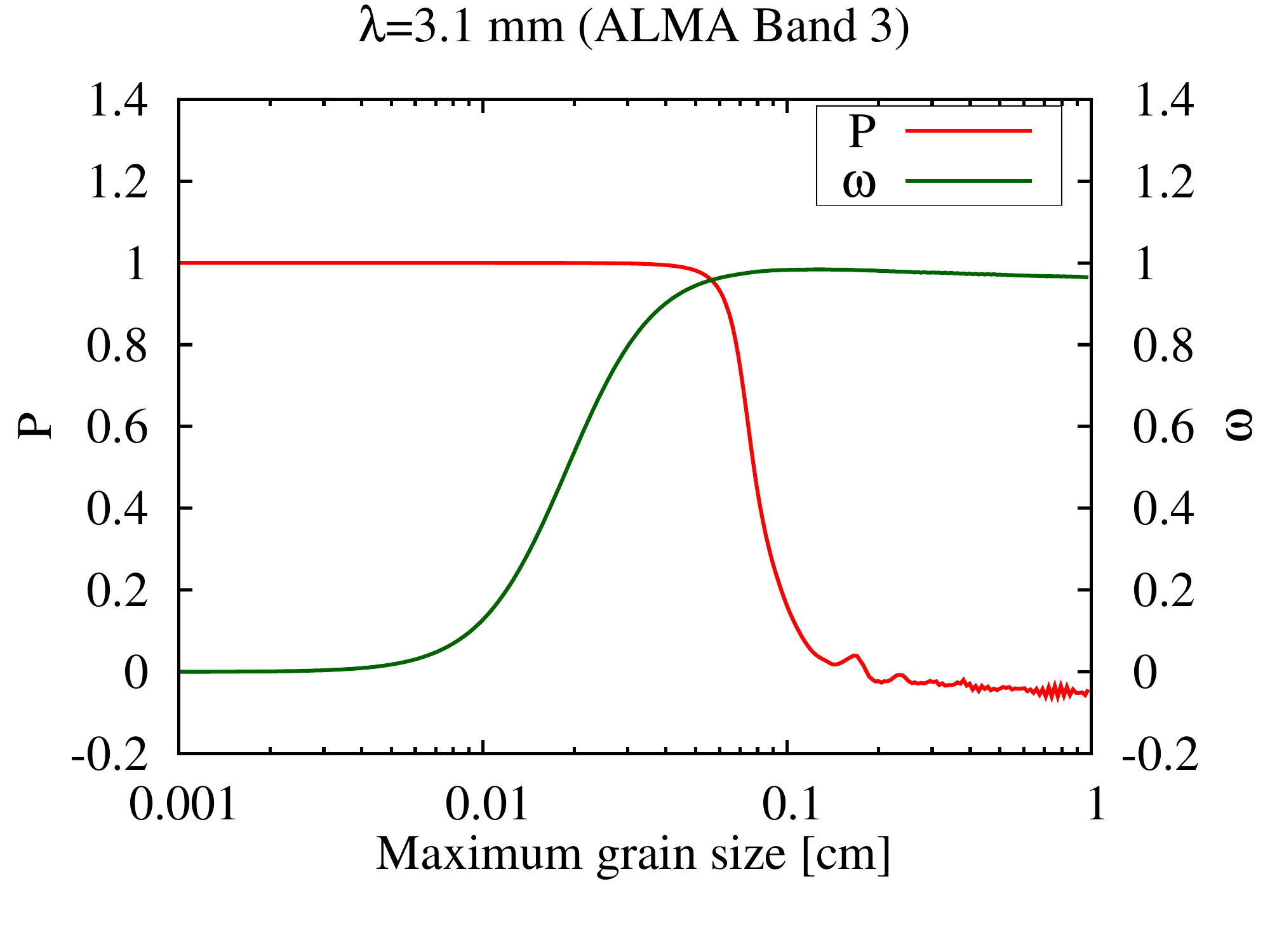}
  }
 \subfigure{
    \includegraphics[width=75mm]{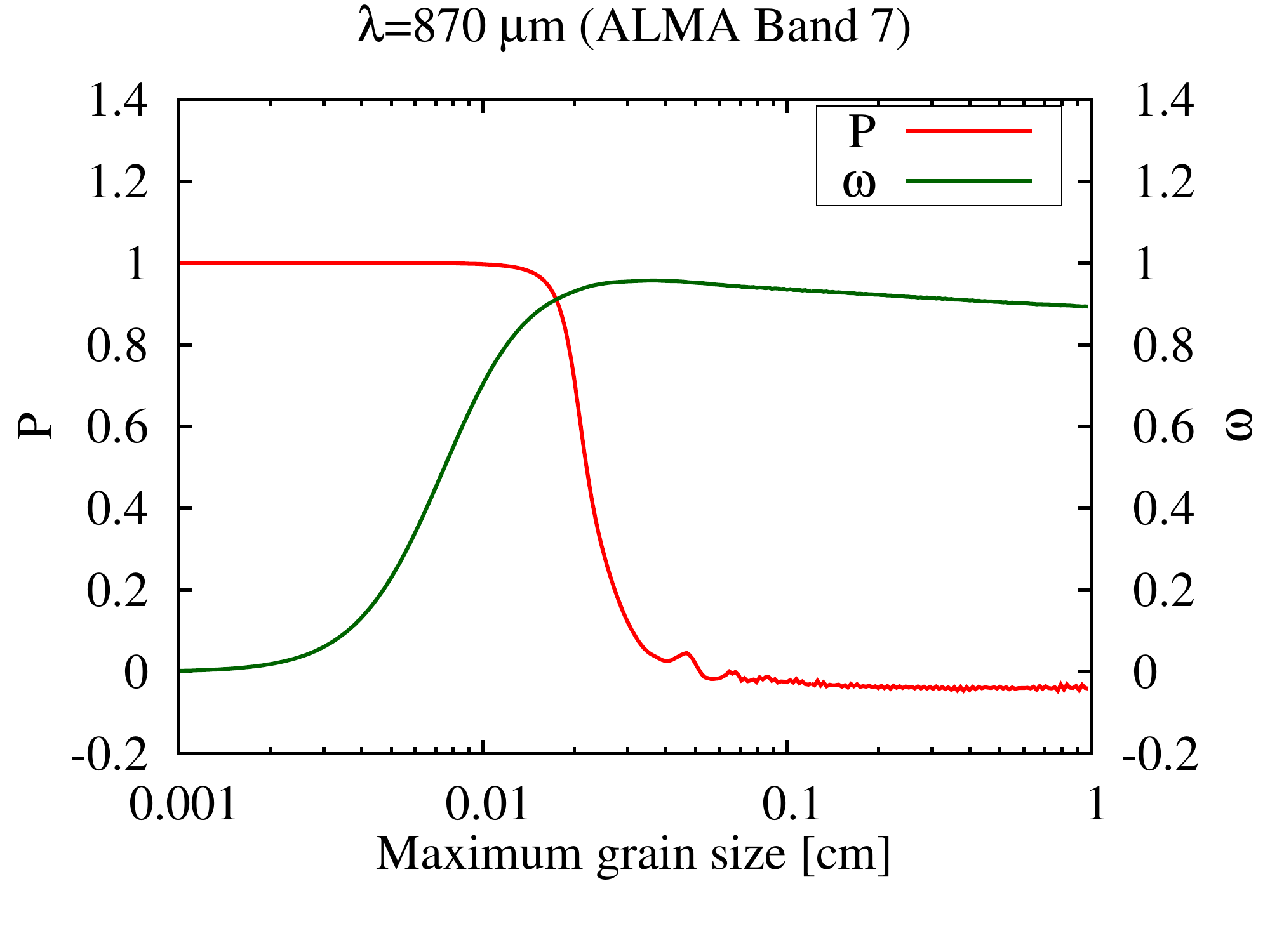}
  }
 \caption{
 The polarization $P$ at the scattering angle of $90^\circ$ and the albedo $\omega=\kappa_{\rm sca}/(\kappa_{\rm abs}+\kappa_{\rm sca})$ as a function of maximum grain size.
 The size distribution is assumed to have power law of $n(a)\propto a^{-3.5}$.
 The wavelengths are assumed to be 3.1 mm for the left panel and $870 {\rm~\mu m}$ for the right panel.
 The arrows indicate the maximum grain size which has the most efficient polarization by $90^\circ$ scattering.
 }
 \label{fig:peak}
\end{figure*}

The polarization by dust scattering strongly depends on its scattering angle.
We investigate the dependence of the polarization on the grain size, the scattering angle, and the wavelengths.
Figure \ref{fig:polari} shows the degree of polarization due to single scattering averaged with the size distribution ($-Z_{12,{\rm eff}}/Z_{11,{\rm eff}}$) for several maximum grain sizes in the cases of $\lambda=3.1{\rm~mm}$ and $\lambda=870{\rm~\mu m}$.
In the cases of $a_{\rm max}=10 {\rm~\mu m}$ and $100 {\rm~\mu m}$, the polarization degree shows a peak at the scattering angle of $\theta=90^\circ$, which is the case for the Rayleigh scattering regime that is realized when the grain size is smaller than the wavelength.
More specifically, the size parameter $x=2\pi a/\lambda$ is less than unity \citep{BohrenHuffman83}.
In the cases of $a_{\rm max}=1 {\rm~mm}$ and $1 {\rm~cm}$, by contrast, the polarization degree is 0 for almost all the scattering angle except at around $135^\circ$, where the polarization degree is around $-0.2$.
Note that at the minimum, the sign of the polarization has been changed, which is known as the polarization reversal \citep{Daniel80, BastienMenard88, Fischer94, Murakawa10, KirchschlagerWolf14}.
Thus, no polarization is expected if the maximum grain size is larger than roughly the wavelengths, except for the polarization reversal.

\subsection{Detectable grain size for each wavelength}
\label{sec:detectable}
We now discuss at what grain size the polarization due to scattering is significant.
On one hand, we have demonstrated that the scattering by dust grains is significant only when the grain size exceeds roughly the wavelength as shown in Section \ref{sec:albedo}.
On the other hand, the grain size should be less than $\sim \lambda/2\pi$ to have significant polarization in the scattered light as shown in Section \ref{sec:polarization}.
Thus, there is a grain size which contribute most to the polarization.

\begin{deluxetable}{cclcccc}
\tabletypesize{\scriptsize}
\tablecaption{The detectable grain size for observed wavelengths}
\tablewidth{0pt}
\tablehead{
wavelengths $\lambda$ & the detectable grain size $a_{\rm max}$             
}
\startdata
$7 {\rm~mm}$ &$ 1 {\rm ~mm}$\\
$3.1 {\rm~mm}$& $500{\rm ~\mu m}$ \\
$870 {\rm~\mu m}$& $150{\rm ~\mu m}$ \\
$340 {\rm~\mu m}$& $70{\rm ~\mu m}$ 
\enddata
\label{table:grain}
\end{deluxetable}

We investigate the dependence of polarization efficiency on grain size especially in the cases that wavelengths are $870 {\rm~\mu m}$ and $3.1 {\rm~mm}$, which correspond to ALMA Band 7 and 3, respectively.
Figure \ref{fig:peak} shows both the albedo $\omega=\kappa_{\rm sca}/(\kappa_{\rm abs}+\kappa_{\rm sca})$ and the polarization $P=-Z_{12,\mathrm{eff}}/Z_{11,\mathrm{eff}}$ at $90^\circ$.
If the grain size is small, the scattered light is perfectly polarized in the case of 90$^{\circ}$ scattering.
When $a_{\max} \gtrsim \lambda/2\pi$, the polarization starts decreasing to 0.
By contrast, the albedo $\omega$ increases with increasing grain size.

\begin{figure}[htbp]
 \begin{center}
 \includegraphics[width=80mm]{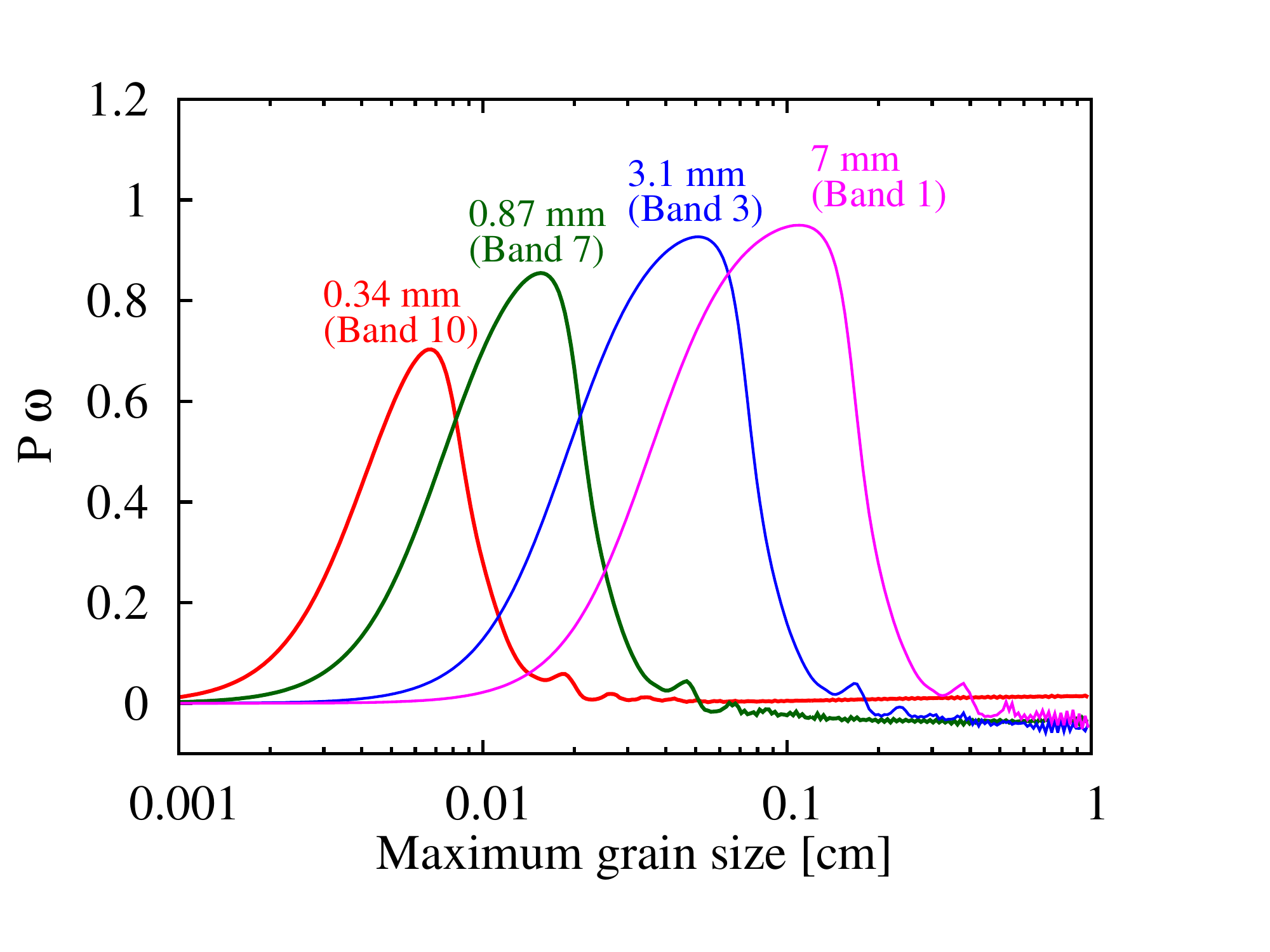}
 \end{center}
\caption{
The polarization degree times the albedo $P\omega$ against the maximum grain size.
This figure represents a grain size which contributes most to the polarized intensity.
Each line corresponds to the wavelengths of 0.34 mm, 0.87 mm, 3.1 mm, and 7 mm.
The band numbers correspond to the ALMA band numbers for each wavelength.
}
\label{fig:window}
\end{figure}

The product of the polarization and the albedo, $P\omega$, gives the grain size that contributes most to the polarization at any observing wavelength because the actual observed quantities are Stokes parameters $Q$ and $U$, which is proportional $P \omega$ in the case of single scattering.
In other words, $P\omega$ represents a window function for the grain size detectable with polarization observations.
Figure \ref{fig:window} shows $P\omega$ at the wavelengths of $\lambda=340 {\rm~\mu m}$, $870 {\rm~\mu m}$, $3.1{\rm ~mm}$, and $7 {\rm~mm}$ as a function of $a_{\max}$.
The grain sizes that give the maximum values of $P\omega$ for each observing wavelengths are summarized in Table \ref{table:grain}.
This suggests that detection and non-detection of polarization for a wide range of sub-mm, mm, and cm wavelengths can put a strong constraint on the grain size.

\begin{figure*}[htbp]
 \begin{center}
 \subfigure{
 \hspace{-100pt}
\includegraphics[width=100mm]{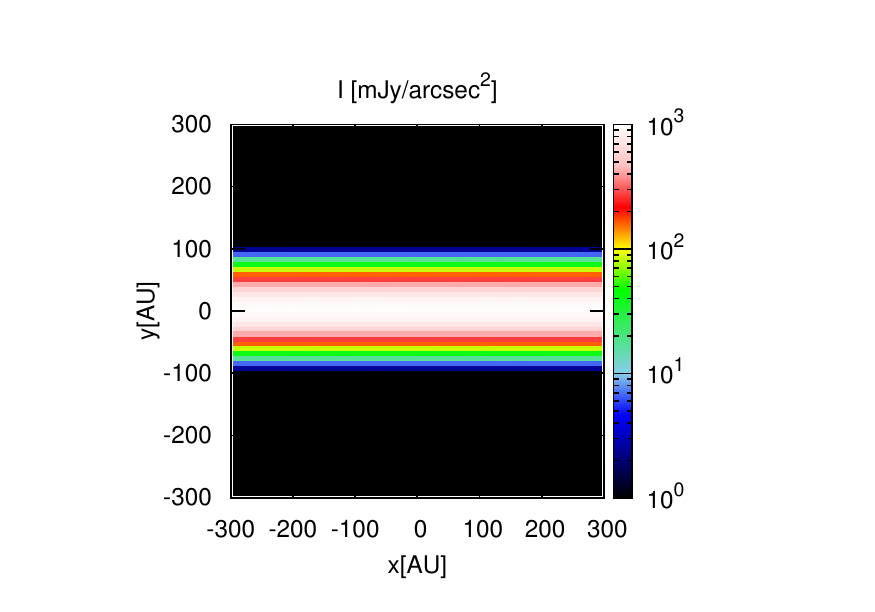}
}
 \subfigure{
 \hspace{-120pt}
\includegraphics[width=100mm]{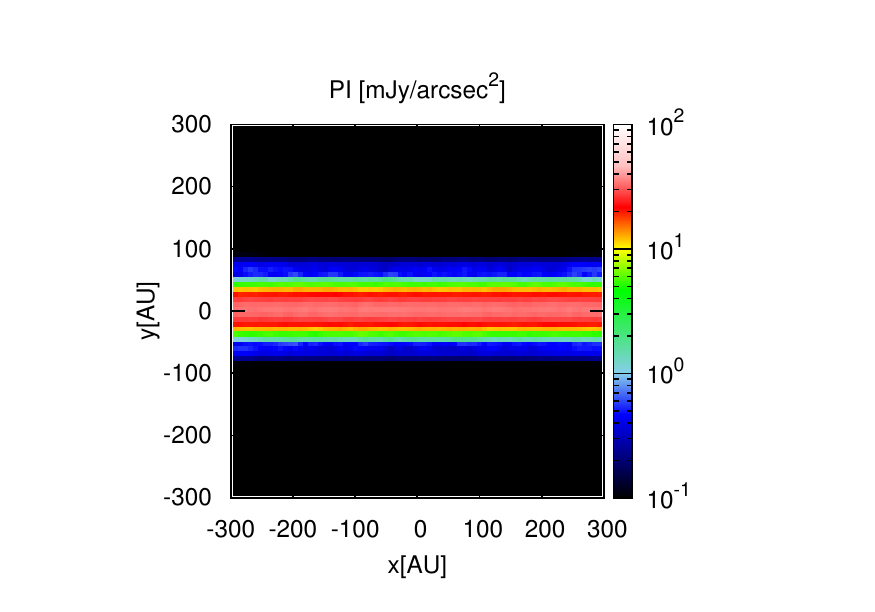}
}
 \subfigure{
  \hspace{-120pt}
\includegraphics[width=100mm]{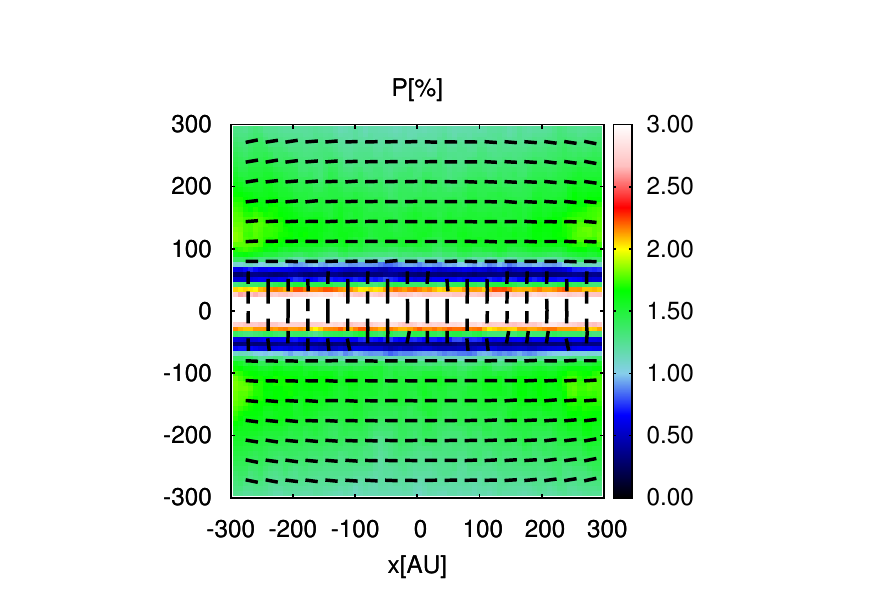}
 \hspace{-100pt}
}
 \end{center}
\caption{
The intensity, polarized intensity, and polarization degree with polarization vectors in the case of the tube model.
}
\label{fig:2D}
\end{figure*}

\subsection{Anisotropic radiation field}
\label{sec:toymodel}
The second condition for producing the polarization due to scattering is the asymmetry in the distribution of the light source.
In the case of sub-mm emission from protoplanetary disks, the light source is the thermal emission of dust particles.
Therefore, its spatial distribution should be anisotropic to produce polarized scattered light.
For example, if the incident radiation from one direction is stronger than that from the direction different by 90$^\circ$, the final scattering is partially polarized.
Such asymmetry or anisotropy may occur in protoplanetary disks showing ring-like or lopsided surface brightness \citep[e.g.,][]{Casassus13, vanderMarel13, Fukagawa13, Isella13, Perez14}.
In this section, we demonstrate that polarization due to self-scattering occurs in such anisotropic radiation field by using a simple toy model.
Note that the polarization due to scattering of anisotropic radiation field has been discussed in the context of E-mode polarization of the cosmic microwave background radiation \citep[e.g.,][]{Rees68, Seljak97, Kamionkowski97}.

Hereafter, we fix the maximum grain size and wavelengths to be $a_{\rm max}=100{\rm~\mu m}$ and $\lambda=870{\rm~\mu m}$, which is one of the best combinations of the efficient polarization, in order to investigate possibilities of detecting mm-wave polarization from protoplanetary disks.
Note that the calculated absorption and scattering opacities are $\kappa_{\rm abs}=0.51{\rm~cm^2/g}$ and $\kappa_{\rm sca}=1.22{\rm~cm^2/g}$.
We stress here that we consider the same size distribution of dust grains of $n(a)\propto a^{-3.5}$ in the following discussion.

Some protoplanetary disks show that radially Gaussian but azimuthally smooth surface brightness (see Section \ref{sec:radmc} for the protoplanetary disk models).
To investigate a possible polarization in that location, we consider a tube-like density distribution as a toy model.
The surface density has a Gaussian form in two directions ($y,z$) as
\begin{eqnarray}
 \rho_{\rm d}= \rho_{0} \exp\biggl[{-\frac{1}{2}\left(\frac {y}{y_{\rm 0}}\right)^{2}}\biggr] \exp\biggl[ -\frac{1}{2} \left(\frac{z}{z_{\rm 0}}\right)^{2}\biggr]. 
\end{eqnarray}
The adopted values are $y_{0}=z_{0}=27{\rm~AU}$ and $\rho_{0}=10^{-13}{\rm~g~cm^{-2}}$.
The observer is in the $z$ direction.
Radiative transfer calculations are performed with RADMC-3D\footnote{RADMC-3D is an open code of radiative transfer calculations developed by Cornelis Dullemond. The code is available online: http://www.ita.uni-heidelberg.de/\~{}dullemond/software/radmc-3d/}.
We confirm the validity of RADMC-3D by performing a benchmark test proposed by \citealt{Pinte09} (see Appendix).
Note that we include the multiple scattering and solve the Stokes parameters for each single scattering.

Figure \ref{fig:2D} shows that the intensity, polarized intensity, and polarization degree overlaid with polarization vectors \footnote{The polarization degree is higher at $x\sim-300{\rm~AU}$ and $x\sim300{\rm~AU}$, which is due to the boundary effects but it does not affect the discussion.
Also, the polarization vectors have different lengths at $y=0$, which is due to drawing program but this does not affect the discussion.
}.
The map of the polarization degree shows that it is peaked around $y=0$, which coincides with the peak of the surface density.
The polarization degree has two dips around the tube at $y\sim60{\rm~AU}$ and $y\sim-60{\rm~AU}$, then it goes up to a constant value in the outside.
In addition, polarization vectors are in the $y$ direction at the peak of the surface density, $y=0$, but it is in the $x$ direction in the outer side, $y\gtrsim70{\rm~AU}$.

\begin{figure}[htbp]
\centering
\includegraphics[width=80mm]{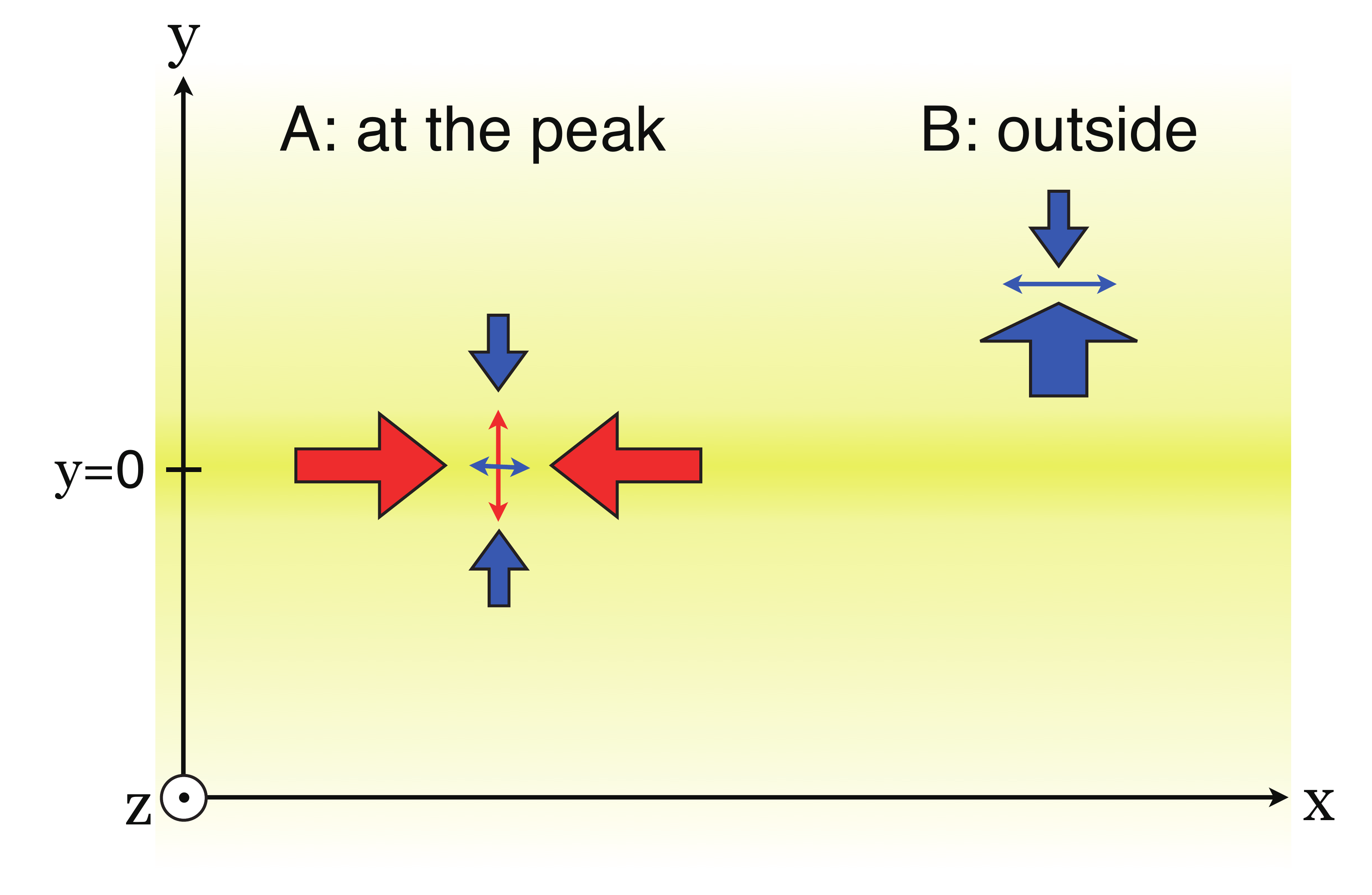}
\caption{
A schematic picture of the mechanism of the polarization due to scattering of anisotropic radiation field.
}
\label{fig:schematic}
\end{figure}
Figure \ref{fig:schematic} illustrates the polarization mechanism schematically.
First we consider the last scattering at point A (at $y = 0$) before the light reaches the observer in the $z$ direction.
If the incident photons propagate in the $x$ direction before the last scattering (the thick red arrows), they are polarized in the $y$ direction after the scattering (the thin red arrow). 
If the incident photons are in the $y$ direction (the thick blue arrows), they are polarized in the $x$ direction (the thin blue arrow). 
The former photons dominate over the latter because point A is at the peak surface density and therefore, the radiation is partially polarized in the $y$ direction.
The polarization degree depends on the degree of incident radiation anisotropy, or, in particular, the quadrupole moment.
Next, we consider the last scattering at point B in Figure \ref{fig:schematic} which corresponds to the periphery of the tube.
Most of the incident photons of the last scattering come from the tube region ($y\sim 0$) and therefore are in the $y$ direction.
Thus, the radiation from point B is polarized in the $x$ direction.

In summary, we have shown that the polarization degree is significantly high when the following conditions are met.
One is that dust grains have a maximum size of $a_{\rm max}\sim\lambda/2\pi$.
The other is that flux coming from two opposite directions are stronger than that from the $90^\circ$ different two directions (point A in Figure \ref{fig:schematic}) or that the radiation field has a strong gradient (point B in Figure \ref{fig:schematic}).

\section{Implications for polarization observations of protoplanetary disks}
\label{sec:radmc}
In this section, we investigate whether the polarization due to dust scattering is effective by using protoplanetary disk models.
In Section \ref{sec:ring}, we discuss the possibility of detection of the polarization using an axisymmetric ring-shaped disk.
In Section \ref{sec:lopsided}, we use lopsided disk models to calculate the polarization.

\begin{figure*}[htbp]
 \begin{center}
 \subfigure{
 \hspace{-100pt}
\includegraphics[width=100mm]{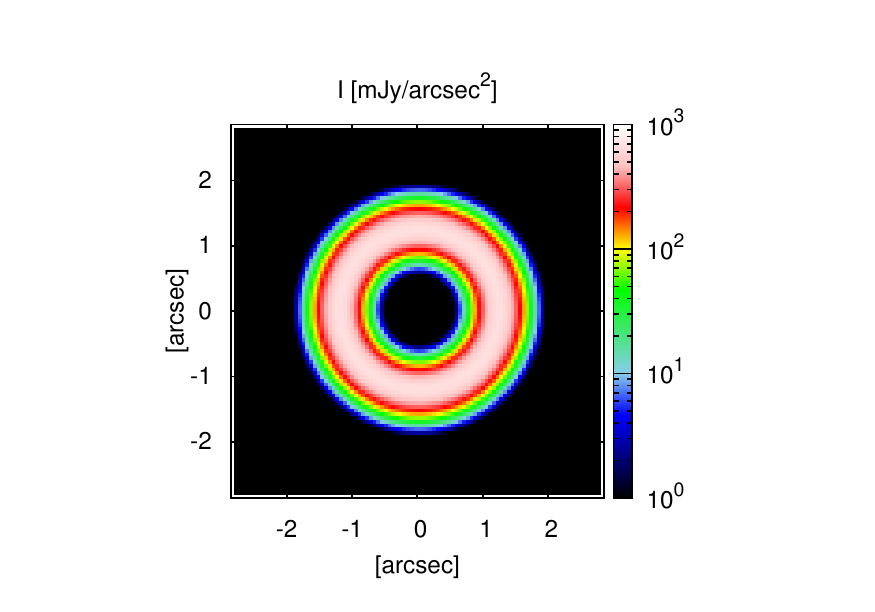}
}
 \subfigure{
 \hspace{-120pt}
\includegraphics[width=100mm]{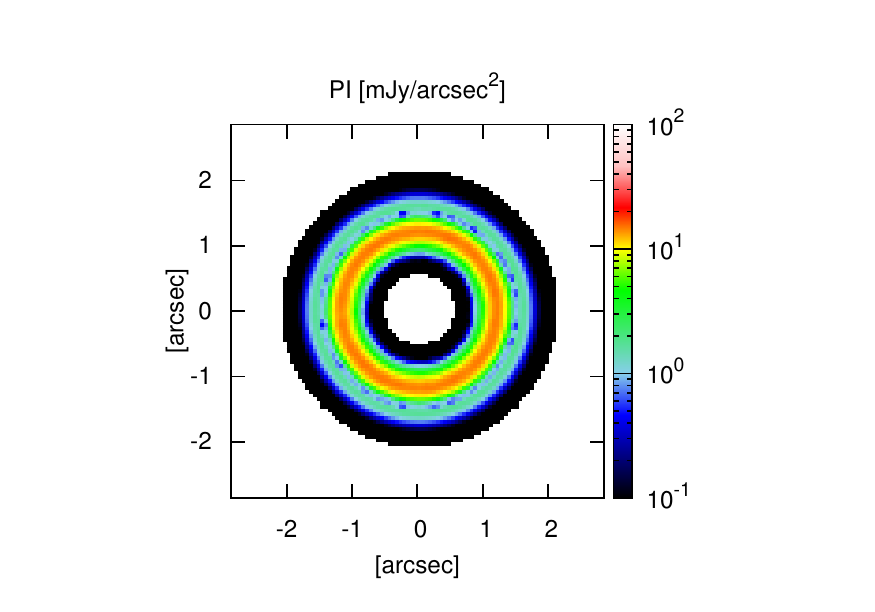}
}
 \subfigure{
  \hspace{-120pt}
\includegraphics[width=100mm]{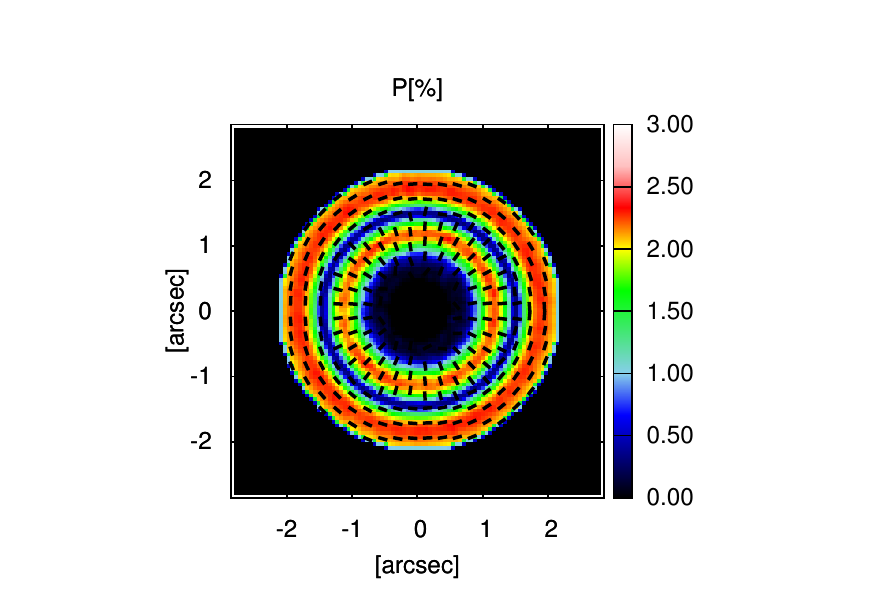}
 \hspace{-100pt}
}
 \end{center}
\caption{
The same as Figure \ref{fig:2D} but in the case of the ring-shaped protoplanetary disk.
}
\label{fig:3D}
\end{figure*}

\subsection{Polarization from a ring-shaped protoplanetary disk}
\label{sec:ring}
We assume a protoplanetary disk where the radial profile of dust continuum emission is described by Gaussian, which is realized in some observations \citep[e.g.,][]{Isella12}.
The grain density distribution is assumed to be 
\begin{eqnarray}
 \rho_{\rm d}= \Sigma_{0} \exp\biggl[{-\frac{1}{2}\left(\frac {r-r_{\rm d}}{w_{\rm d}}\right)^{2}}\biggr] \frac{1}{\sqrt{2\pi} h_{\rm g}}\exp\biggl[ -\frac{1}{2} \left(\frac{z}{h_{\rm g}}\right)^{2}\biggr],
\end{eqnarray}
where $r$ is the orbital radius. 
The mass of the central star is taken to be $1.9\thinspace M_\odot$.
The adopted values are $\Sigma_{0}=0.6{\rm~g~cm^{-2}}$, $r_{\rm d}=173$ AU, $w_{\rm d}=27$ AU, and $h_{\rm g}=19.8~{\rm AU}~(r/173{\rm~ AU})^{1.5}$, which corresponds to the isothermal disk of 36 K.
The dust mass of the disk is $5.0 \times 10^{-3}~M_\odot$.
We confirm that the results in the case of a power-law temperature distribution do not show any significant difference from the constant temperature adopted here (see Appendix \ref{sec:temperature} for more details).

These choices of parameters are motivated by recent results of the modeling of HD 142527 \citep{Muto15}, although we use different dust models.
In addition, the dust density is assumed to be zero if $R<70$ AU or if $R>300$ AU.
We assume that the target is at 140 pc and thus 1 arcsec = 140 AU.
Note that the optical depth at the peak is $\tau= \Sigma_0 \times \kappa_{\rm abs} = 0.6{\rm~g~cm^{-2}} \times 0.51 {\rm~cm^{2}~g^{-1}}= 0.31$.
Thus, this object is optically thin.

Figure \ref{fig:3D} shows the intensity, the polarized intensity, and the polarization degree overlaid with polarization vectors.
The polarization degree has a double-ring structure.
The polarization vectors are orientated to totally opposite directions in the two rings.
The vectors in the outer polarization ring are in the azimuthal direction.
This is because the background thermal emission has a strong radial gradient at the location of the outer polarized ring.
This corresponds to the point B in Figure \ref{fig:schematic}.
By contrast, the vectors are in the radial direction in the inner polarization ring.
This is due to the net flux from the azimuthal direction is larger than the net flux in the radial direction.
This corresponds to the point A in Figure \ref{fig:schematic}.
This double-ring pattern is a unique feature of the polarization due to dust scattering, and thus this will be a clue to distinguish the polarization mechanism.

\subsection{Polarization from lopsided protoplanetary disks}
\label{sec:lopsided}
We now calculate the expected polarization from a lopsided disk.
To mimic the lopsided disk structure observed with the ALMA observations \citep[e.g.,][]{Fukagawa13}, we further add an azimuthally Gaussian distribution \citep[e.g.,][]{Perez14} as follows:\begin{eqnarray}
 \rho_{\rm d}= \left(\Sigma_{\rm max}\exp\biggl[{-\frac{1}{2}\left(\frac {(\theta - \theta_{\rm d})}{\phi_{\rm d}}\right)^{2}}\biggr] + \Sigma_{\rm min}\right)\\
 \times \frac{1}{\sqrt{2\pi} h_{\rm g}}\exp\biggl[ -\frac{1}{2}    \left( \frac{z}{h_{\rm g}}\right)^{2}\biggr]  \exp\biggl[{-\frac{1}{2}\left(\frac {r - r_{\rm d}}{w_{\rm d}}\right)^{2}}\biggr].\nonumber
\end{eqnarray}

Since broad range of azimuthal contrast of dust continuum emission has been reported so far (1.5 for SAO206462; \citealt{Perez14}, 24 for HD 142527; \citealt{Fukagawa13}, 130 for IRS 48; \citealt{vanderMarel13}), we consider two cases: model A for low azimuthal contrast and model B for high azimuthal contrast.
In the model A, we use the model where the disk is entirely optically thin, while in the model B, the disk is partially optically thick, and therefore, we can also investigate how the difference of optical depth affects the polarization.

Specific parameters are as follows. 
In model A, the ratio of the surface densities between the top and the bottom is 4.
The adopted values are $\Sigma_{\rm max}=0.6{\rm~g~cm^{-2}}$, $\Sigma_{\rm min}=0.2{\rm~g~cm^{-2}}$, $r_{\rm d}=173$ AU, $w_{\rm d}=27$ AU, $h_{\rm g}=19.8~{\rm AU}~(r/173{\rm~ AU})^{1.5}$, $\theta_{\rm d}=0$, and $\phi_{\rm d}=\pi/3$. 
The dust mass of the disk is $3.7 \times 10^{-3}~M_\odot$.
The radial peak of the dust surface density is located at $r=r_d$, and it is maximized at $\theta=\theta_d$ and minimized at $\theta=\theta_d+\pi$.  We call the location of the former peak top peak and the latter bottom peak.
The optical depth at the top peak is $\tau= (\Sigma_{\rm max}+\Sigma_{\rm min}) \times \kappa_{\rm abs} = 0.8{\rm~g~cm^{-2}} \times 0.51 {\rm~cm^{2}~g^{-1}}= 0.41$ , and the optical depth at the bottom peak is  $\tau= 0.21{\rm~g~cm^{-2}} \times 0.51 {\rm~cm^{2}~g^{-1}} = 0.11$.
Thus, model A is showing totally optically thin case.

In model B, the ratio of the surface densities between the top and bottom is taken to be roughly 70.
This kind of the strong enhancement is realized in HD 142527, for example \citep{Muto15}.
The adopted values are $\Sigma_{\rm max}=3.3{\rm~g~cm^{-2}}$, $\Sigma_{\rm min}=0.046{\rm~g~cm^{-2}}$, and the other values are the same as model A.
The dust mass of the disk is $1.2 \times 10^{-2}~M_\odot$.
The difference between model A and B is the optical depth.
In model B, the optical depth at the top peak is $\tau= (\Sigma_{\rm max}+\Sigma_{\rm min}) \times \kappa_{\rm abs} = 3.3 {\rm~g~cm^{-2}} \times 0.51 {\rm~cm^{2}~g^{-1}}= 1.68$, which is optically thick.

\begin{figure*}[htbp]
\centering
 \subfigure{
 \hspace{-100pt}
\includegraphics[width=100mm]{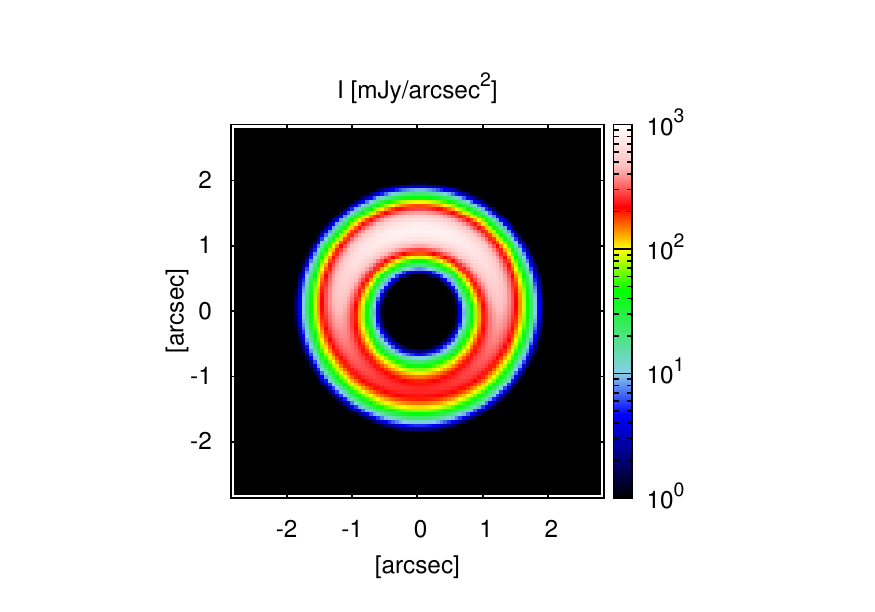}
}
 \subfigure{
 \hspace{-120pt}
\includegraphics[width=100mm]{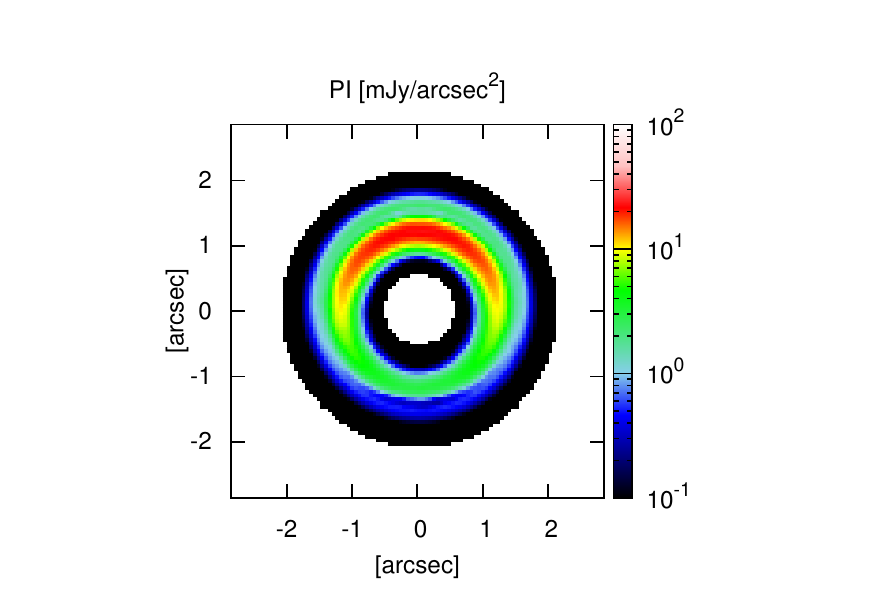}
}
 \subfigure{
  \hspace{-120pt}
\includegraphics[width=100mm]{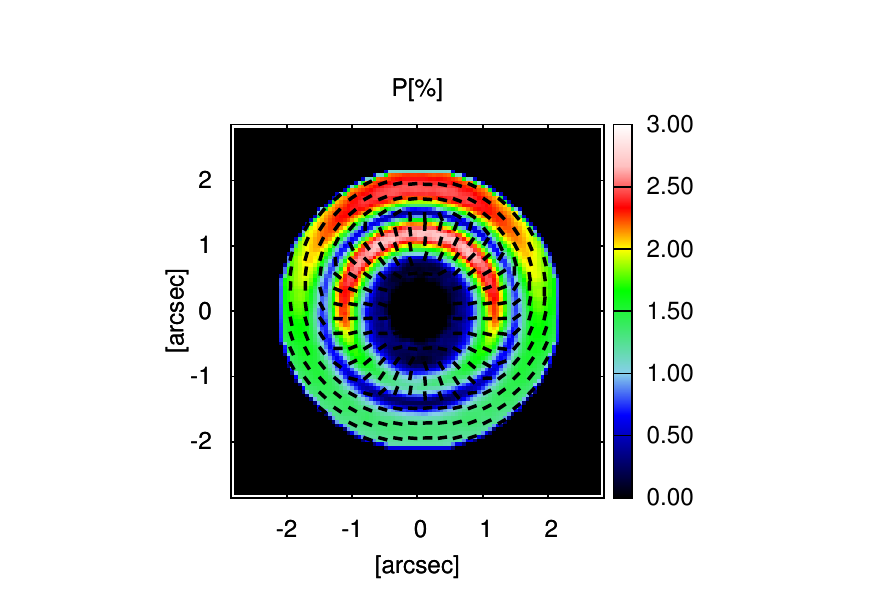}
 \hspace{-100pt}
}
\caption{
The same as Figure \ref{fig:2D} but for model A in the case of the lopsided protoplanetary disk.
The object is optically thin everywhere.
}
\label{fig:lopsidedA}
\end{figure*}

Figure \ref{fig:lopsidedA} shows the intensity, polarized intensity, and polarization degree with polarization vectors of the lopsided disk model A.
The polarization degree mainly shows a double-peaked structure.
The polarization degree is higher in the brighter region.
The polarization vectors are in the azimuthal direction in the outer ring, which corresponds to the point B in Figure \ref{fig:schematic}, and are in the radial direction in the inner ring, which corresponds to the point A in Figure \ref{fig:schematic}.
This feature is the same as  the ring model.

\begin{figure*}[htbp]
\centering
 \subfigure{
 \hspace{-100pt}
\includegraphics[width=100mm]{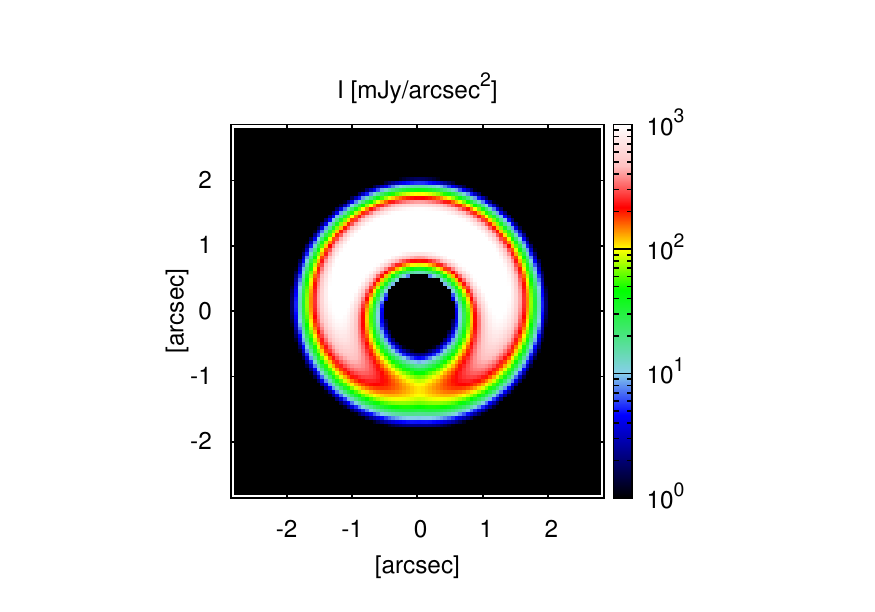}
}
 \subfigure{
 \hspace{-120pt}
\includegraphics[width=100mm]{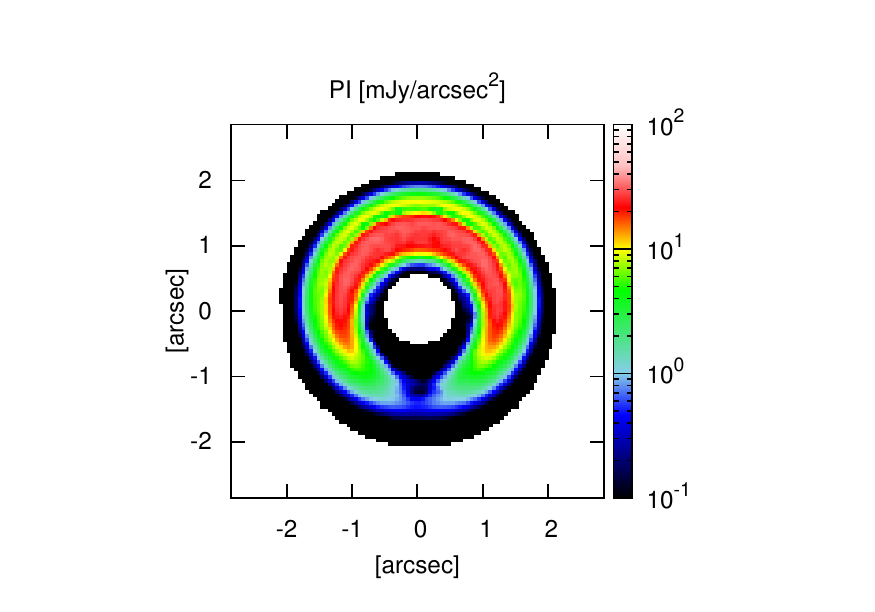}
}
 \subfigure{
  \hspace{-120pt}
\includegraphics[width=100mm]{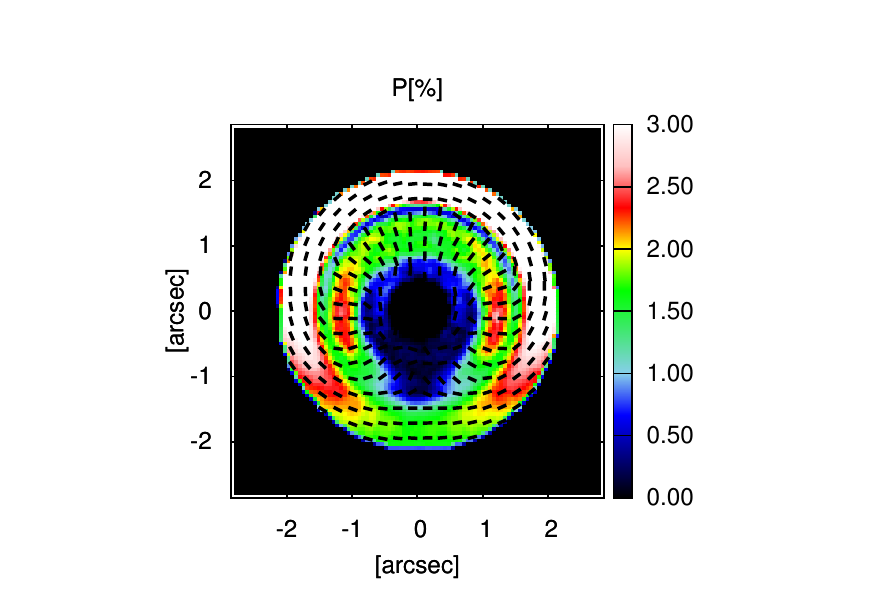}
 \hspace{-100pt}
}
\caption{
The same as Fig. \ref{fig:2D} but for model B.
At the top peak, the optical depth exceeds unity, which is realized in HD 142527, for example.
Note that the polarization degree is saturated at the outer ring in the right figure to emphasize the polarization degree of the inner ring.
The outer ring is unlikely to be detected because the polarized intensity is weak.
}
 \vspace{25pt}
\label{fig:lopsidedB}
\end{figure*}

Figure \ref{fig:lopsidedB} is the same as Figure \ref{fig:lopsidedA} but for model B.
The polarization degree maps of the two models are qualitatively similar except for that the polarization degree is lowered in the region of the intensity peak in model B.
This feature is realized due to the difference in the polarization degree between optically thick and thin regions. 
The polarization degree at a specific location of the disk is determined by the anisotropy of the incident radiation field before the last scattering, which corresponds to the region within the range of $\tau \lesssim 1$ around that location.
If the disk is optically thick, the geometric distance corresponding to $\tau=1$ is small and therefore the incident radiation field is nearly isotropic.
This is the reason why the polarization degree decreases at the optically thick region.
This reduction of the polarization degree results in two peaks of the polarization degree in the inner ring of model B, which correspond to the location of $\tau \sim 1$.
This case can be tested observationally with HD 142527 for example.

\section{Discussion}
\label{sec:discussion}
\subsection{Required spatial resolution and sensitivity to detect the polarization}
We have seen that the $\sim$ 2.5 $\%$ of polarization degree is expected from the dust continuum emission from protoplanetary disks in which the dust particles are grown to  sufficiently large size and are distributed in a ring-like or lopsided shape.
Here, we discuss the required spatial resolution to detect the polarization due to dust scattering with actual observations.

First, we consider the case that we observe a disk without spatially resolving the entire disk.
In the case of the ring disk model, no polarization is expected because the polarization is totally axisymmetric and the polarized flux is cancelled out.
In the case of the lopsided disks, by contrast, the net polarization pattern is expected even without spatially resolving the whole disk because the polarization pattern is not axisymmetric. 
However, the net polarization degree is $\sim$ 0.1 \% for model A and $\sim$ 0.06 \% for model B, respectively.
Such low polarization degree is unlikely to be detected with instruments to date.
Therefore, without spatially resolving the entire disk, no polarization is expected.

\begin{figure*}[htbp]
\centering
 \subfigure{
 \hspace{-100pt}
\includegraphics[width=100mm]{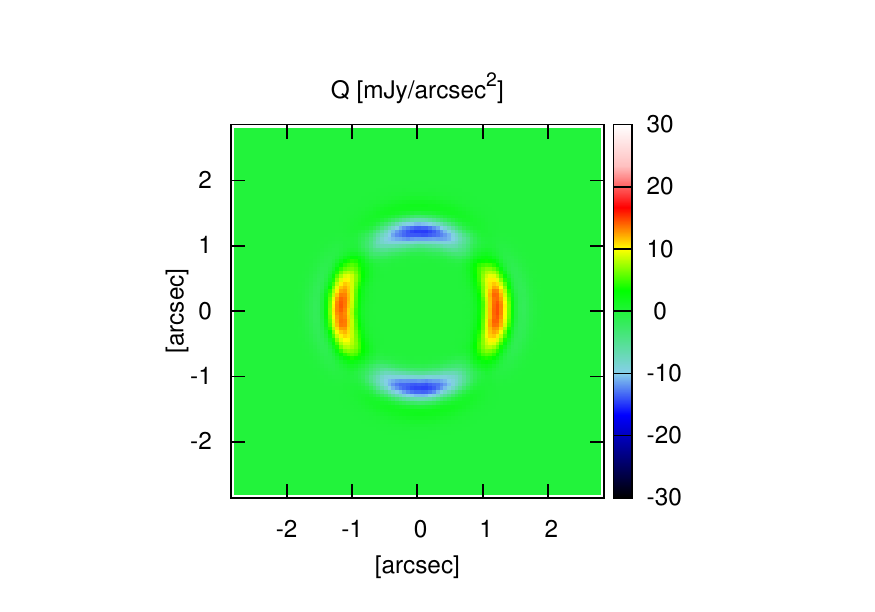}
}
 \subfigure{
 \hspace{-120pt}
\includegraphics[width=100mm]{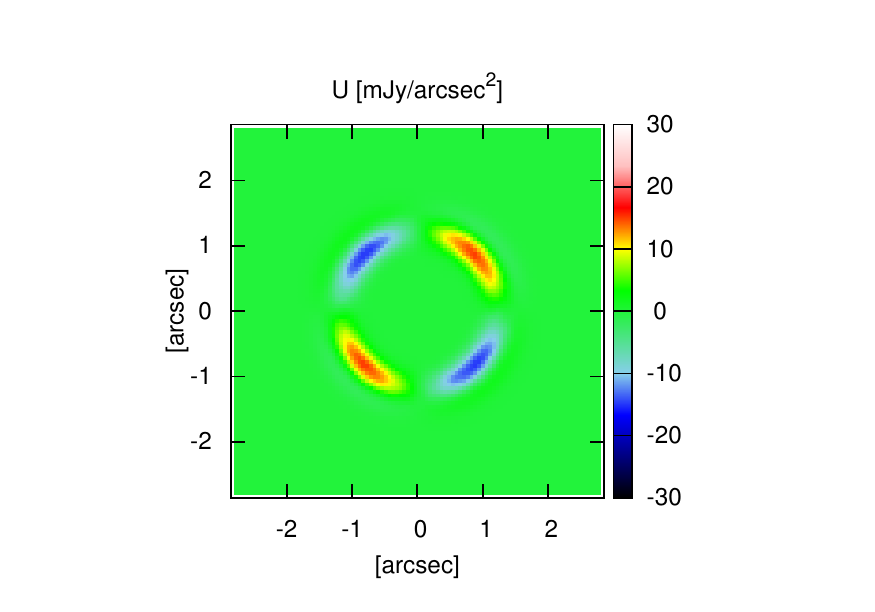}
}\\
 \subfigure{
 \hspace{-100pt}
\includegraphics[width=100mm]{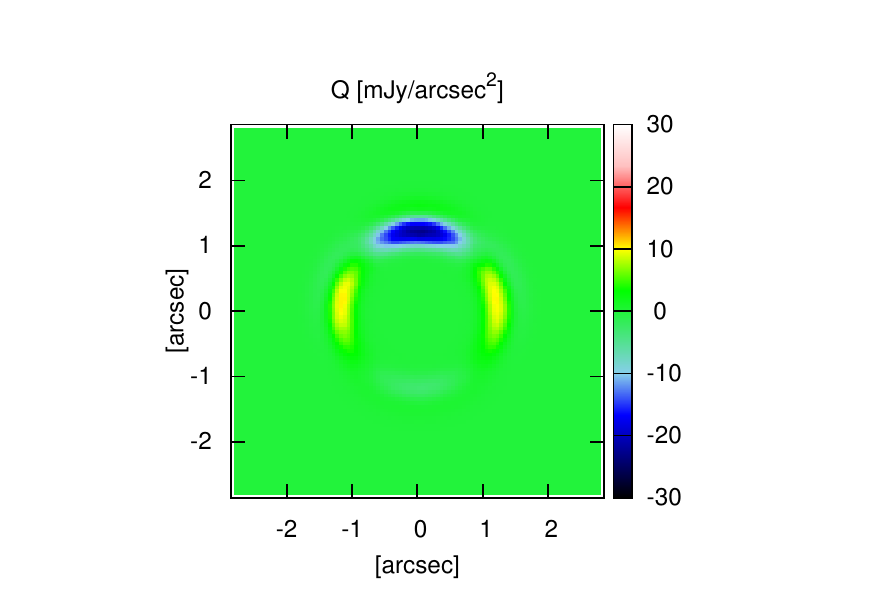}
}
 \subfigure{
 \hspace{-120pt}
\includegraphics[width=100mm]{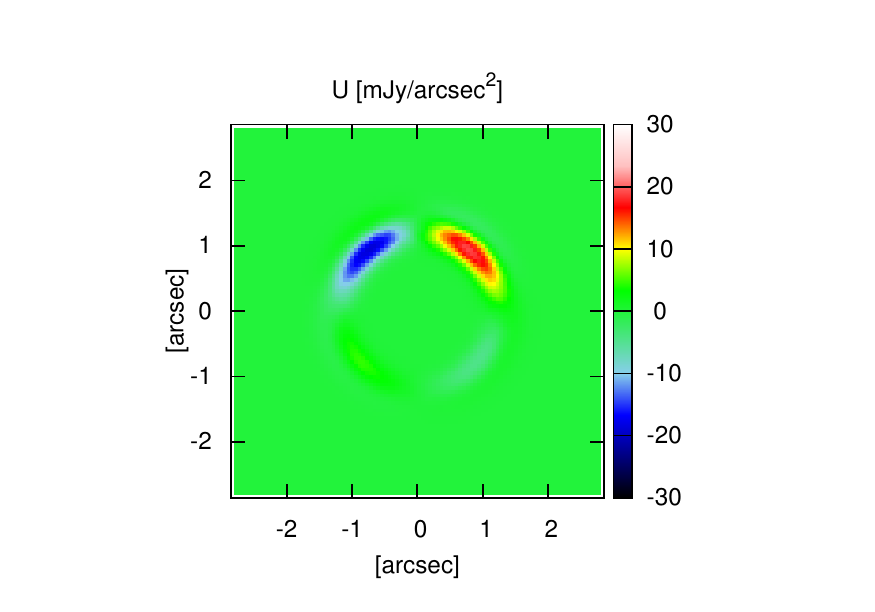}
}
\caption{
The Stokes Q and U maps.
The upper panels are in the case of the ring-shaped disk model and the lower panels are in the case of the model A of the lopsided disk model. 
}
\label{fig:QU}
\end{figure*}

Then, how small should be the spatial resolution to detect the polarization?
Figure \ref{fig:QU} shows the Stokes Q and U images of the ring disk model and the model A of the lopsided disk model.
In the case of the ring disk model, the Stokes Q and U maps show butterfly-like signatures.
To detect the polarization, we have to resolve the two positive peaks and the two negative peaks.
Especially in this case, the peaks have a length of $\sim1$ arcsec in the azimuthal direction and $\sim0.3$ arcsec in the radial direction.
Thus, the spatial resolution should be high enough to safely resolve a $0.3$ arcsec structure. 
Moreover, the dust continuum emission should be detected with good signal-to-noise ratio in order to detect the Stokes Q and U.
Since the polarization degree is at most $\sim 2~\%$, the Stokes I should be detected with at least $\sim 250\sigma$ level to detect Stokes Q and U with $5\sigma$ level.
Such requirements of high spatial resolution and good sensitivity can be met in near future ALMA observations.
The same discussion is applicable to the case of the lopsided disk model.

\subsection{Grain size constraints}
\label{sec:grainsizediscussion}

The polarization due to self-scattering is an independent way to constrain the grain size from the opacity index $\beta$ \citep[e.g.,][]{BeckwithSargent91}.
Although $\beta \approx 1$ at millimeter wavelengths has widely been regarded as an indication of dust growth, in the case of HD 142527 for example, the small $\beta$ can also be reproduced by dust model of $\sim3{\rm~\mu m}$ in size with an irregular shape \citep[e.g.,][]{Min05, Verhoeff11}.
However, the irregular shape grains are not responsible for the polarization because of their small scattering opacity at millimeter wavelengths due to its small size.
Therefore, polarization observations clearly determine whether the grains are large or small with irregularly shape.

In addition, the dust polarization is a unique and powerful way of determining the grain size at optically thick regions.
The spectral index at millimeter wavelengths has no information of dust grains if the emission is optically thick.
By contrast, although the polarization degree has a maximum value at $\tau\sim1$, the polarization can be still detectable even at the optically thick regions (see Figure \ref{fig:lopsidedB}).
Therefore, the polarization due to dust scattering is a promising indicator of grain size at shorter wavelengths or at the inner region of disks where the emission is expected to be optically thick.

\subsection{Other polarization mechanisms}
We have shown that the dust scattering can produce the polarization of the dust continuum emission.
This is a totally different mechanism from commonly used interpretation of the polarization as coming from the dust alignment.
Our models indicate some observational characteristics.
First of all, our mechanism requires the asymmetry of the radiation field, which comes from, for example, ring-like or lopsided distribution of continuum emission.
Secondly, the polarization pattern shows multiple-ring structures.  
Thirdly, the polarization vector is different by 90$^{\circ}$ at the two polarization rings.
These are unique features of the polarization due to dust scattering, and may be a clue to distinguish this mechanism from dust alignment.
For example, if the polarization of disks is due to the grain alignment with the magnetic field, there is a prediction that the polarization degree is $\sim 2 \%$ and the polarization vectors are directed in radial direction if the magnetic field is dominated by toroidal components \citep{ChoLazarian07}.
This characteristics is basically the same as the self-scattering mechanism.
Therefore, it is still difficult to distinguish between the two mechanisms and thus further and detailed comparisons should be necessary.

\subsection{Effects of dust settling}
We also investigate the effects of dust settling.
If dust grains are large enough to be decoupled from the gas, they settle to the midplane.
However, dust grains are stirred up by the gas turbulence.
Then, the vertical distribution of the dust grains are approximated by the Gaussian function with a dust scale height, $h_{\rm d}$, which is determined by the balance of settling and stirring.
The dust scale height is related to the gas scale height as $h_{\rm d}=\sqrt{\alpha/{\rm St}}h_{\rm g}$, where $\alpha$ is the viscous parameter, St is the Stokes number of the dust grains, and $h_{\rm g}$ is the scale height of the gas \citep[e.g.,][]{YoudinLithwick07}.
The Stokes number is given by ${\rm St}\sim a \rho_{\rm int}/\Sigma_{\rm gas}$, where $\rho_{\rm int}$ is the internal density of the dust grains and $\Sigma_{\rm gas}$ is the gas surface density.
The equation is valid if ${\rm St}\ll1$ and $\alpha<{\rm St}$.
In this paper, the grain size is taken to be $100{\rm~\mu m}$, which has a Stokes number of ${\rm St}\sim10^{-3}\times({\Sigma_{\rm gas}}/10 {\rm~g~cm^{-2}})^{-1}$.
If we assume the gas surface density of $\Sigma_{\rm gas}=10 {\rm~g~cm^{-2}}$, dust settling is effective if $\alpha < 10^{-3}$.
Therefore, the dust settling is important only if the turbulence of the gas disk is as weak as $\alpha\sim10^{-3}$.

\begin{figure*}[htbp]
\centering
 \subfigure{
 \hspace{-100pt}
\includegraphics[width=100mm]{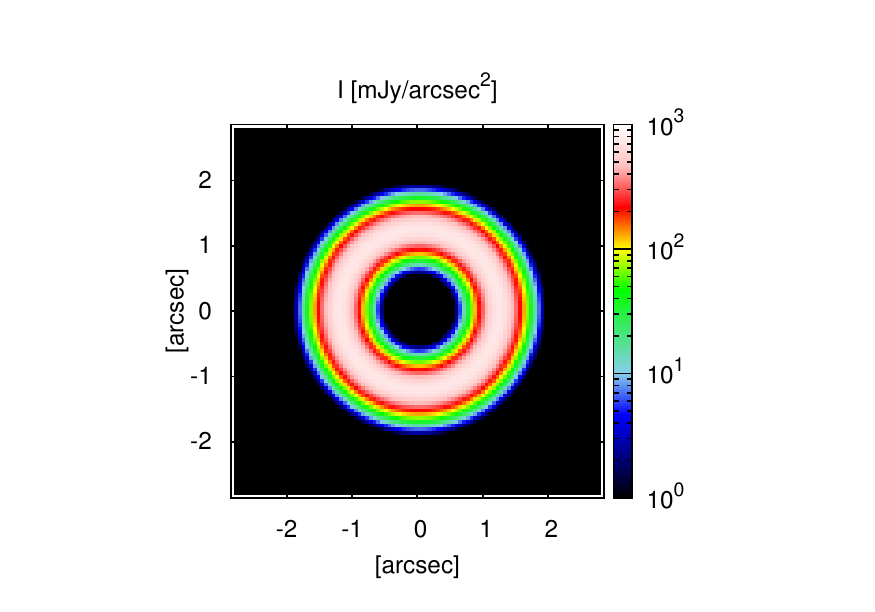}
}
 \subfigure{
 \hspace{-120pt}
\includegraphics[width=100mm]{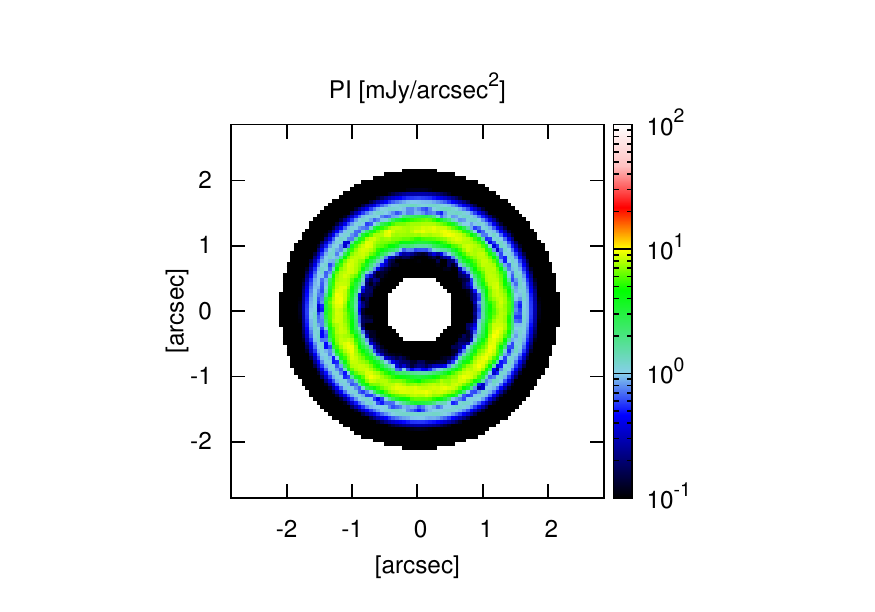}
}
 \subfigure{
  \hspace{-120pt}
\includegraphics[width=100mm]{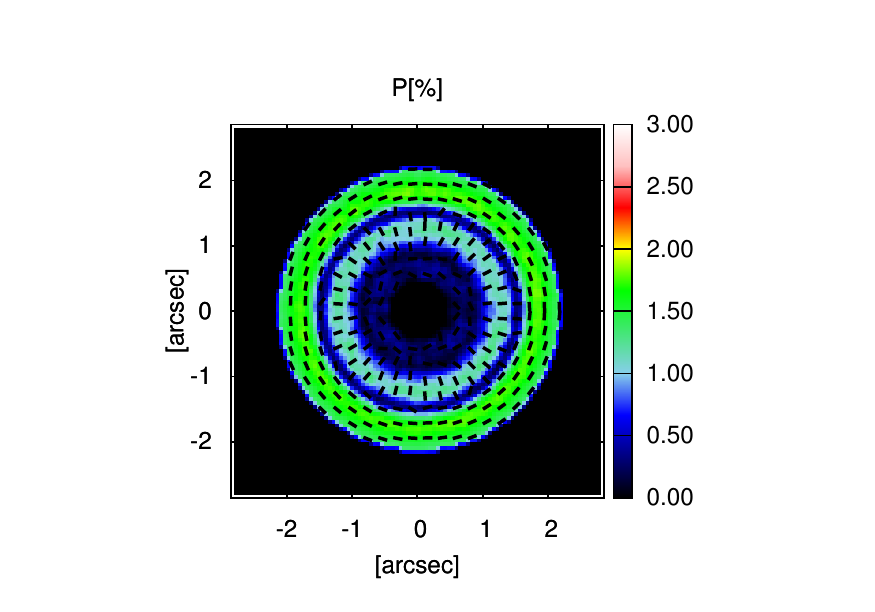}
 \hspace{-100pt}
}
\caption{
To check the effects of grain settling, we artificially reduce the scale height of the disk as $H=H_{\rm g} /3$.
This results in less polarization degree than the disk model without grain settling.
}
 \vspace{25pt}
\label{fig:settling}
\end{figure*}

Although the settling is not expected to occur in this case, it is important to investigate the effects of dust settling.
We did the same simulations discussed in Section \ref{sec:ring} except for the scale height, which we artificially reduce with a factor of 3.
Figure \ref{fig:settling} shows the results of the geometrically thin disk.
The polarization degree is 1.2 \% in the inner ring of the polarization map although it is 2.2 \% in the case of no settling (see Figure \ref{fig:3D}).
Thus, the dust settling reduces the polarization degree.

The physical reason is as follows.
Although the optical depth in the line of sight does not change if the scale height changes, the optical depth perpendicular to the line of sight increases because the dust density increases.
Thus, the incoming flux to be scattered by dust grains becomes more isotropic with decreasing scale height.
This is the reason of the less polarization degree in the case of the thinner disk.

\section{Summary and outlook}
\label{sec:conclusions}
We have presented a new mechanism of millimeter-wave polarization of protoplanetary disks, which is due to dust scattering of anisotropic radiation field, resulting from a ring-like or lopsided dust continuum emission.
Our main findings are as follows:

\begin{itemize}

\item{
The polarization degree due to dust scattering is the highest at $a_{\rm max}\sim \lambda/2\pi$, where $a_{\rm max}$ is the maximum grain size with a size distribution of $n(a)\propto a^{-3.5}$ and $\lambda$ is the observing wavelengths.
This is because dust grains should be sufficiently large to have a high albedo but should be sufficiently small to be in the Rayleigh scattering regime in order to make the emission polarized.
The grain sizes that contribute most to the polarization are shown in Figure \ref{fig:window} and Table \ref{table:grain}.
The strong dependence of the dust size on polarization implies that we can use the polarization signature as an indicator of dust size in protoplanetary disk.
}

\item{
Even without a specific light source, the continuum emission is polarized due to self-scattering if the dust emission exhibits quadrupole anisotropy around each location of the disk (see Figure \ref{fig:schematic}).
To demonstrate the self-scattering, we used a tube-like density distribution and performed radiative transfer calculations.
As a result, the polarization vectors are perpendicular to the tube at the peak due to the strong flux from the tube.
By contrast, the polarization vectors are parallel to the tube at the periphery of the tube because the net flux is in the direction of the gradient of the surface brightness.
}

\item{
In the case of a ring disk which has a radial surface density distribution expressed by a Gaussian function, the polarization is expected due to the dust scattering.
The polarization map shows a double-ring structure.
The polarization vectors are in the azimuthal direction in the outer disk and in the radial direction in the inner disk.
The polarization degree has a peak value just inside the peak of the Stokes I emission.
The polarization degree is up to $\sim 2.5\%$ in the specific case.
}

\item{
We also performed radiative transfer calculations on lopsided protoplanetary disks.
We use two models: totally optically thin model (model A) and the partially optically thick model (model B).
The model B is motivated by recent millimeter-wave observations on HD 142527.
They also show the double-ring structure of the polarization.
In the case of the model B, the emission is polarized even at the optically thick region.
In addition, the peak polarization is $2.5\%$, which is likely to be detected with ALMA observations.
The polarization degree decreases if the dust is settled because the optical depth increases with decreasing dust scale height, which makes the incoming flux more isotropic.
}

\end{itemize}

We have shown the results of calculations mainly in the case of $a_{\max}=100~\mu$m and $\lambda=870~{\mu}$m as an illustrative example.
This set of parameters is one of the most efficient case for producing the polarized light.
The maximum grain size that is responsible for producing polarization is given by $a_{\max}\sim \lambda/2\pi$ (see also Figure \ref{fig:window}) so we expect that we can apply our results to observations at different wavelengths, but further study should be necessary for observations at different wavelengths.
In addition, we have assumed that the disk is face-on for simplicity.
Since the polarization degree depends on the scattering angle, which is determined by the disk inclination and the position angle of the major axis of the disk, further parameter study is necessary to apply our results to actual observations.

The polarization due to scattering has a potential to determine the porosity of dust grains.
\citet{Kataoka14} have shown that to constrain the porosity of dust aggregates, we need both information of the absorption and scattering opacities.
The continuum and polarized intensity have the information of both opacities.
Therefore, the proposed mechanism should be further tested with including the porosity to constrain the porosity of dust aggregates in protoplanetary disks.
This is a good observational test on the planetesimal formation scenario via fluffy dust aggregates \citep[e.g.,][]{Kataoka13b}.

As discussed in Section \ref{sec:grainsizediscussion}, the polarization is a powerful technique to constrain the grain size compared with the opacity index.
This work suggests that spatially resolved polarization observations toward protoplanetary disks opens a new window on grain size constraints in protoplanetary disks.

\acknowledgments
We thank Satoshi Okuzumi, Hidekazu Tanaka, Ryo Tazaki, Taku Takeuchi, and Luca Ricci for fruitful discussions.
A.K. is financially supported by JSPS through Research Fellowship for Young Scientists ($24\cdot2120$) and by JSPS Postdoctral Fellowship for Study Abroad ($27\cdot213$)
This work is supported by MEXT KAKENHI No. 23103004 and No. 26800106 and by JSPS KAKENHI Grant Number 15K17606. 

\bibliography{cite}

\clearpage
\appendix
\section{Scattering matrix}
\label{sec:app:matrix}
The scattering properties of dust grains are described by the scattering matrix $Z_{ij}$.
We briefly review the scattering matrix to understand which matrix element determines the polarization due to scattering.
Let us consider the case when an incident light, which has Stokes parameters $(I_{i}, Q_{i}, U_{i}, V_{i})$, is scattered by a single dust grain.
We assume that the dust grains are spherical grains, and then the independent elements are reduced to 4 elements (see \citealt[e.g.,][]{BohrenHuffman83}).
Choosing certain observational axes, the Stokes parameters of the scattered light $(I_{s}, Q_{s}, U_{s}, V_{s})$ are given by
\begin{equation}
\left(
\begin{array}{c}
I_s\\
Q_s\\
U_s\\
V_s
\end{array}
\right)
=
\frac{m_{\rm grain}}{d^2}
\left(
\begin{array}{cccc}
Z_{11}&Z_{12}&0&0\\
Z_{12}&Z_{11}&0&0\\
0&0&Z_{33}&Z_{34}\\
0&0&-Z_{34}&Z_{33}
\end{array}
\right)
\left(
\begin{array}{c}
I_i\\
Q_i\\
U_i\\
V_i
\end{array}
\right),
\label{eq:matrix}
\end{equation}
where $d$ is the length from the observer, and $m_{\rm grain}$ the grain mass.
We use the scattering matrix of $Z_{ij}$, which follows the notations of the RADMC-3D.
The traditional notation $S_{ij}$ \citep{BohrenHuffman83} is related as $Z_{ij}=S_{ij}/(k^2 m_{\rm grain})$, where $k$ is the wave number.
Note that $Z_{ij}$ is a function of scattering angle and dust size.

\label{sec:app:zeff}
We derive the effective scattering matrix with a power-law size distribution of dust grains.
To address the dependence of the scattering matrix on the grain size $a$ and the scattering angle $\theta$, we write $Z_{ij}=Z_{ij}(a,\theta)$.
 
The total scattering opacity $\kappa_{\rm sca}(a)$ is given by the integration of $Z_{11}(a,\theta)$ over the solid angle as
\begin{equation}
\kappa_{\rm sca}(a) = \oint Z_{11}(a,\theta) d\Omega.
\label{eq:ksca}
\end{equation}

If we consider a size distribution of dust grains with $n(a)\propto a^{-q}$, the total scattering mass opacity $\kappa_{\rm sca,total}$ is given by
\begin{eqnarray}
\kappa_{\rm sca,tot} 
&=& \frac{1}{\rho_{\rm d}}\int^{a_{\rm max}}_{a_{\rm min}} \kappa_{\rm sca}(a) m_{\rm grain}n(a)da \nonumber \\
&=& \int^{a_{\rm max}}_{a_{\rm min}} \kappa_{\rm sca}(a)  a^{3-q}da \big/ \int^{a_{\rm max}}_{a_{\rm min}}  a^{3-q}da.\nonumber\\
\end{eqnarray}
Substituting Eq.(\ref{eq:ksca}), we obtain
\begin{eqnarray}
\kappa_{\rm sca,tot} &=& \int^{a_{\rm max}}_{a_{\rm min}} \left( \oint Z_{11}(a,\theta) d\Omega \right)  a^{3-q}da \big/ \int^{a_{\rm max}}_{a_{\rm min}}  a^{3-q}da \nonumber\\
 &=&  \oint\left( \int^{a_{\rm max}}_{a_{\rm min}} Z_{11}(a,\theta) a^{3-q}da \big/ \int^{a_{\rm max}}_{a_{\rm min}}  a^{3-q}da \right)d\Omega  \nonumber\\
 &=& \oint Z_{11, {\rm eff}}(\theta) d\Omega,
\end{eqnarray}
where 
\begin{eqnarray}
Z_{11, {\rm eff}}(\theta)&=&\int^{a_{\rm max}}_{a_{\rm min}} Z_{11} (a,\theta)a^{3-q}da \big/ \int^{a_{\rm max}}_{a_{\rm min}}  a^{3-q}da.
\end{eqnarray} 
We use Mie theory to calculate the effective phase function $Z_{11,{\rm eff}}(\theta)$ in this paper.
The other scattering matrix elements $Z_{ij, {\rm eff}}(\theta)$ is derived as the same way.

\label{sec:app:polarization}
If the incident light is unpolarized, that is $Q_i=U_i=V_i=0$, the Stokes parameters of the scattered light can be obtained from Eq.(\ref{eq:matrix}) as
\begin{equation}
I_s=\frac{m_{\rm grain}}{d^2}Z_{11} I_i,
\end{equation}
\begin{equation}
Q_s=\frac{m_{\rm grain}}{d^2}Z_{12} I_i,
\end{equation}
$U_s=0$, and $V_s=0$.
Thus, the degree of polarization is given by
\begin{equation}
P=\frac{ \sqrt{Q_{s}^2 + U_{s}^2}}{I_s} = \left|\frac{Z_{12}}{Z_{11}}\right|.
\end{equation}
Therefore, the ratio of $Z_{12} $ to ${Z_{11}}$ represents the degree of polarization.

\section{Temperature structure}
\label{sec:temperature}

\begin{figure*}[htbp]
\centering
 \subfigure{
 \hspace{-100pt}
\includegraphics[width=100mm]{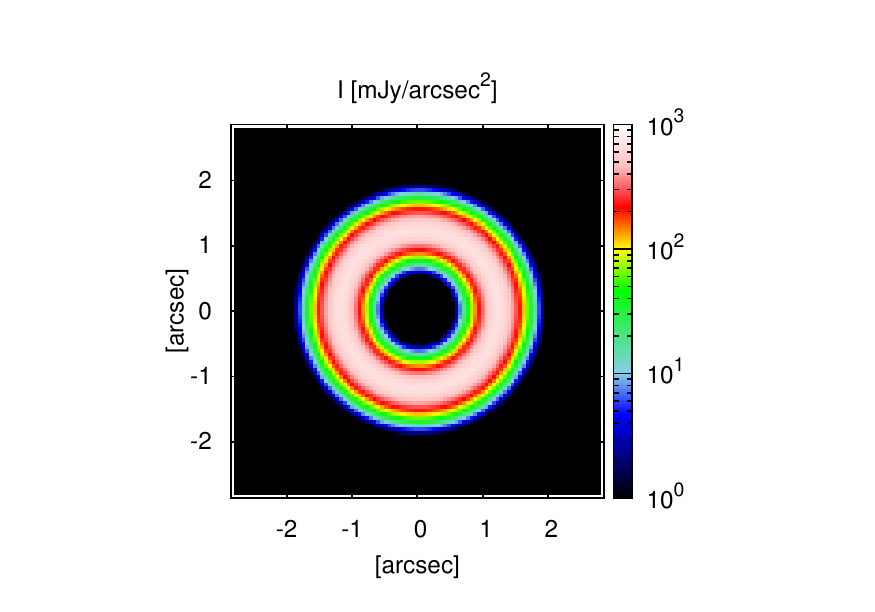}
}
 \subfigure{
 \hspace{-120pt}
\includegraphics[width=100mm]{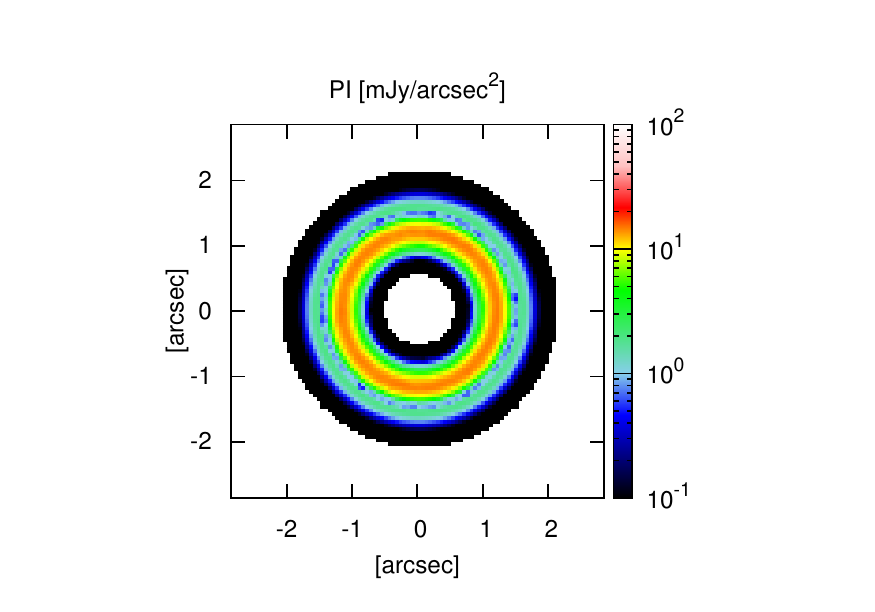}
}
 \subfigure{
  \hspace{-120pt}
\includegraphics[width=100mm]{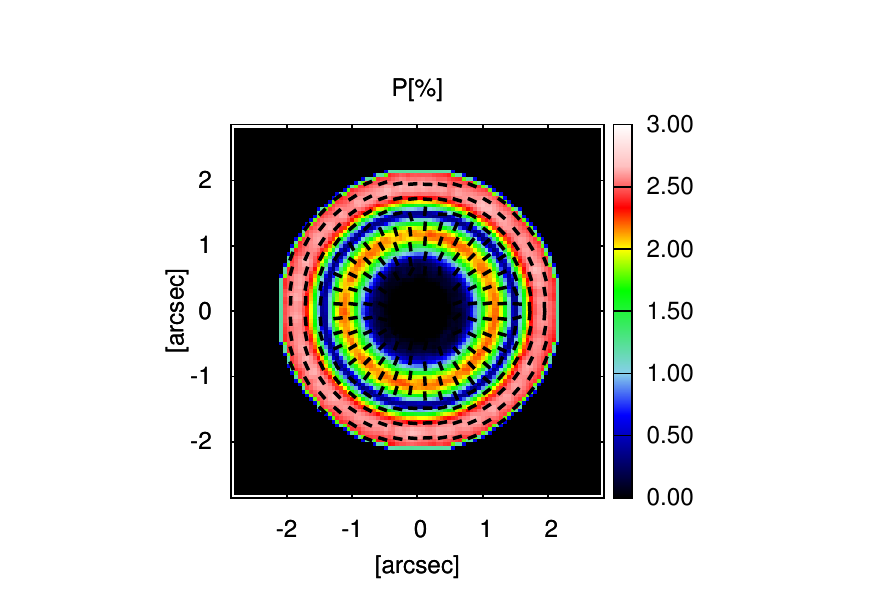}
 \hspace{-100pt}
}
\caption{
The same as Fig. \ref{fig:2D} but in the case of the power-law temperature distribution as $T=36 {\rm~K} \times (R/173{\rm AU})^{-0.5}$.
}
 \vspace{25pt}
\label{fig:tempstructure}
\end{figure*}

To check the validity of the assumption of the constant temperature throughout the disk, we change the temperature distribution as $T=36 {\rm~K} \times (R/173{\rm AU})^{-0.5}$.
Figure \ref{fig:tempstructure} shows the results in the case of the power-law temperature distribution.
We do not see significant change in the polarization degree.
Thus, the assumption of the constant temperature does not significantly affect the results in the case of ring-like disks.

\clearpage

\section{Benchmark test of RADMC-3D}
To test the reliability of the radiative transfer code RADMC-3D, we perform benchmark tests especially focusing on the polarization.
We use a sophisticated numerical model proposed by \citealt{Pinte09} and compared the results with other codes, which are MCFOST \citep{Pinte06}, MCMAX \citep{Min09}, TORUS \citep{Harries00}, and Pinball \citep{WatsonHenney01}\footnote{The results of the other codes are obtained from the webpate: http://ipag.osug.fr/\~{}pintec/benchmark/}.

\subsection{The problem and numerical settings}

The problem is set up as follows.
The dust particles of the disk has a surface density profile of $\Sigma_{d}=\Sigma_0 (r/r_0)^{-1.5}$ with $r_0=100{\rm~AU}$.
$\Sigma_0$ is determined by the disk mass and is discussed later.
The vertical profile is $\rho_{d}= \rho_0 \exp(-z^2/2h(r)^2)$.
The scale height is parametrized as $h(r)=h_0(r/r_0)^{1.125}$ with $h_0=10 {\rm~AU}$.
The inner and outer edges of the disk are $r_{\rm in}=0.1 {\rm~AU}$ and $r_{\rm out}=400 {\rm~AU}$.
The dust opacity is $1 {\rm~\mu m}$ silicate particles with a particle density of $3.5{\rm ~g~cm^{-3}}$, which is different from the opacity model of the main discussion of this paper.
The refractive index is taken from \citealt{WeingartnerDraine01}.

The only parameter is the disk mass $M_{\rm disk}$, and we use the disk mass of $M_{\rm disk}= 3\times10^{-5} ~M_{\odot}$ to compare the results of the polarization with the other radiative transfer codes.
The surface density at 100 AU, $\Sigma_0$, is calculated from $M_{\rm disk}=4\pi \Sigma_{0}(100 {\rm~AU})^{1.5}[r_{\rm out}^{0.5} - r_{\rm in}^{0.5}]$.
To avoid photons from being trapped at optically thick inner region of the disk, we truncate the disk with a radius of $3 {\rm~AU}$.
We use spherical coordinate $(r,\theta,\phi)$.
The radial coordinate $r$ is logarithmically divided into 512 grids between 0.1 AU and 1024 AU.
The polar angle $\theta$ is linearly divided into 512 grids in $0+\epsilon<\theta <\pi-\epsilon$, where $\epsilon=0.01$, which is set to avoid a numerical error exactly on the polar axis.
The azimuthal angle $\phi$ is linearly divided into 512 grids in $0<\phi<2\pi$.
First, we solve the temperature structure with Monte Carlo radiative transfer simulations and then simulate the intensity and polarization with $10^{9}$ photons.
The calculation takes 60 hours with a CPU with 3.20 GHz.
The viewing grid is set to be 251 pixel $\times$ 251 pixel for the square of 900 AU $\times$ 900 AU; the pixel scale is 3.6 AU.

\subsection{Results}
\begin{figure*}[htbp]
\centering
 \subfigure{
    \includegraphics[width=75mm]{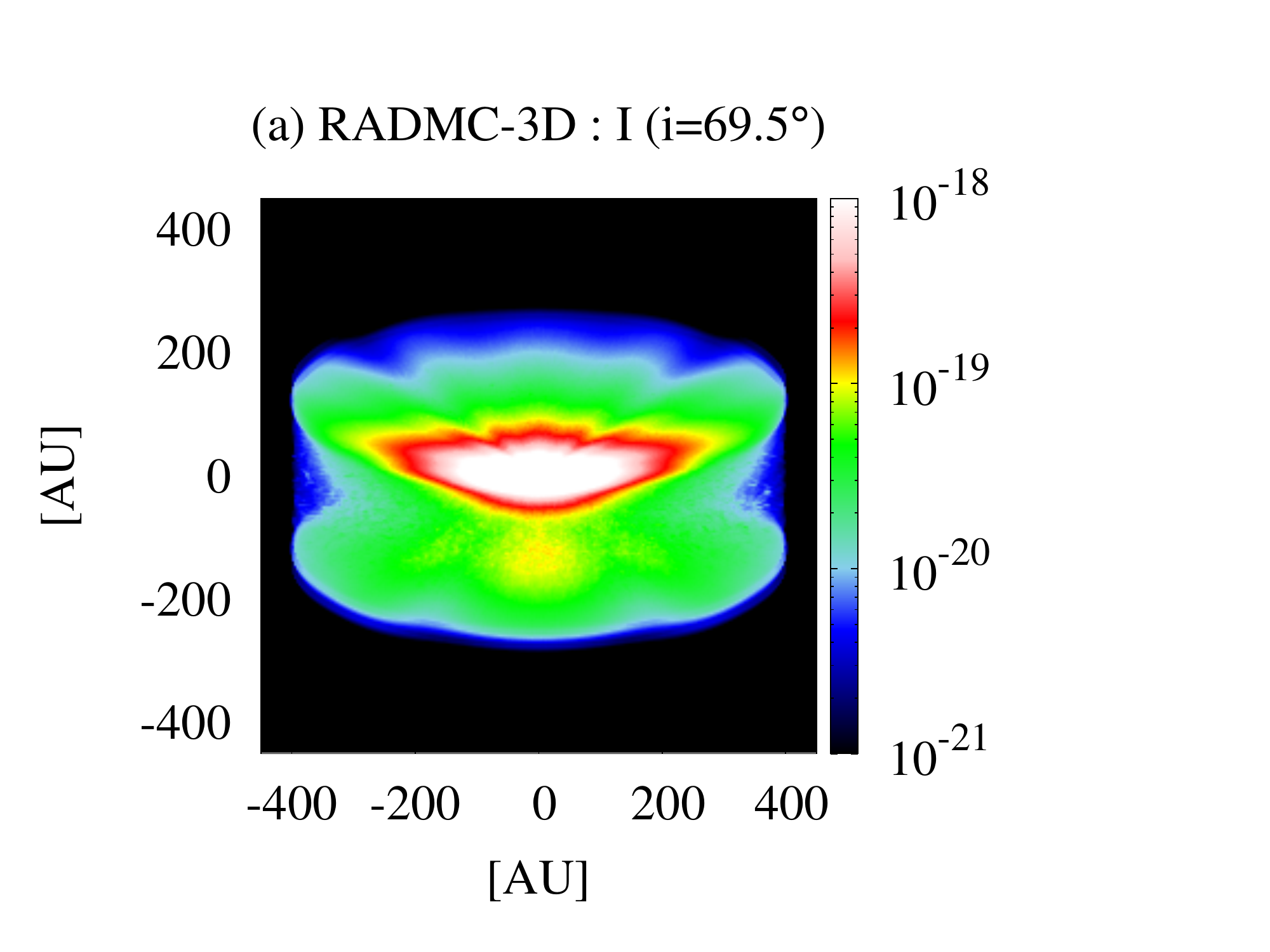}
  }
\subfigure{
  \includegraphics[width=75mm]{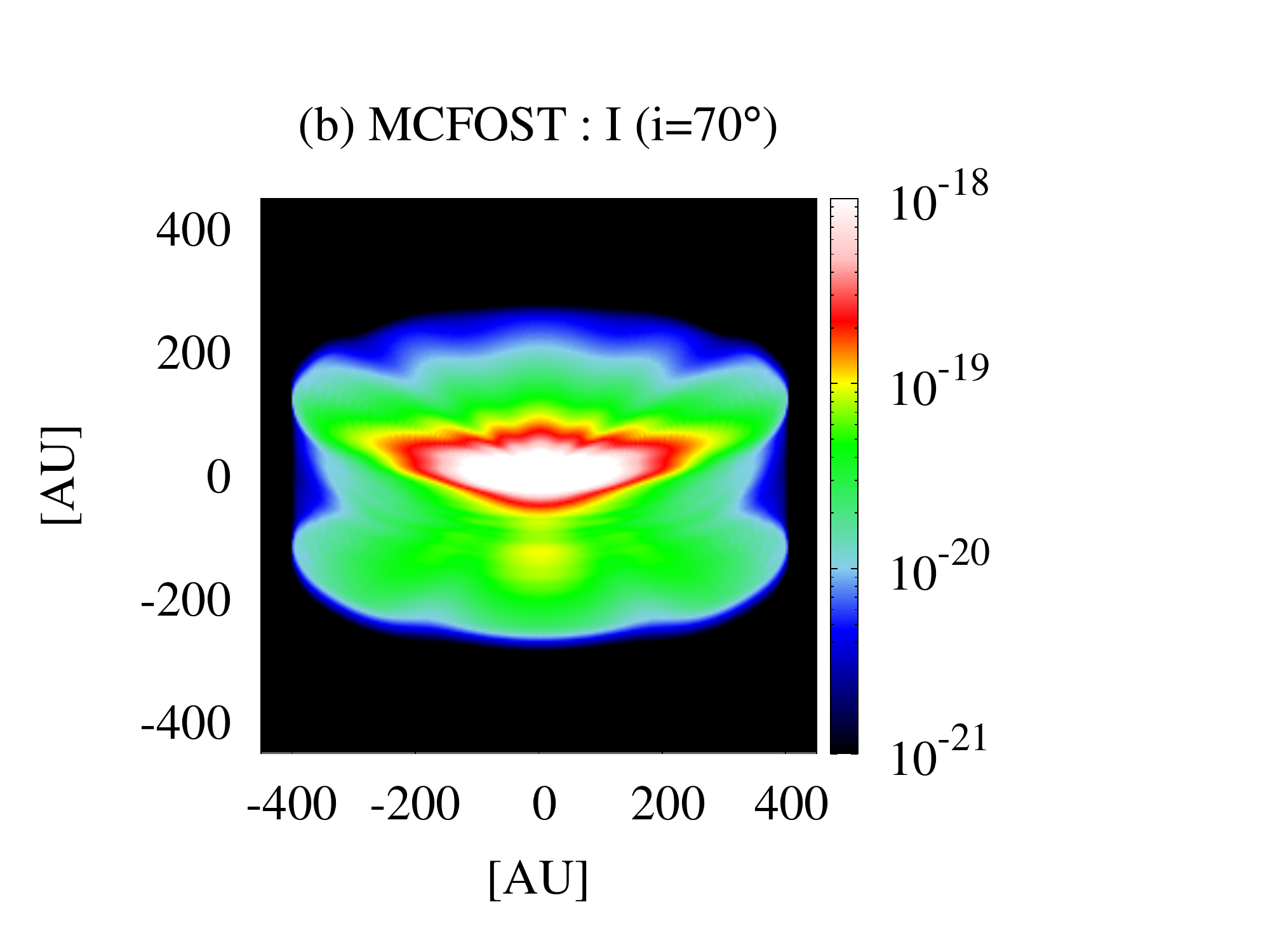}
  }\\
   \vspace{-30pt}
 \subfigure{
    \includegraphics[width=75mm]{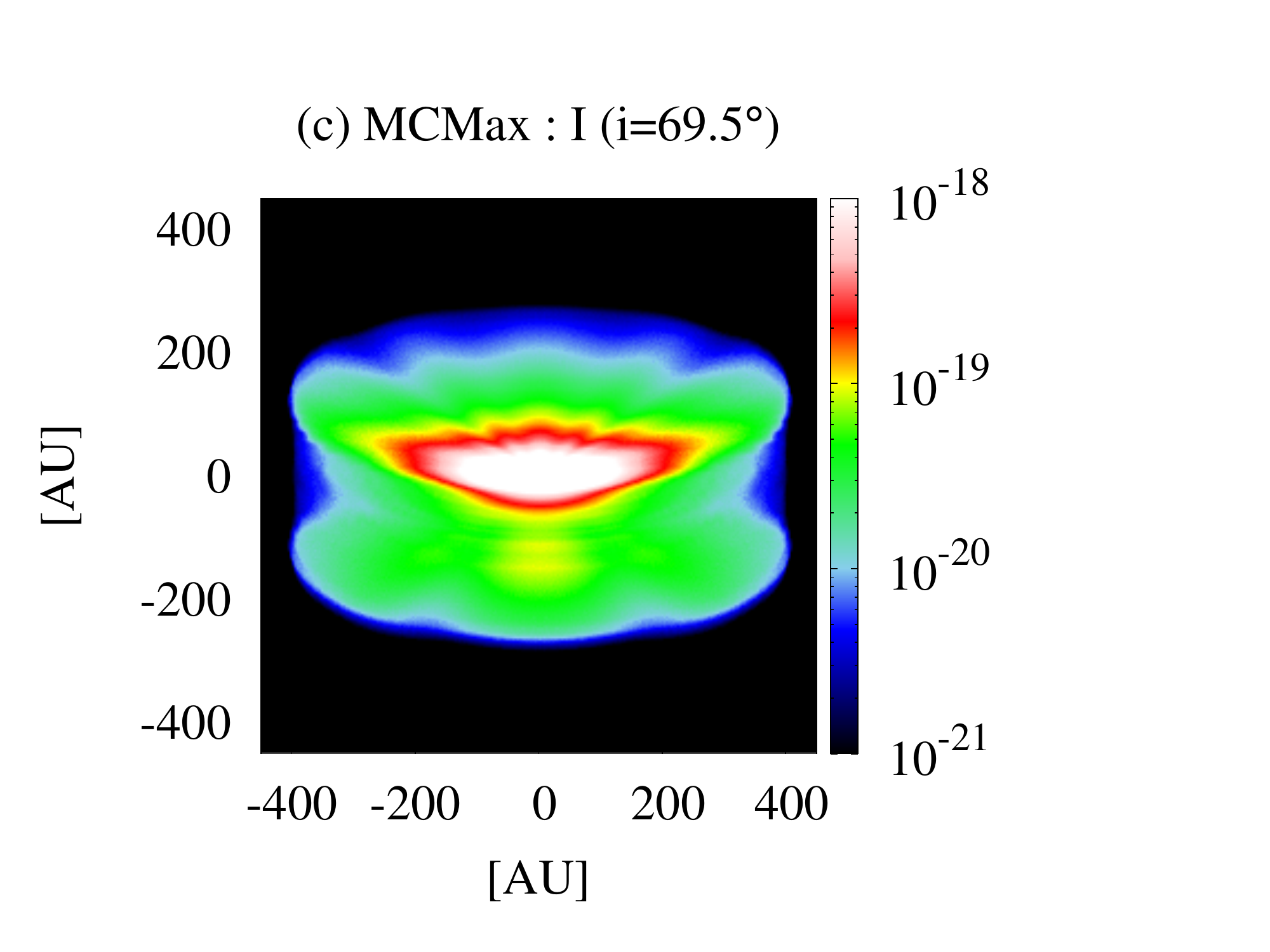}
  }
 \subfigure{
    \includegraphics[width=75mm]{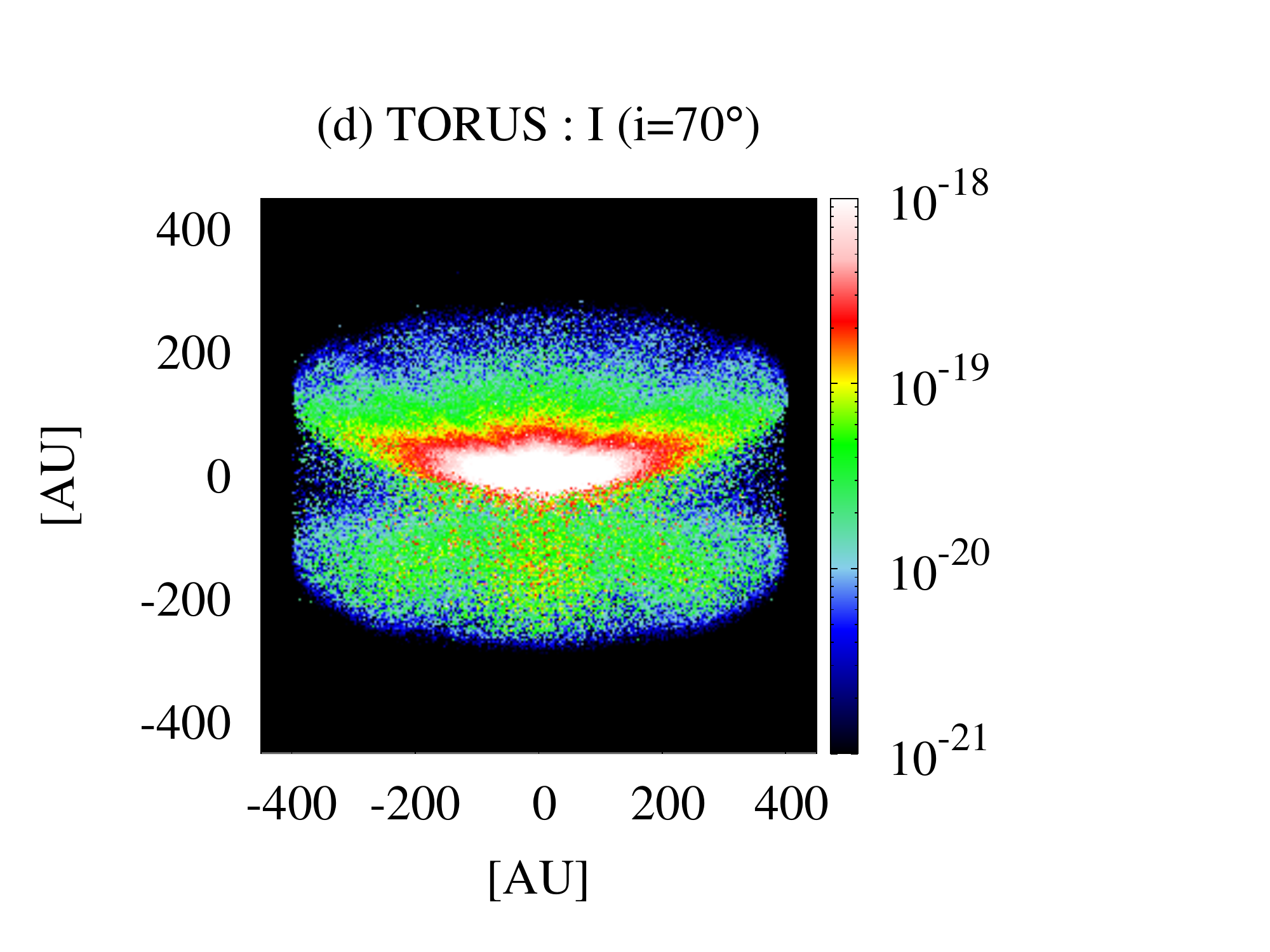}
  }
 \caption{
Intensity maps of other models: (a) RADMC-3D, (b) MCFOST, (c) MCMAX, and (d) TORUS.
The inclination angle is set to be $69.5^\circ$.
 }
 \label{fig:int_i70}
\end{figure*}

\begin{figure*}[htbp]
\centering
\subfigure{
    \includegraphics[width=50mm]{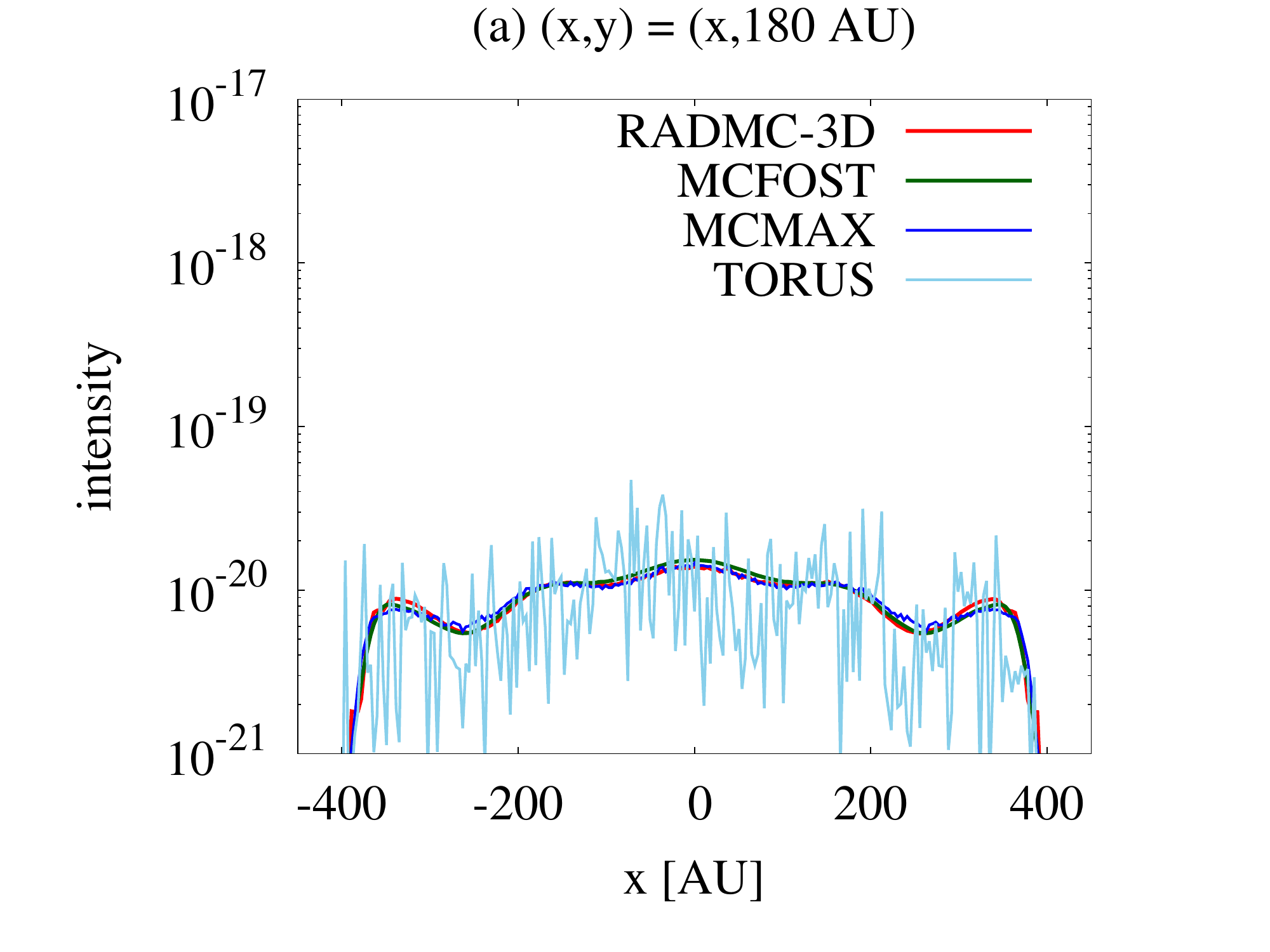}
  }
 \subfigure{
    \includegraphics[width=50mm]{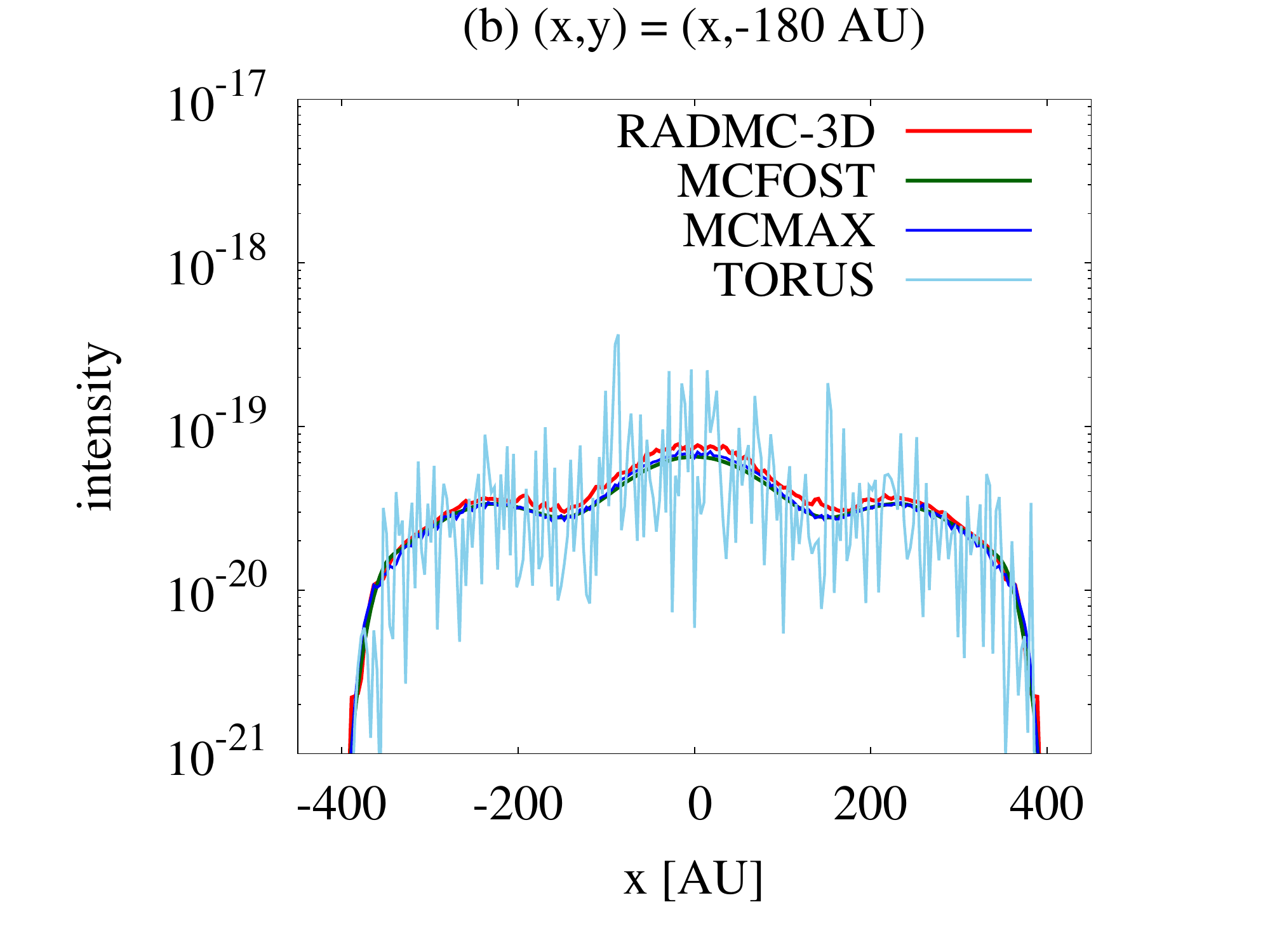}
  }
 \subfigure{
    \includegraphics[width=50mm]{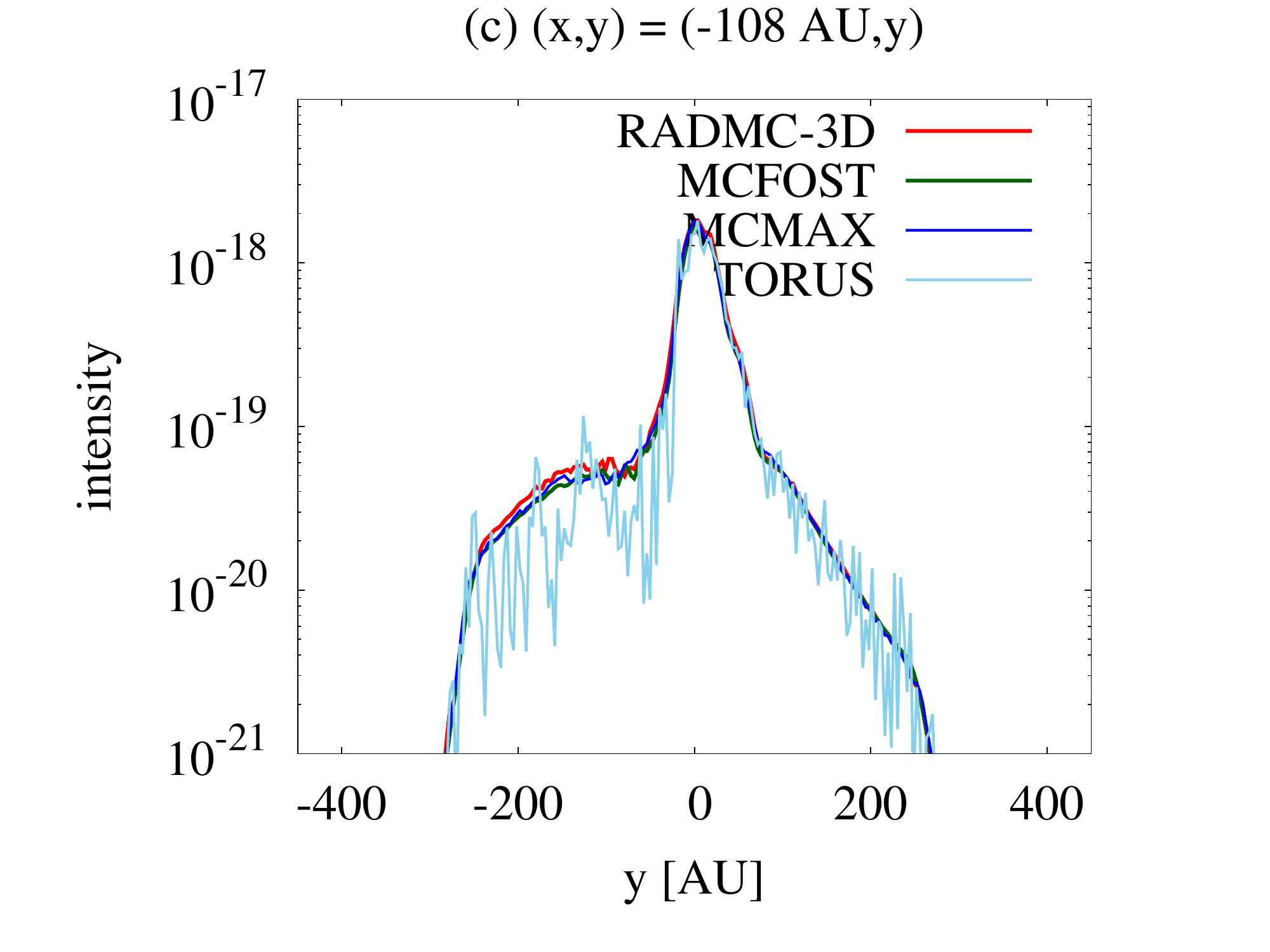}
  }
  \subfigure{
    \includegraphics[width=50mm]{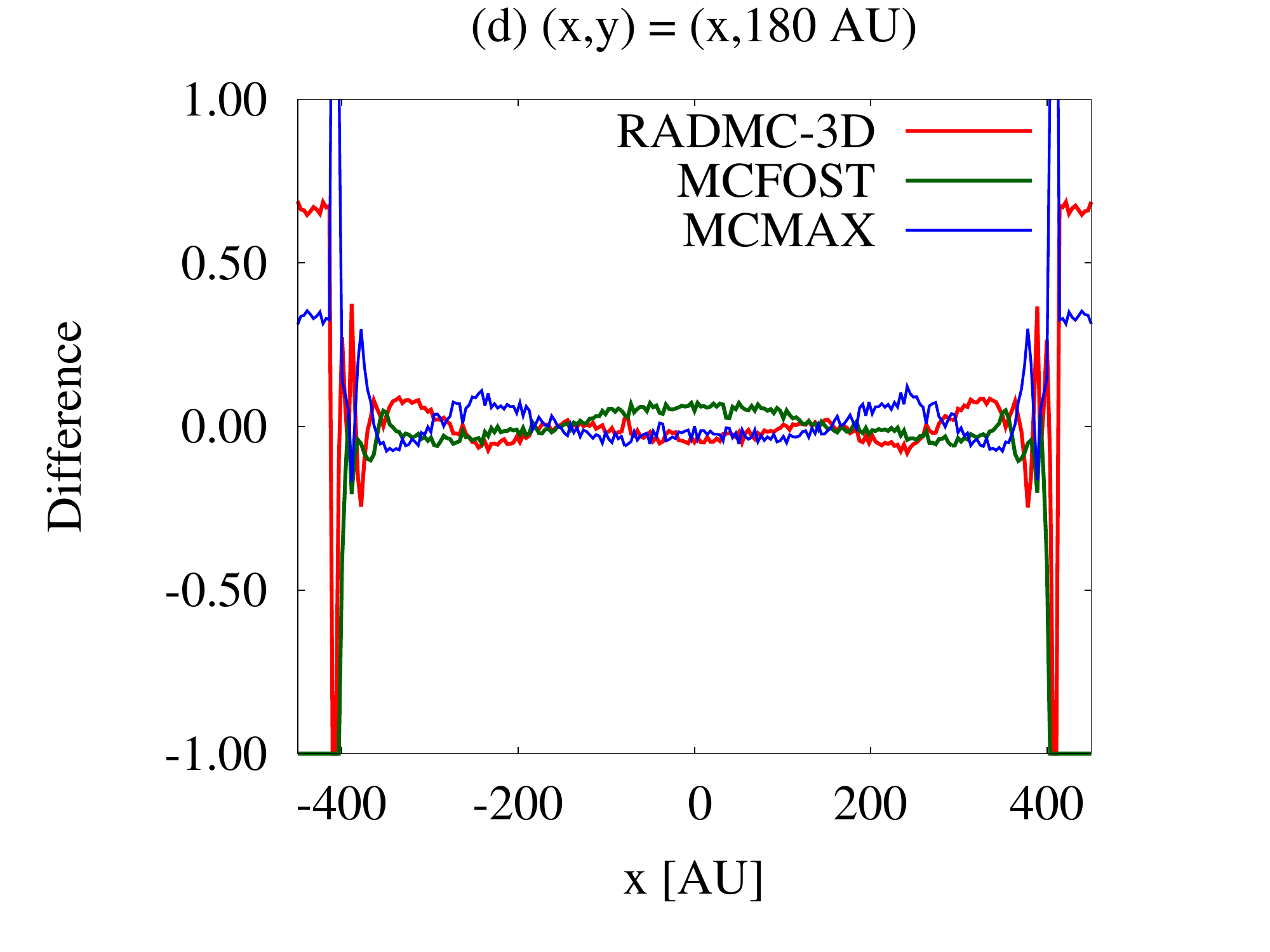}
  }
 \subfigure{
    \includegraphics[width=50mm]{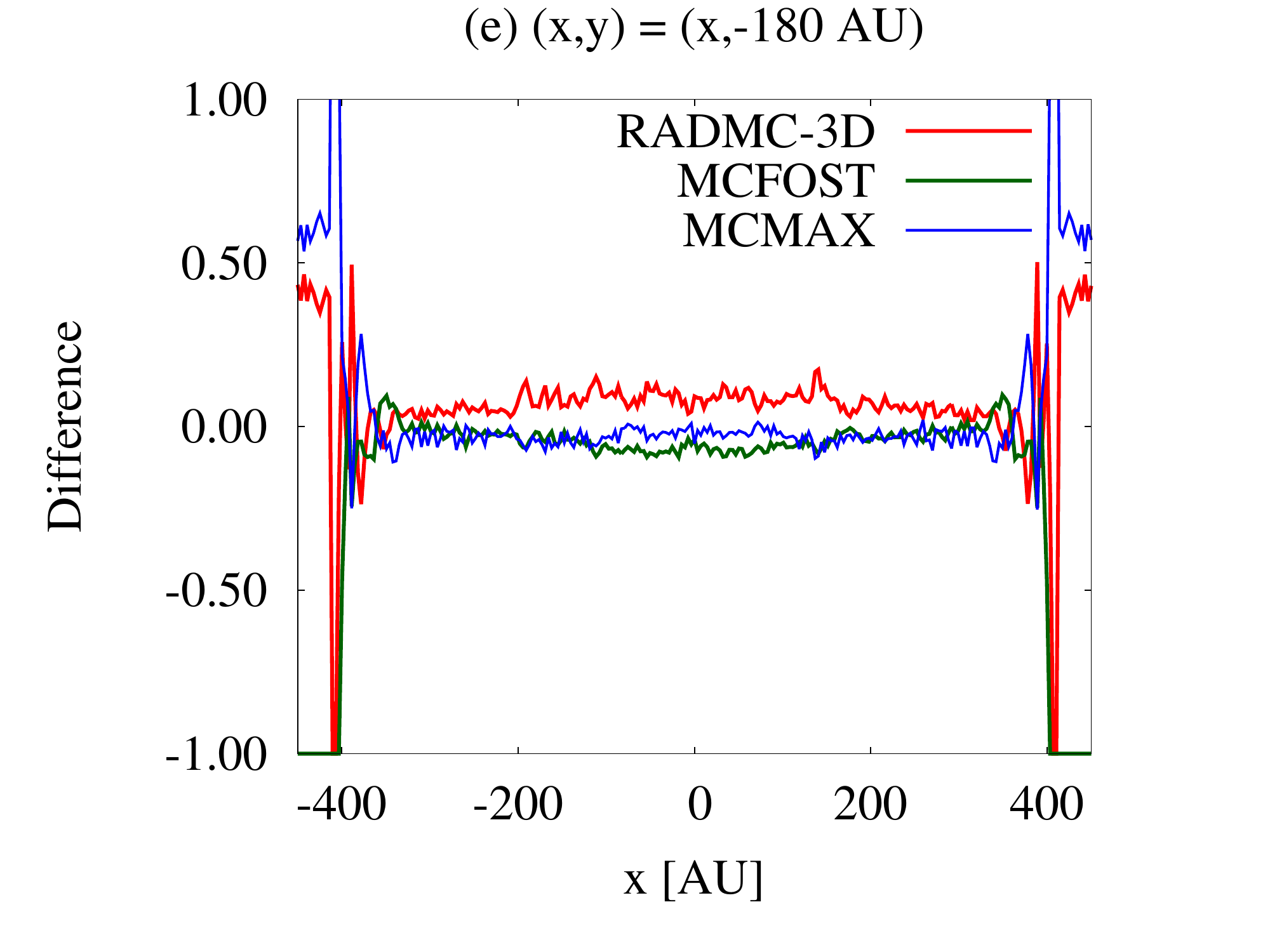}
  }
 \subfigure{
    \includegraphics[width=50mm]{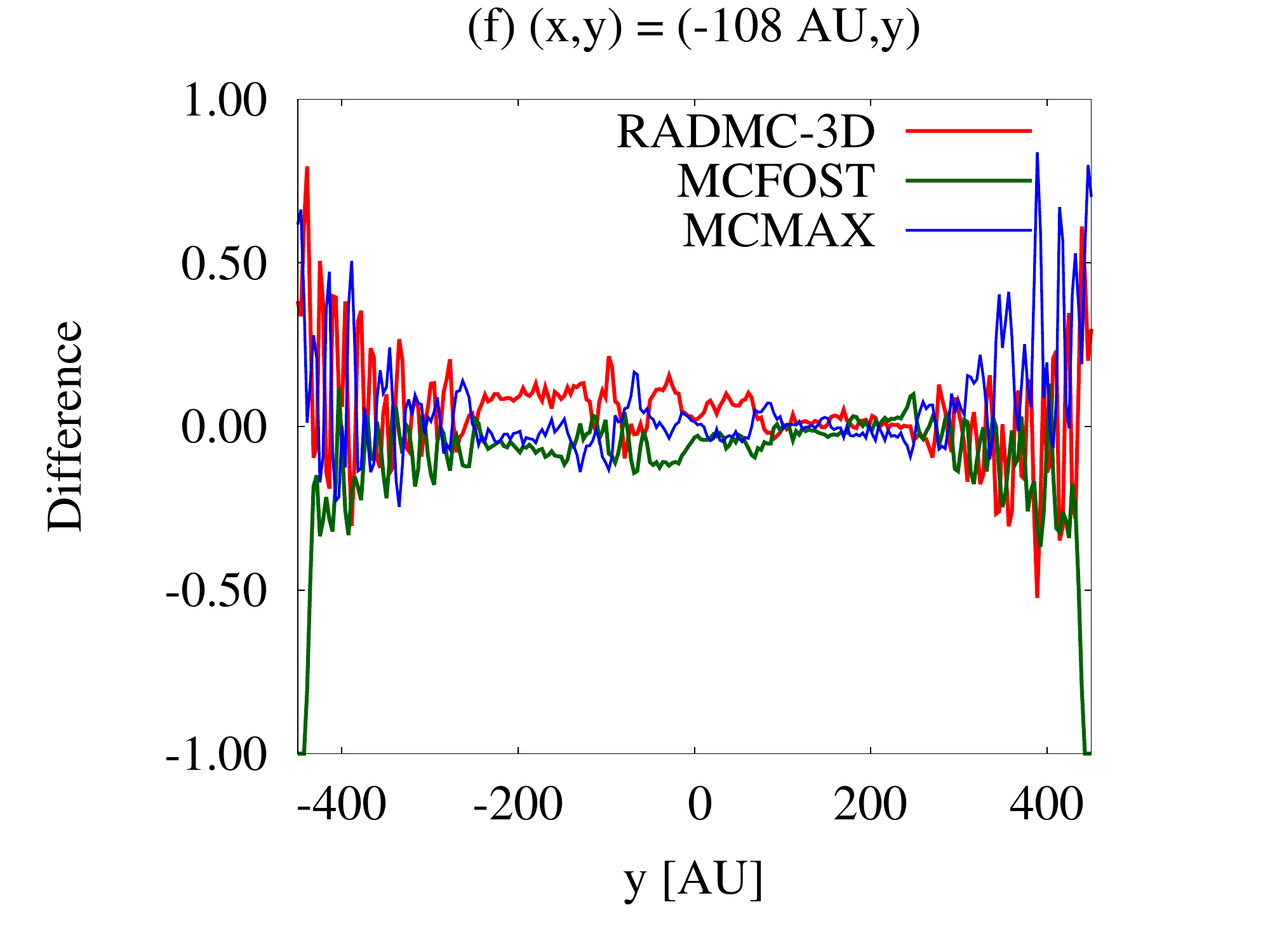}
  }
 \caption{
 The intensity along the three cuts of $(x,y)=(x,180 {\rm~AU})$, $(x,-180 {\rm~AU})$, and $(-108 {\rm~AU},y)$ in the case of $i=70^\circ$.
 The lower panels show the difference against the average.
 Due to the low signal-to-noise ratio, we take the average without the results of TORUS.
 }
 \label{fig:int_i70_cuts}
\end{figure*}

Figure \ref{fig:int_i70} shows that intensity maps with the same set up described above calculated with RADMC-3D, MCFOST, MCMAX, and TORUS\footnote{We do not plot the results of Pinball because the intensity data with the same set up is not provided.}.
The inclination angle is taken to be $69.5^\circ$.
The intensity is written in the unit of ${\rm W~m^{-2}~pixel^{-1}}$, where the distance is assumed to be 140 pc.
The intensity of the different codes are quantitatively match well as shown in Figure \ref{fig:int_i70}.
To quantitatively investigate the difference, we also plot the intensities and their relative differences from the average in Figure \ref{fig:int_i70_cuts} along the three lines on the intensity maps, which is $(x,y)=(x,180 {\rm~AU})$, $(x,-180 {\rm~AU})$, and $(-108 {\rm~AU},y)$.
Note that $108 {\rm~AU}$ corresponds to 30 pixels and $180 {\rm~AU}$ corresponds to 50 pixels.
We take the average without the results of TORUS due to the low signal-to-noise ratio.
The large error at large or small $x$ or $y$ in Figures \ref{fig:int_i70_cuts} (d), (e), and (f) is because there is no data points or a number of photons is too low.
In Figures \ref{fig:int_i70_cuts} (d) and (e), the error is statistical is within 15 \%.
In Figure \ref{fig:int_i70_cuts} (f), the error is larger beyond around 300 AU, which is due to a low number of photons in this inclination set up. 
We should be careful to this point to discuss the difference of the polarization degree in the next discussion.

\begin{figure*}[htbp]
\centering
 \subfigure{
    \includegraphics[width=75mm]{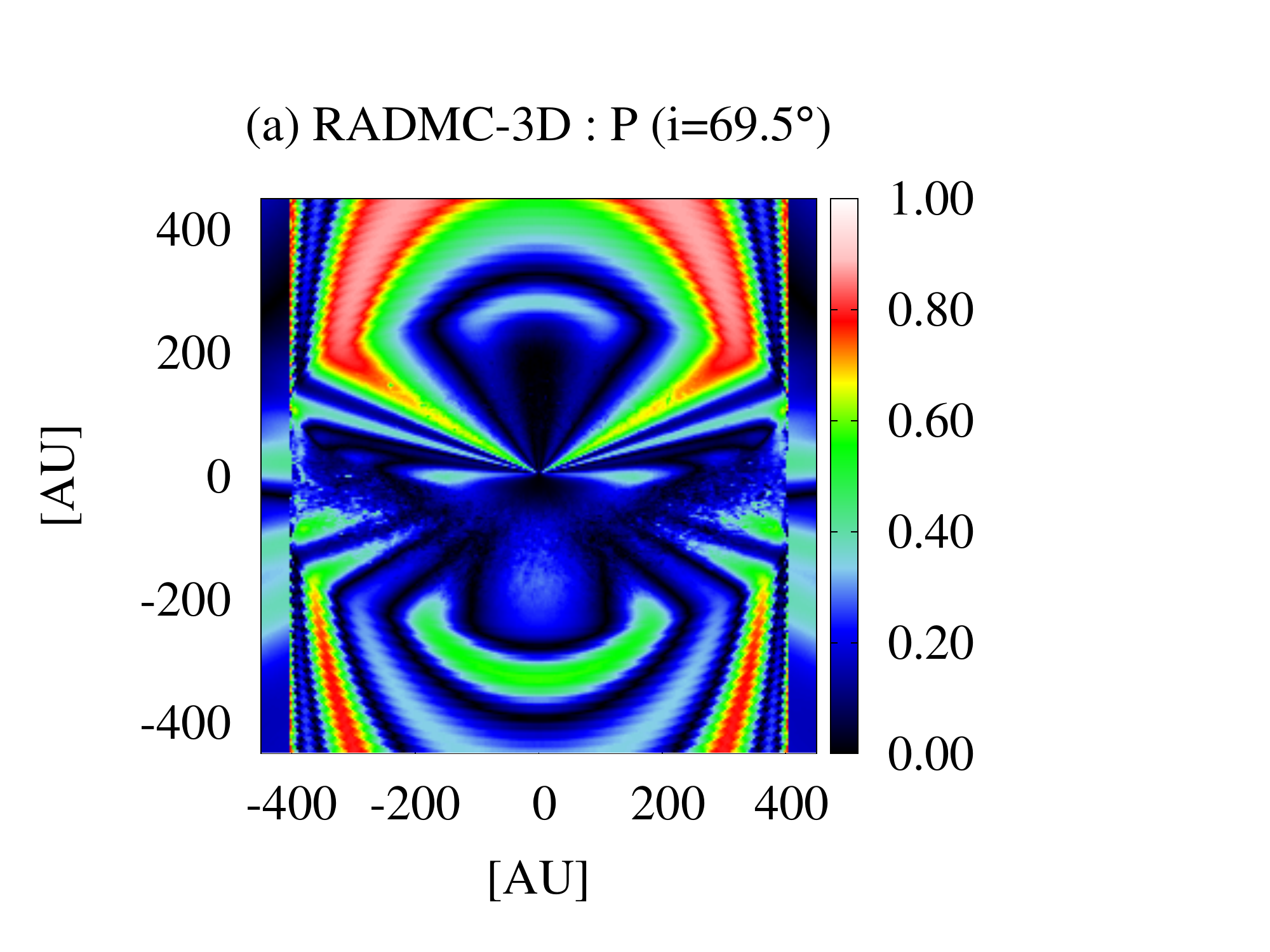}
  }
\subfigure{
  \includegraphics[width=75mm]{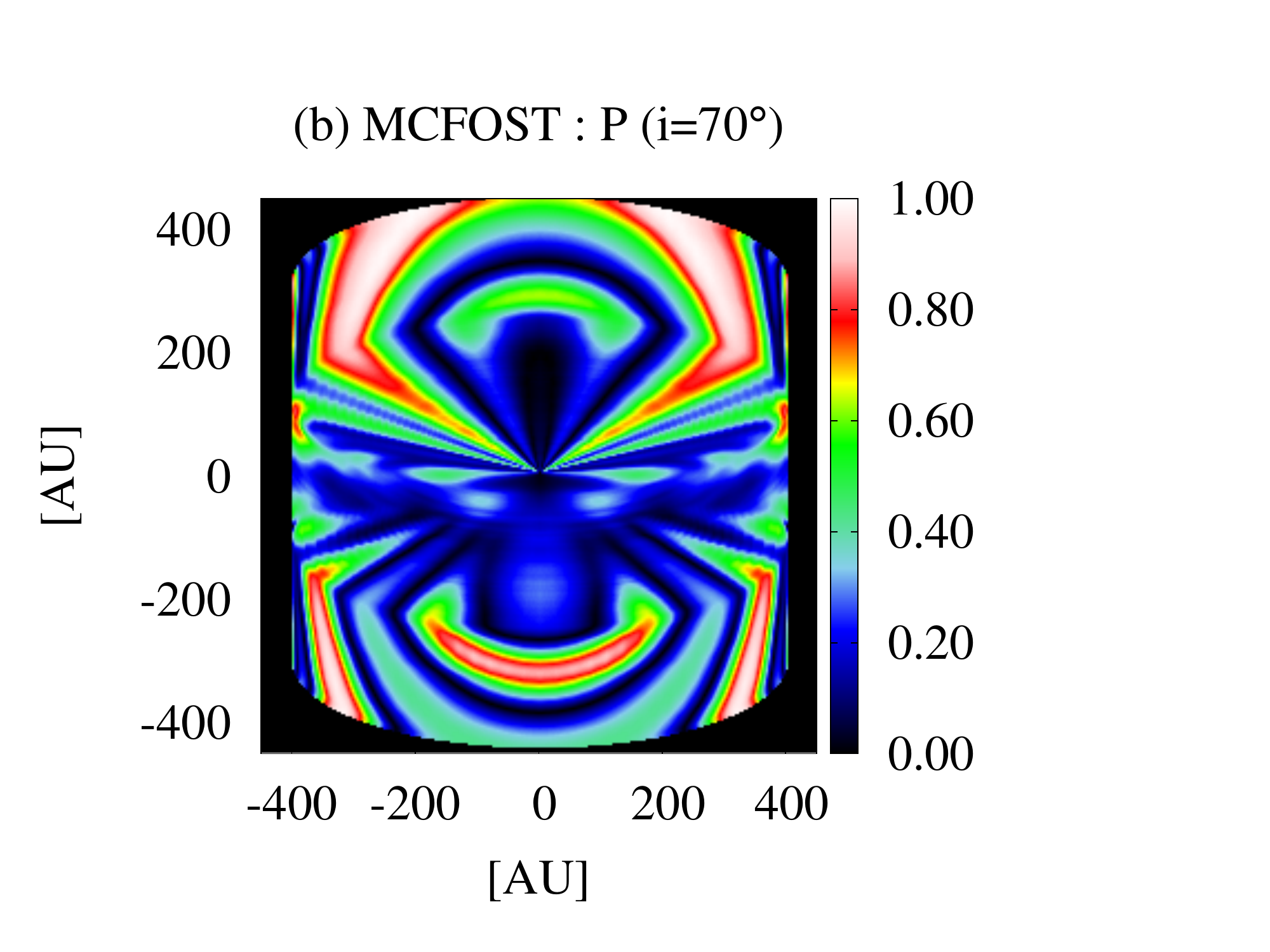}
  }
 \subfigure{
    \includegraphics[width=75mm]{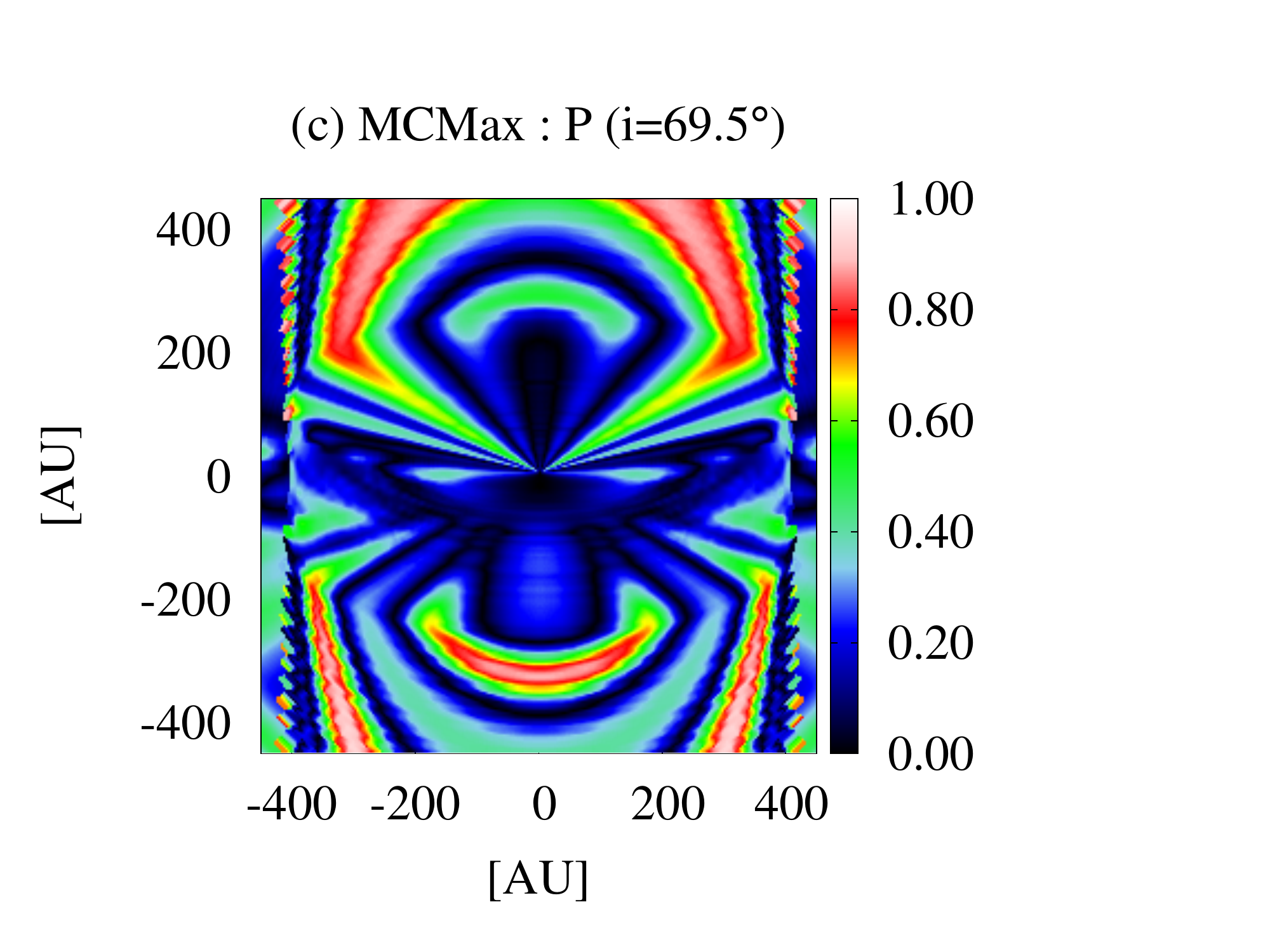}
  }
 \subfigure{
    \includegraphics[width=75mm]{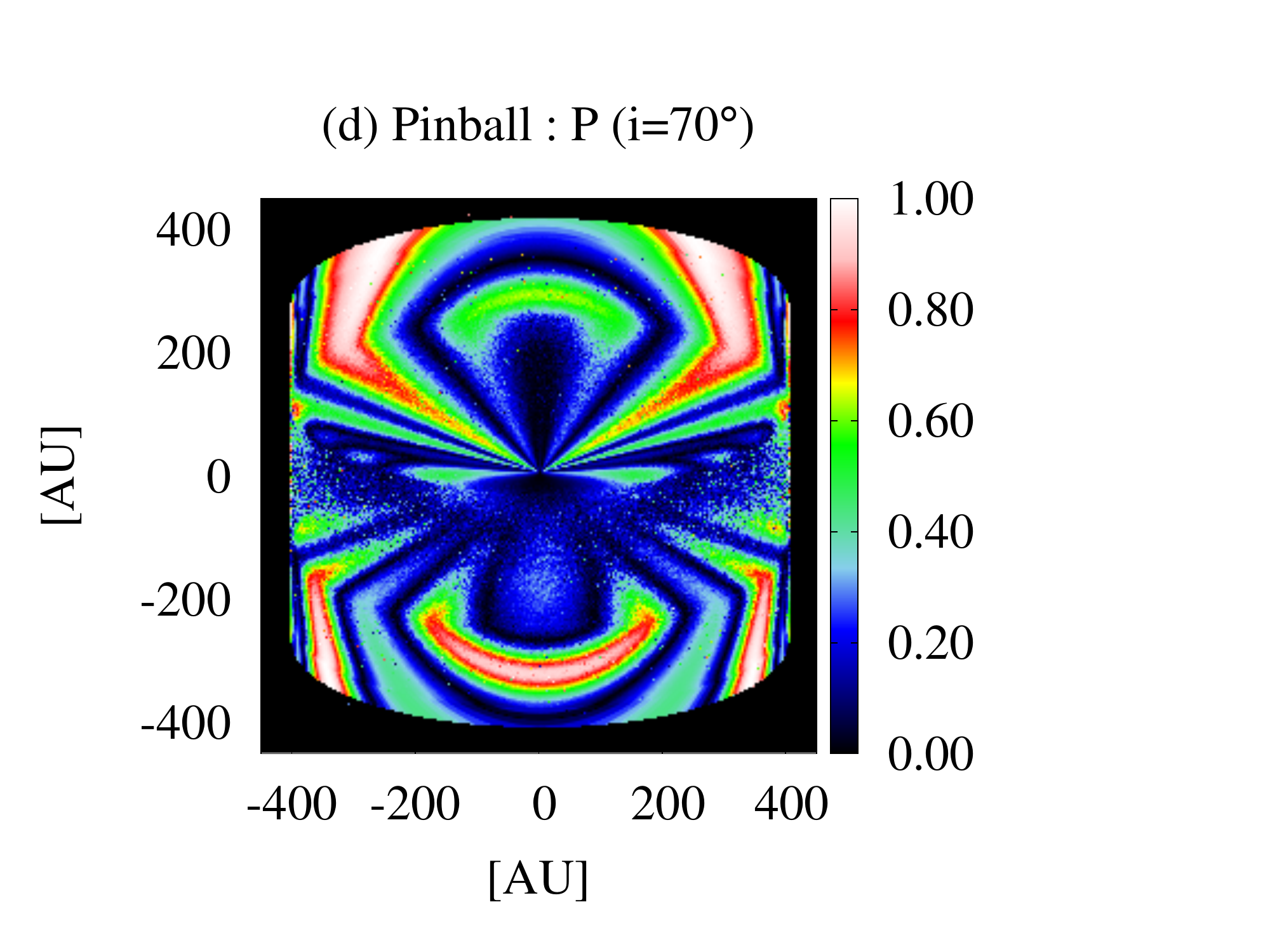}
  }
 \caption{
Polarization maps of other models: RADMC-3D, MCFOST, MCMAX, and Pinball.
The inclination is $69.5^\circ$.
 }
 \label{fig:pol_i70}
\end{figure*}

\begin{figure*}[htbp]
\centering
\subfigure{
    \includegraphics[width=50mm]{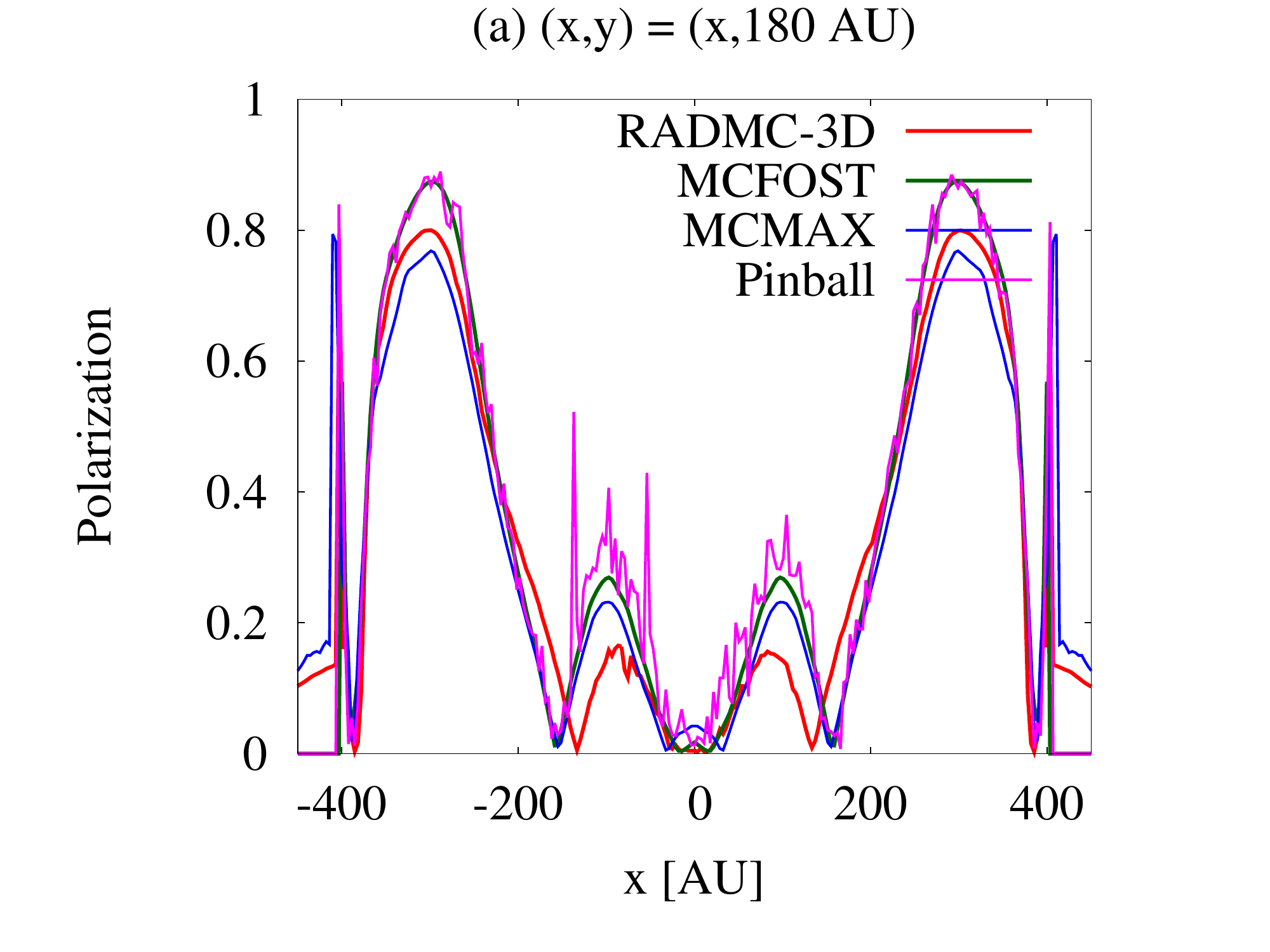}
  }
 \subfigure{
    \includegraphics[width=50mm]{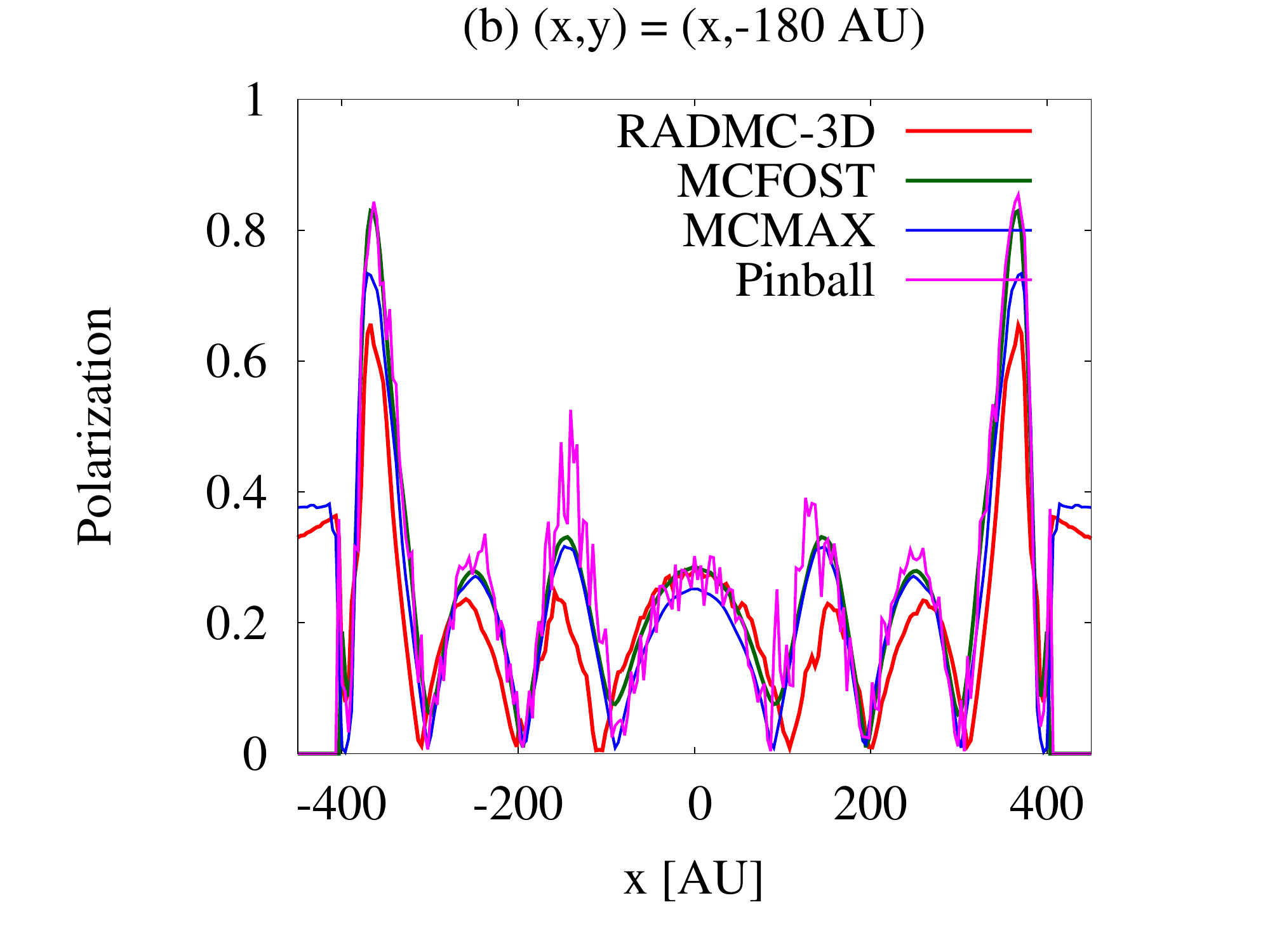}
  }
 \subfigure{
    \includegraphics[width=50mm]{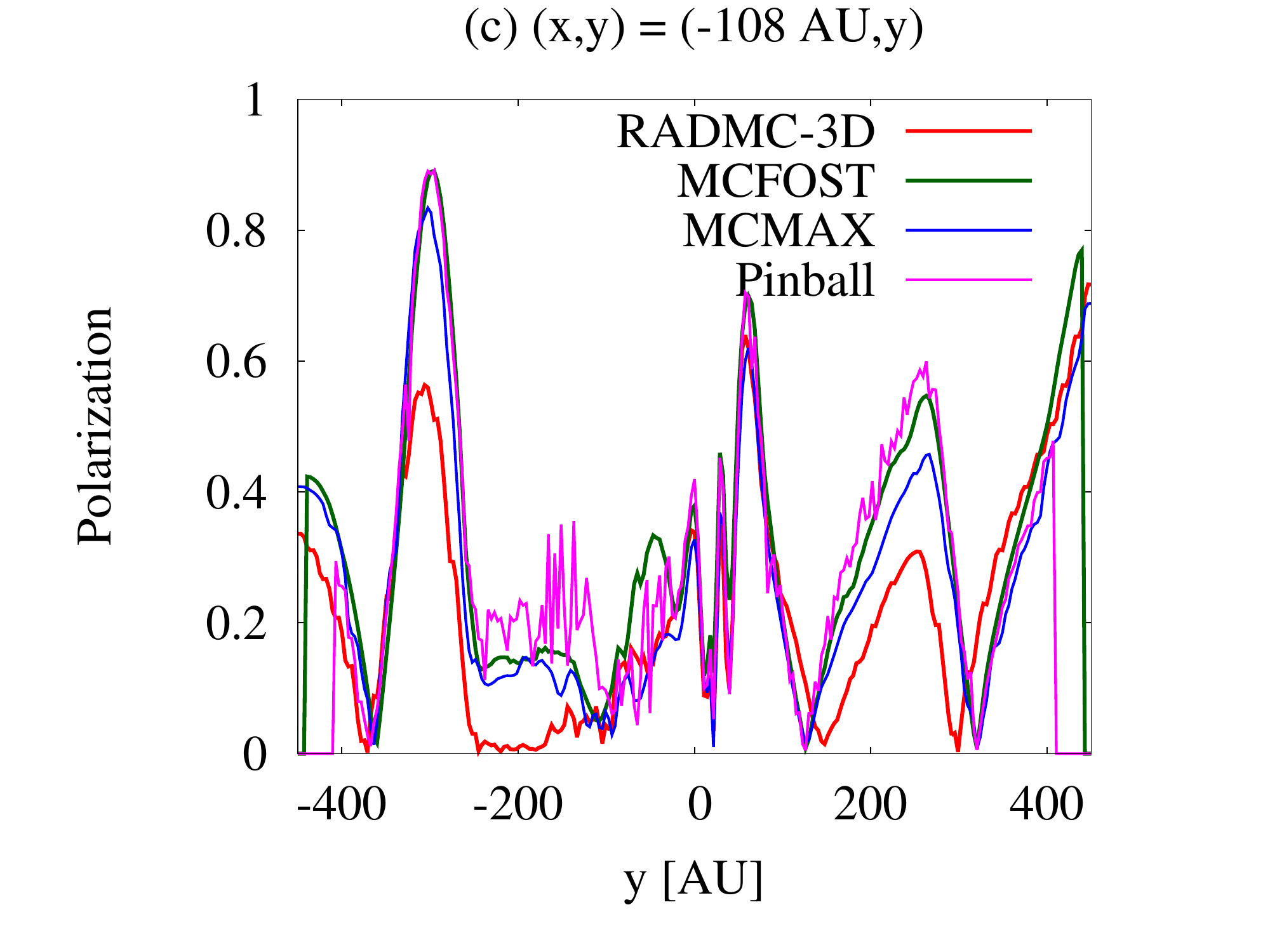}
  }
  \subfigure{
    \includegraphics[width=50mm]{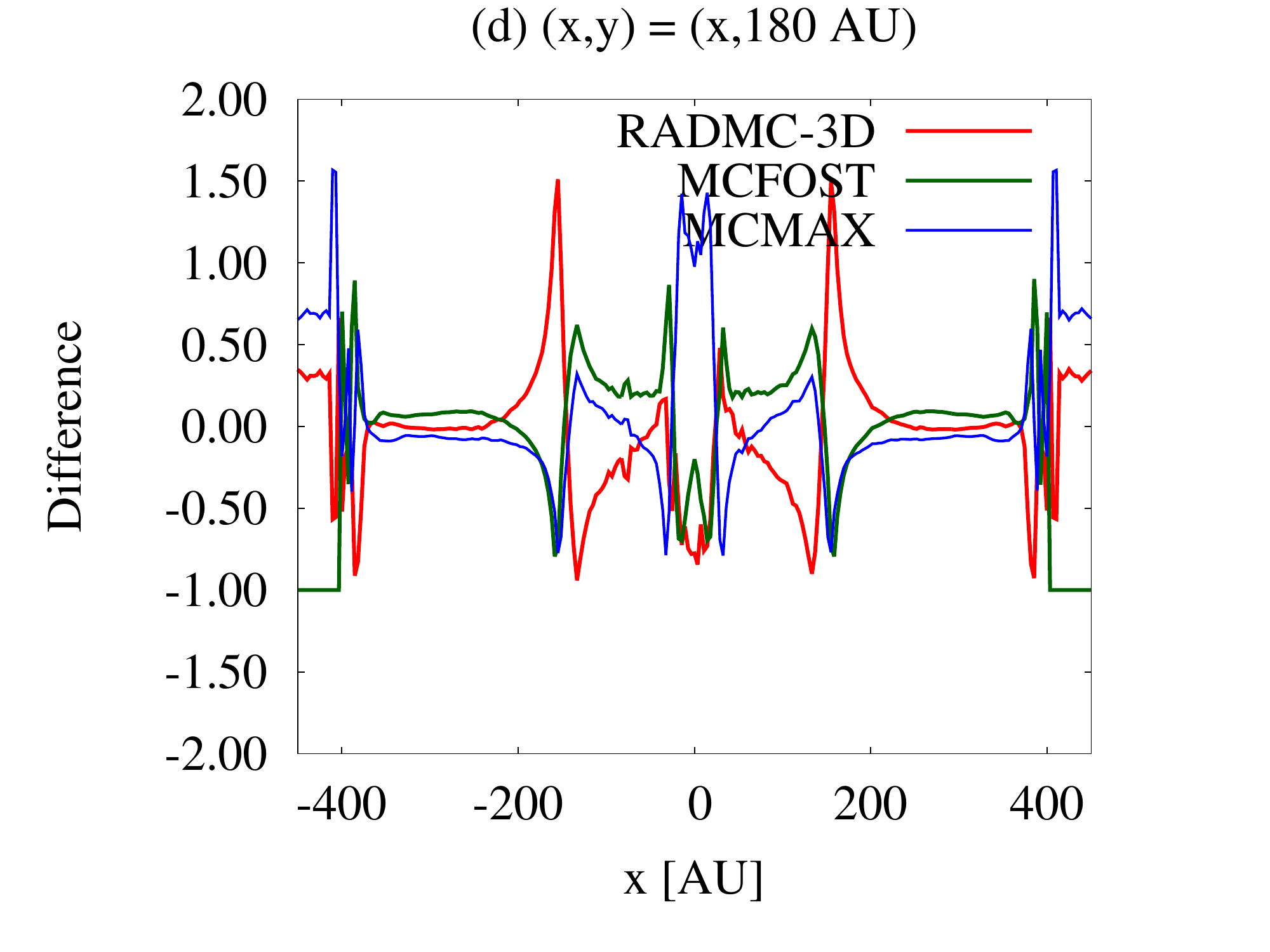}
  }
 \subfigure{
    \includegraphics[width=50mm]{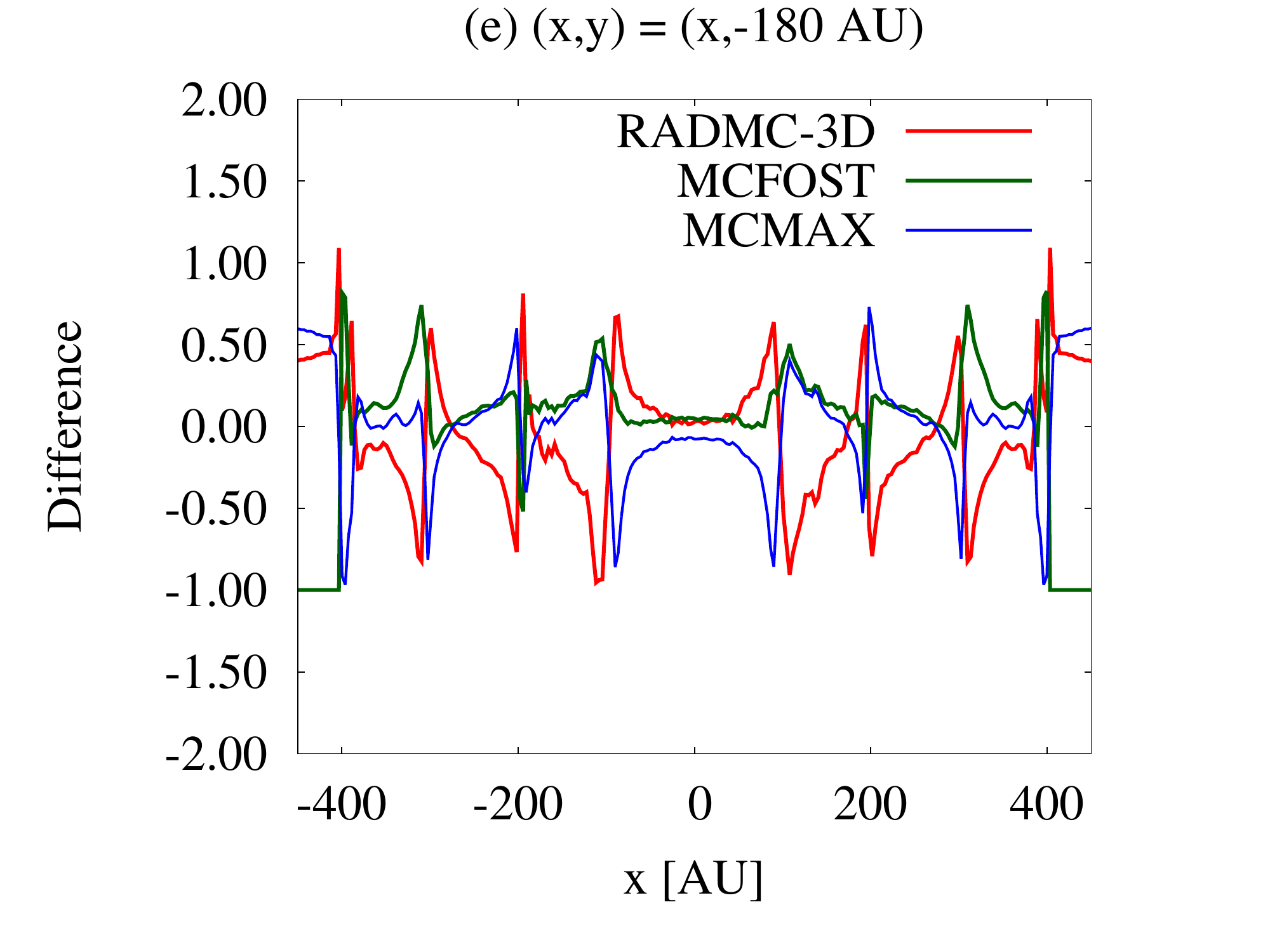}
  }
 \subfigure{
    \includegraphics[width=50mm]{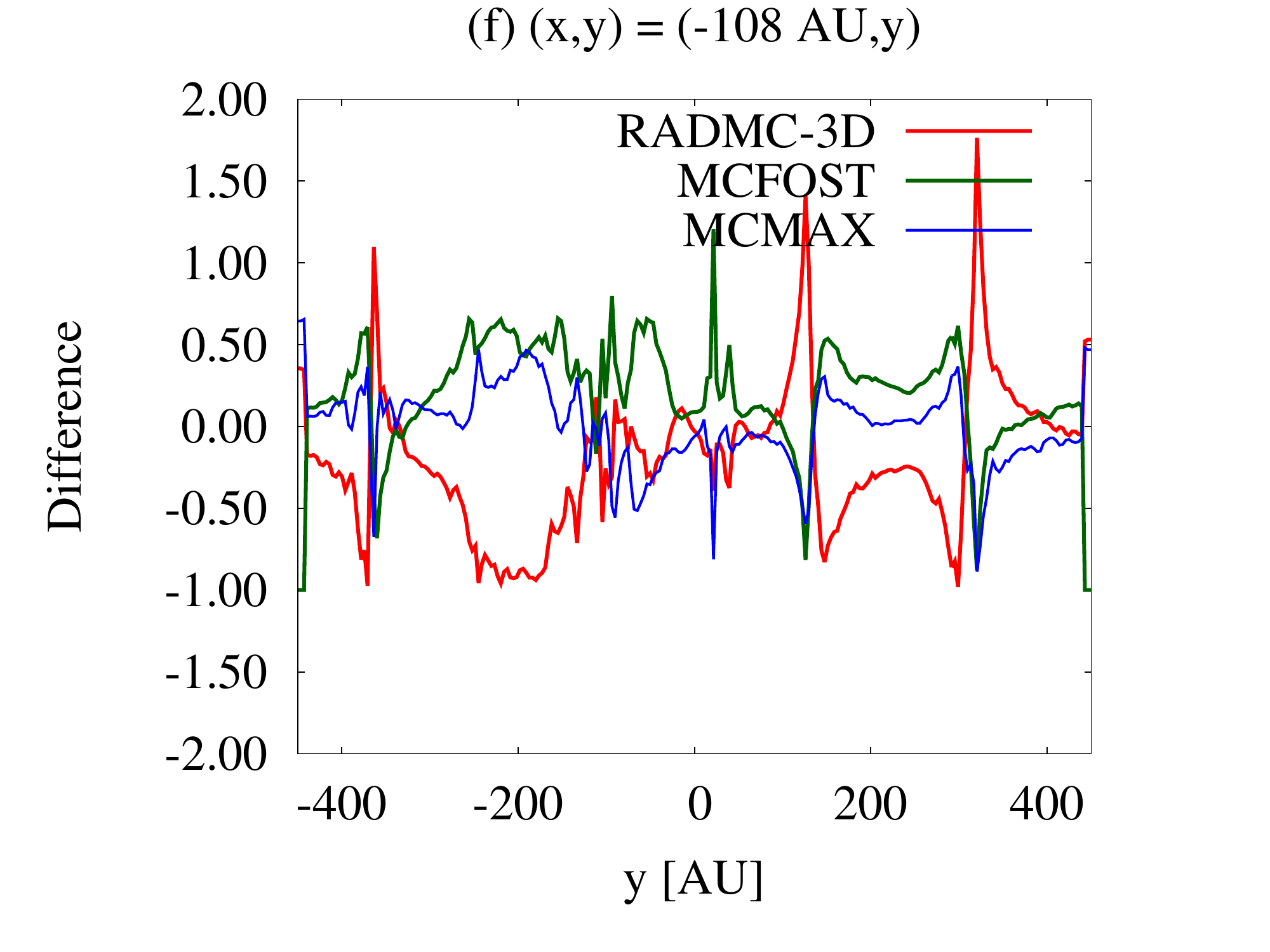}
  }
 \caption{
 The polarization degree along the three cuts of $(x,y)=(x,180 {\rm~AU})$, $(x,-180 {\rm~AU})$, and $(-108 {\rm~AU},y)$ in the case of $i=69.5^\circ$.
 The lower panels show the difference against the average.
 Due to the low signal-to-noise ratio, we take the average without the results of Pinball.
 }
 \label{fig:pol_i70_cuts}
\end{figure*}

Figure \ref{fig:pol_i70} shows that the maps of the polarization degree with the same set up.
We do not plot the results of TORUS due to the low signal-to-noise ratio.
Instead, we plot the results of Pinball.
The maps of the polarization degree have a pattern due to the resonance of the phase function.
The results are quite similar, but it has some differences in the polarization degree.
To quantitatively investigate the difference, we also plot the polarization degrees and their relative differences from the average in Figure \ref{fig:pol_i70_cuts} along the three lines that is the same as the previous discussion.
Note that we take the average without the results of Pinball.
Figures \ref{fig:pol_i70_cuts} (a), (b), and (c) show the polarization degree along the three lines, and they have a good agreement in their shape that corresponds to the resonance.
The relative differences in Figures \ref{fig:pol_i70_cuts} (d), (e), (f) are low where the polarization degree is high, which is for example at the regions around $(250 {\rm~AU} < x< 350 {\rm~AU})$ in Figure \ref{fig:pol_i70_cuts} (d) although the difference is as large as 1.5 where the polarization degree is low.
This is due to the relative difference is enhanced in the polarization degree when the degree is low because the average itself is low.
In addition, the large relative difference around $y<-200{\rm~AU}$ and $y>200{\rm~AU}$ on the line of $(x,y)=(-108{\rm~AU},y)$ is due to a small number of photons (see Figure \ref{fig:int_i70_cuts} (f) and \ref{fig:pol_i70_cuts} (f)).
The polarized intensity is mainly coming from the single scattering on the surface of the disk, and thus the photon noise is larger for increasing of the distance from the central star.
In addition, the disk is optically thick, and thus the emission of the central star hardly reaches the midplane, which also makes a large error at the midplane.

\begin{figure*}[htbp]
\centering
 \subfigure{
    \includegraphics[width=75mm]{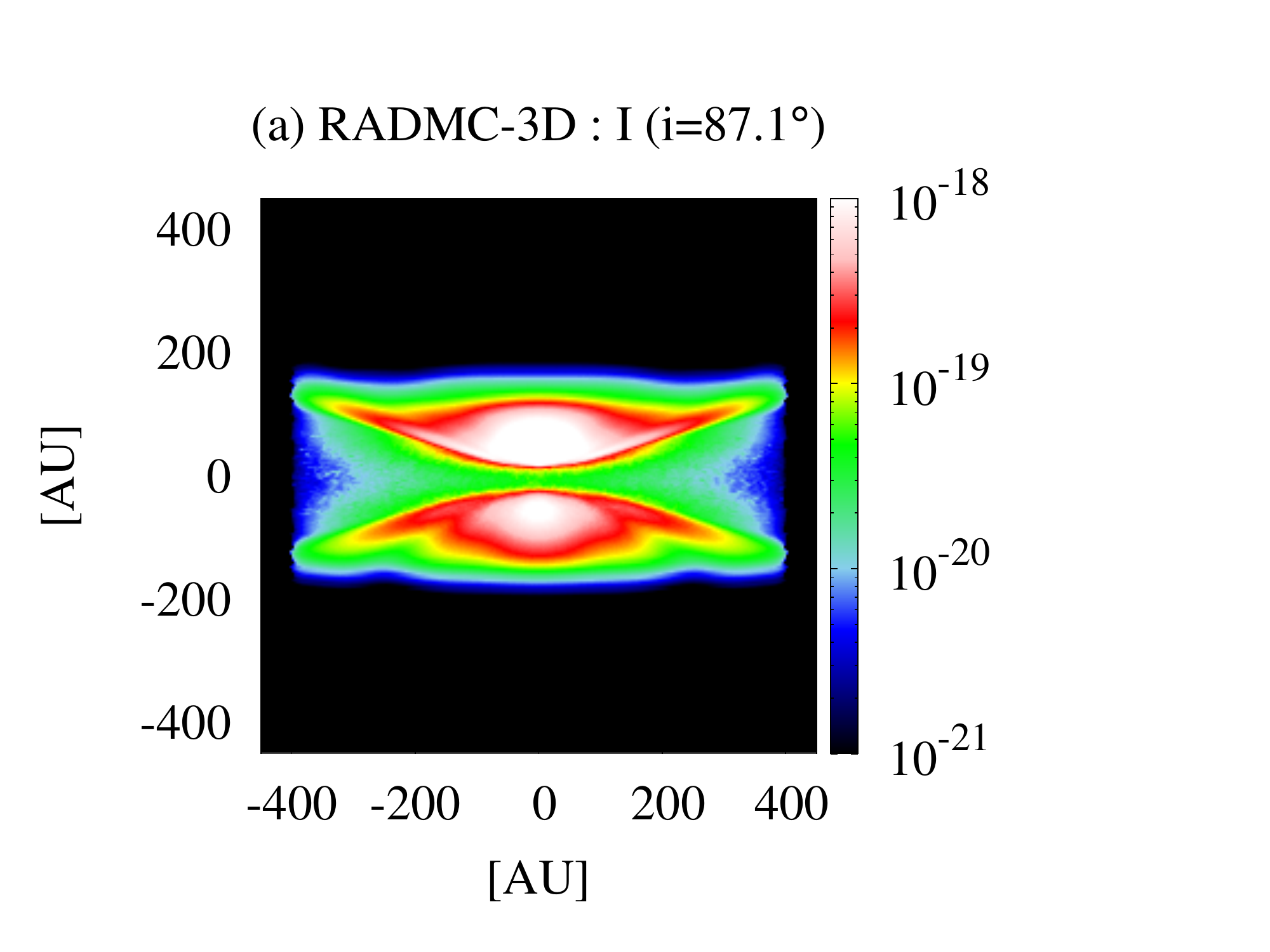}
  }
\subfigure{
  \includegraphics[width=75mm]{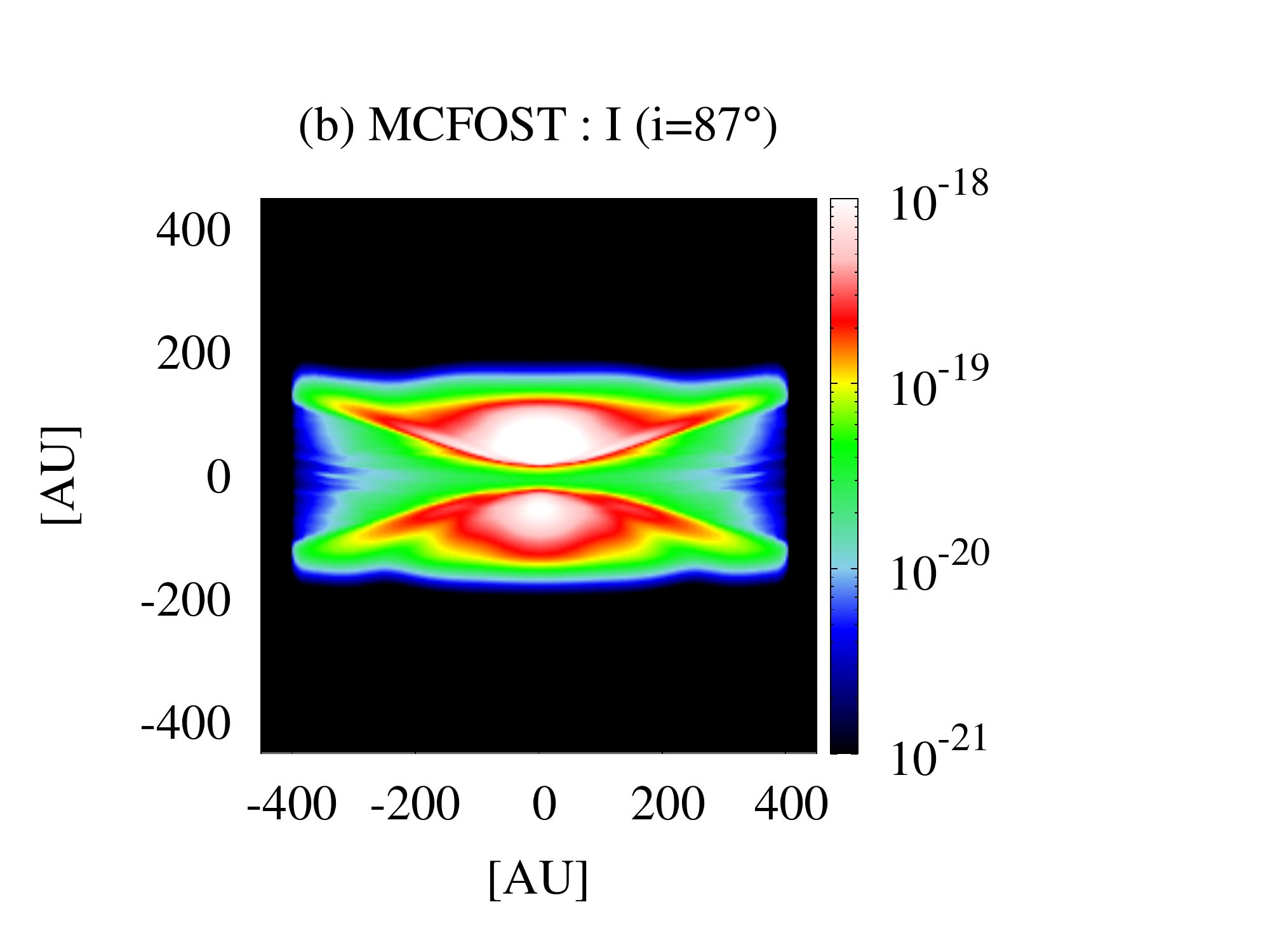}
  }
 \subfigure{
    \includegraphics[width=75mm]{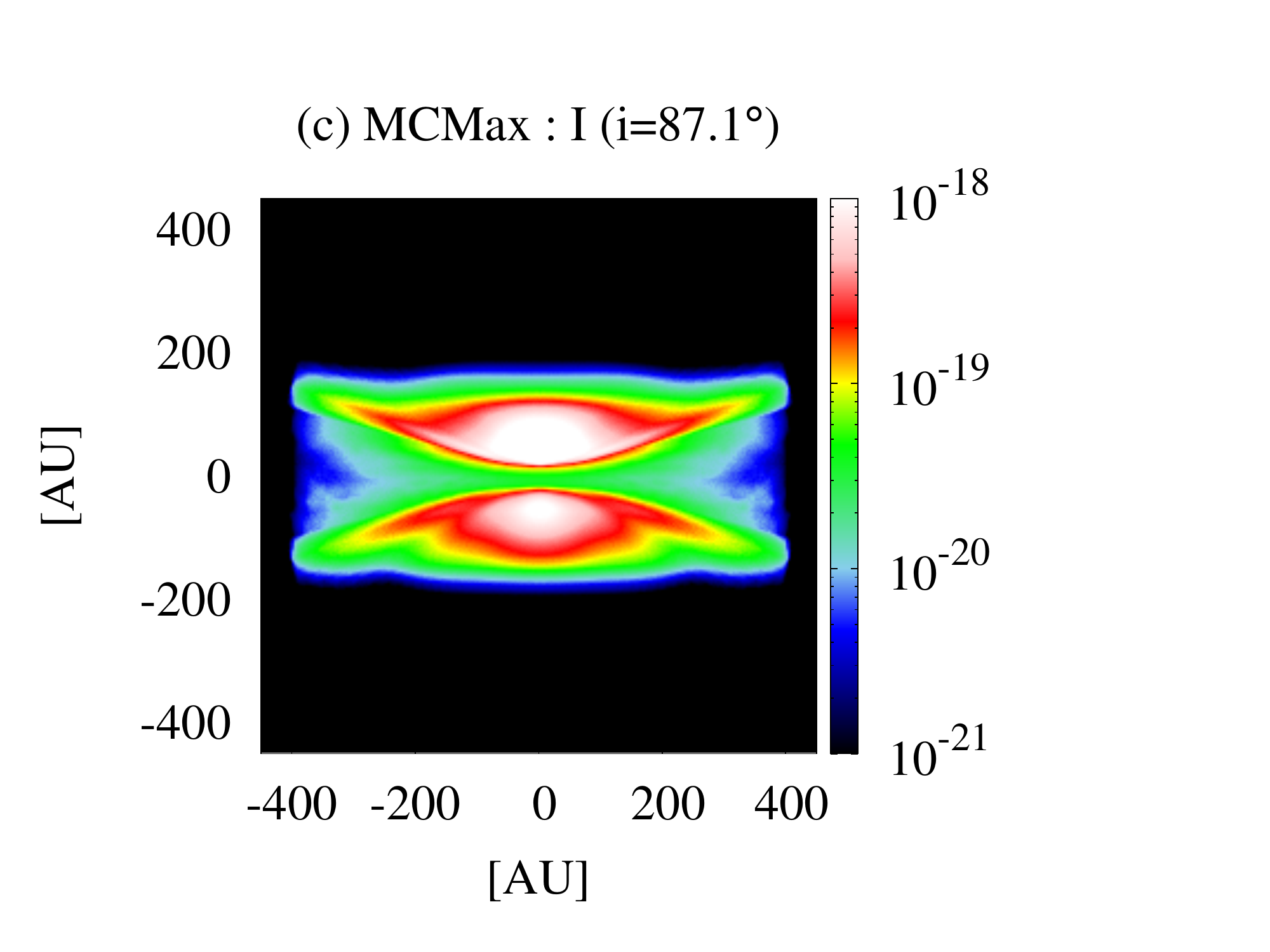}
  }
 \subfigure{
    \includegraphics[width=75mm]{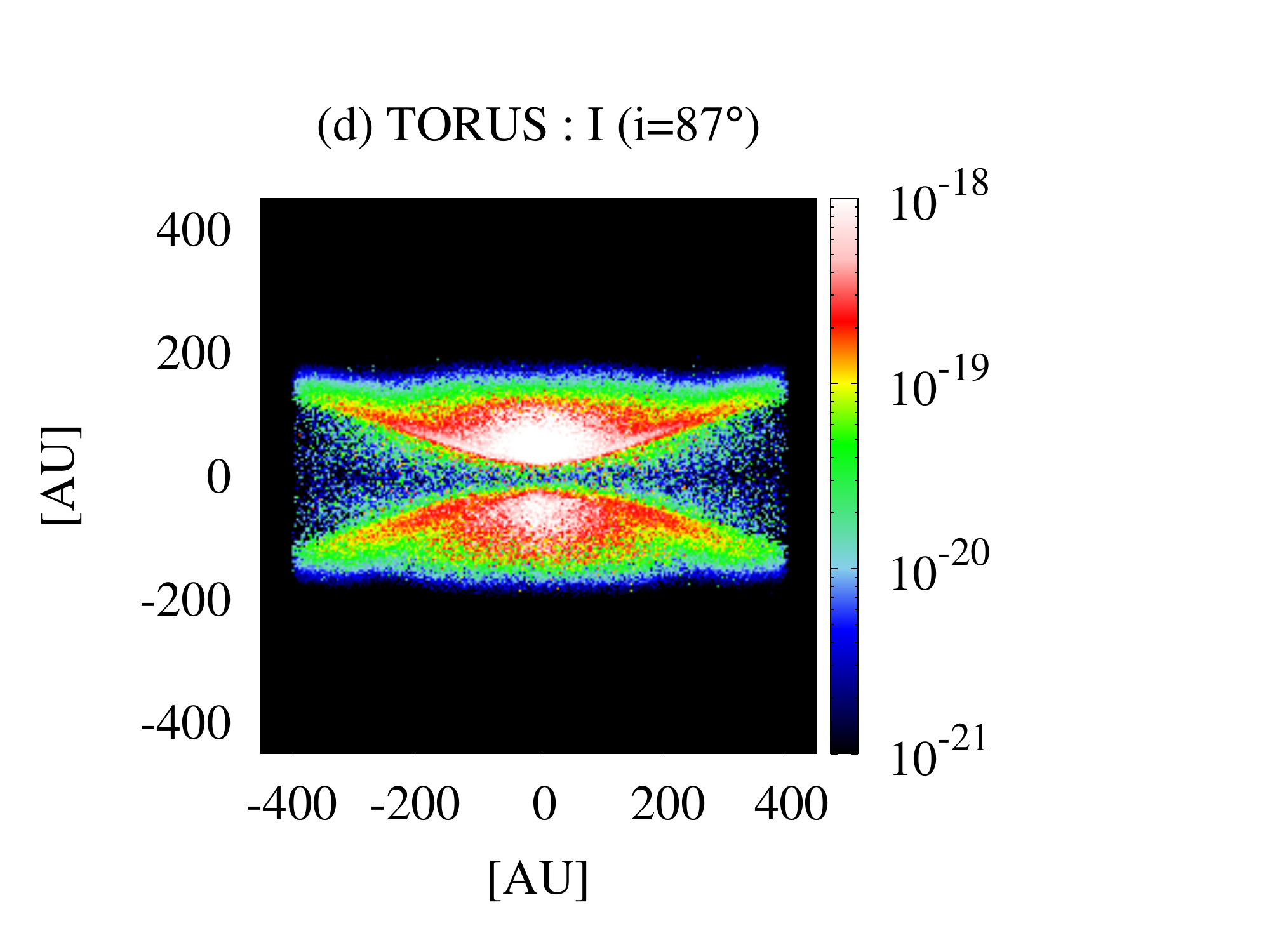}
  }
 \caption{
Intensity maps of other models: (a) RADMC-3D, (b) MCFOST, (c) MCMAX, and (d) TORUS.
The inclination angle is set to be $87.1^\circ$.
 }
 \label{fig:int_i87}
\end{figure*}

\begin{figure*}[htbp]
\centering
\subfigure{
    \includegraphics[width=50mm]{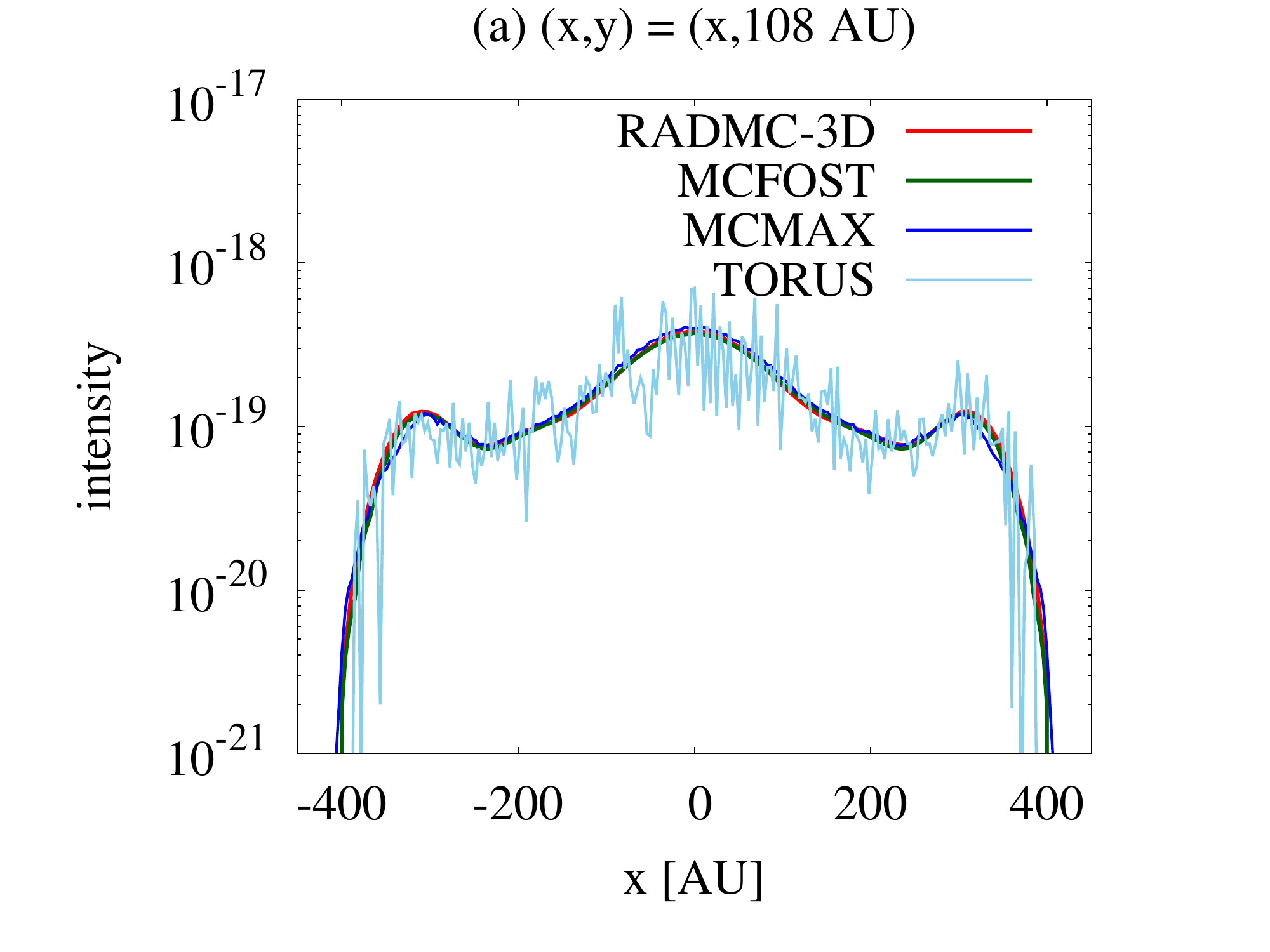}
  }
 \subfigure{
    \includegraphics[width=50mm]{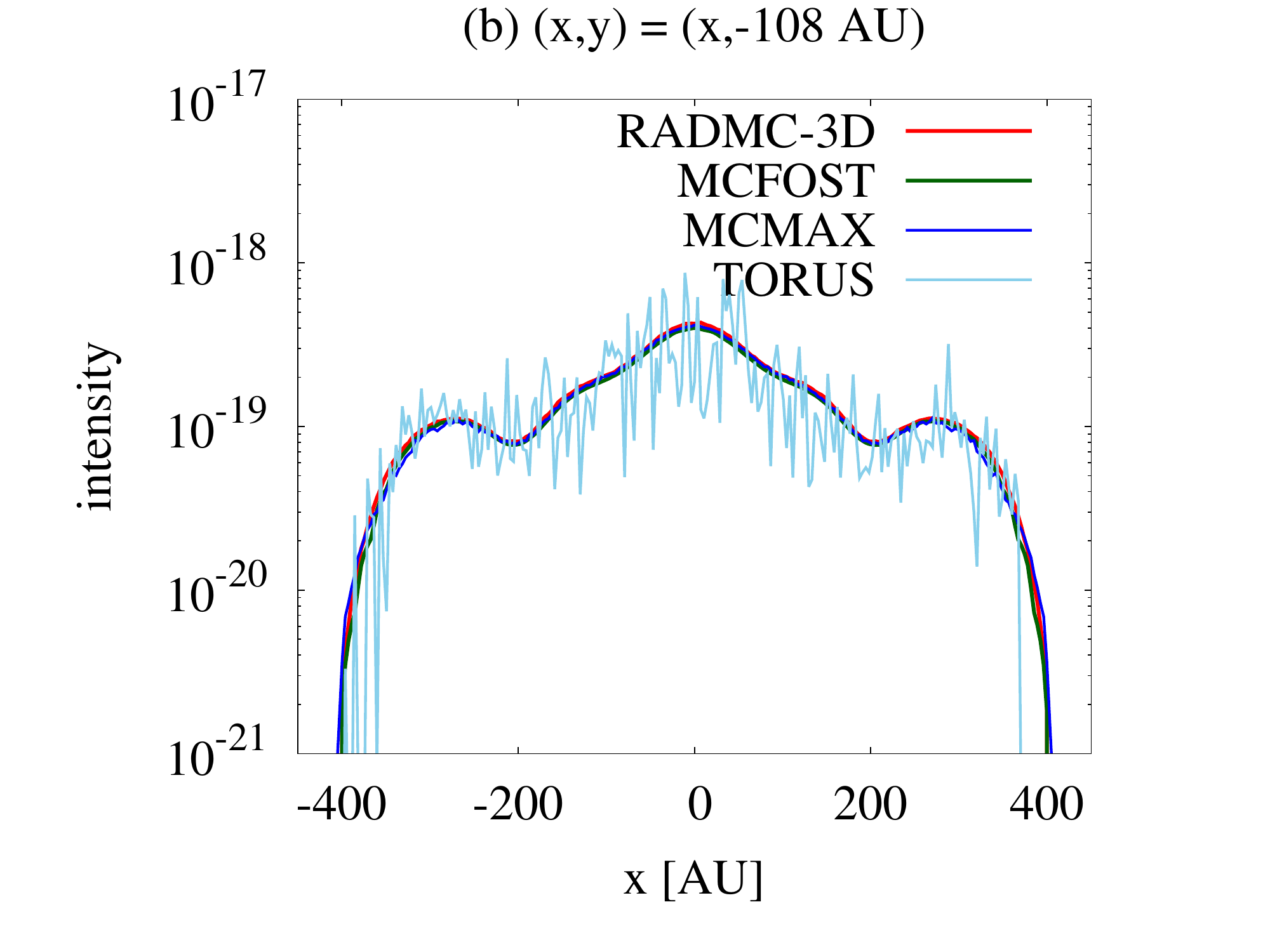}
  }
 \subfigure{
    \includegraphics[width=50mm]{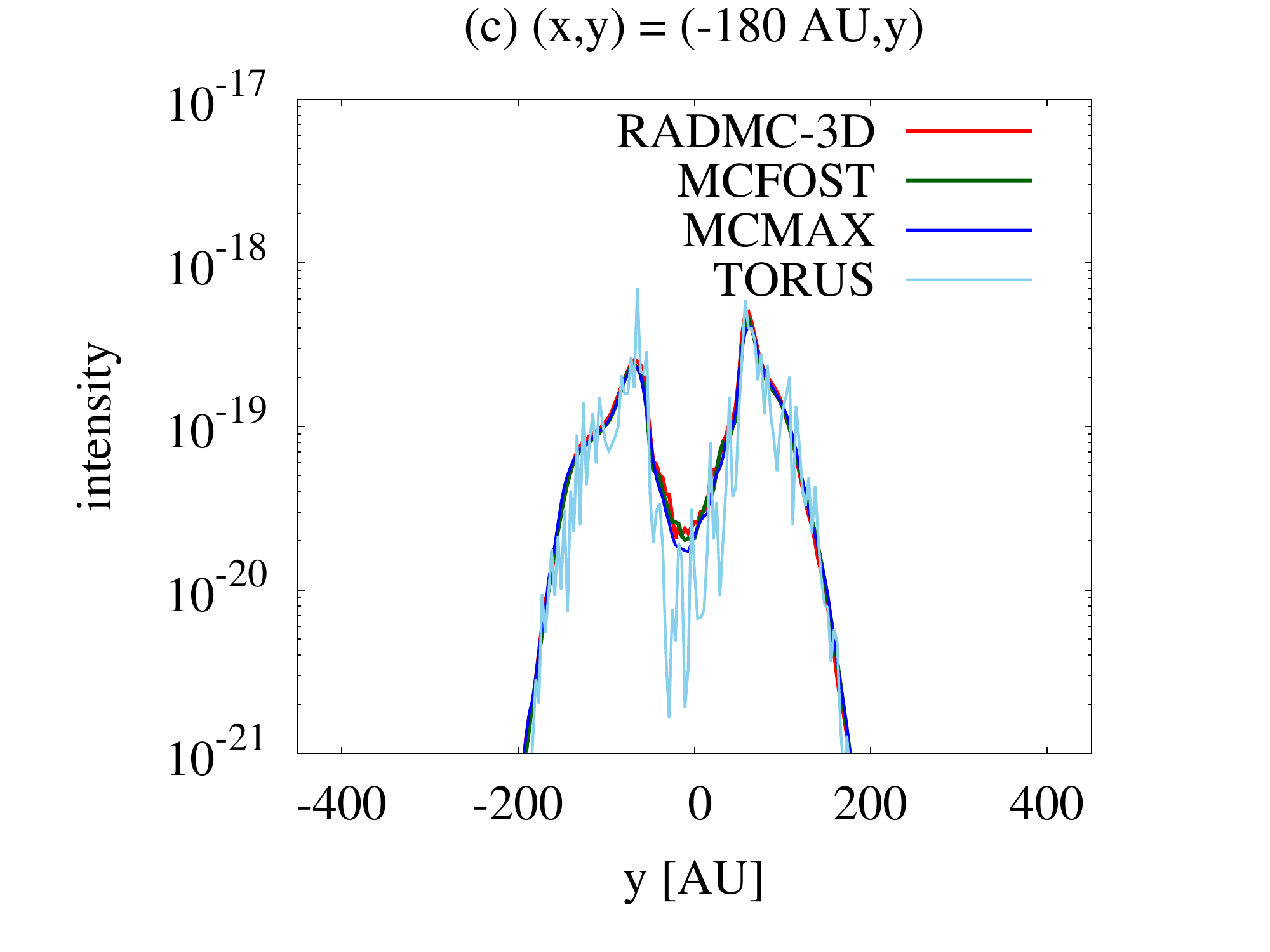}
  }
  \subfigure{
    \includegraphics[width=50mm]{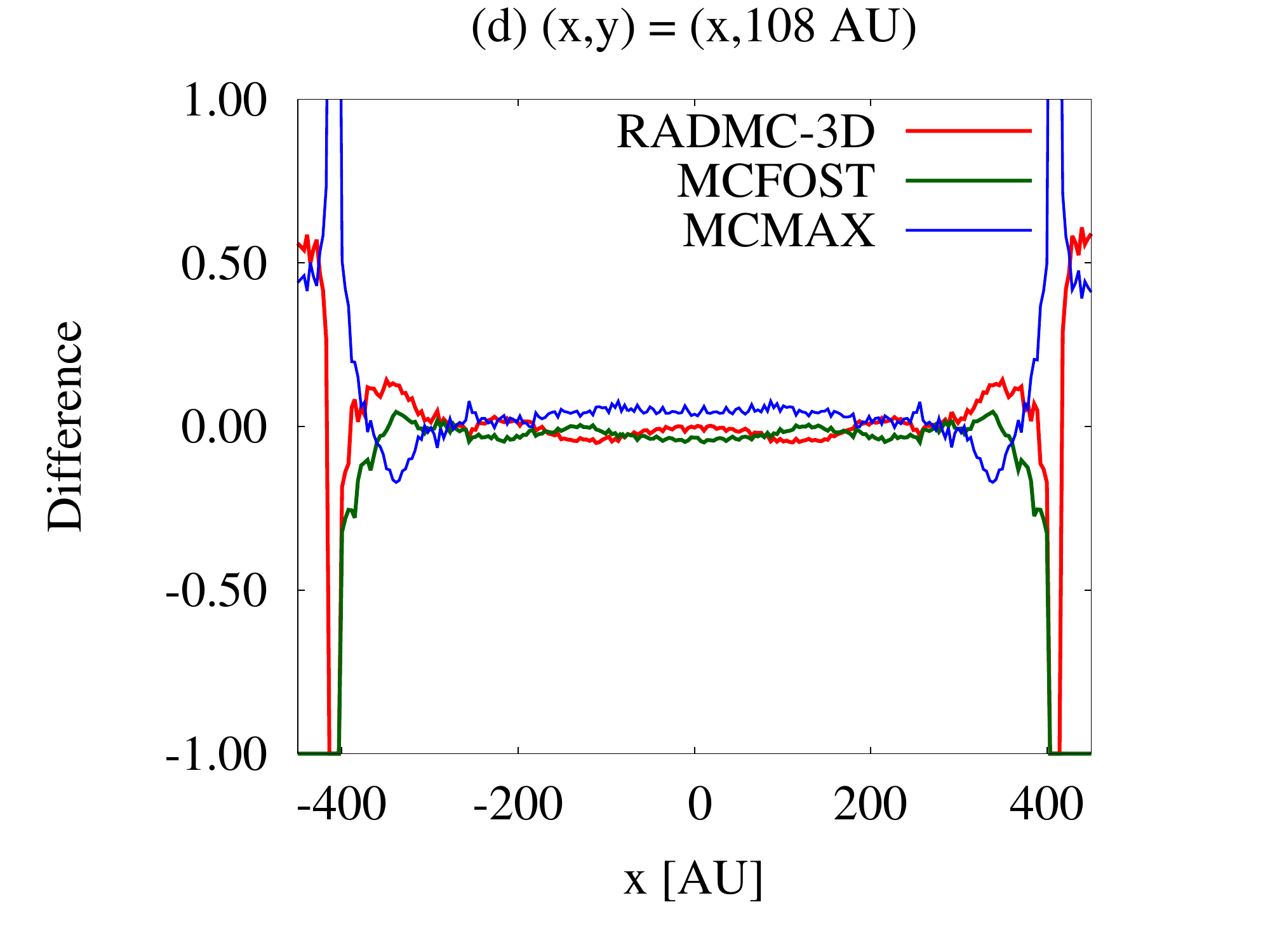}
  }
 \subfigure{
    \includegraphics[width=50mm]{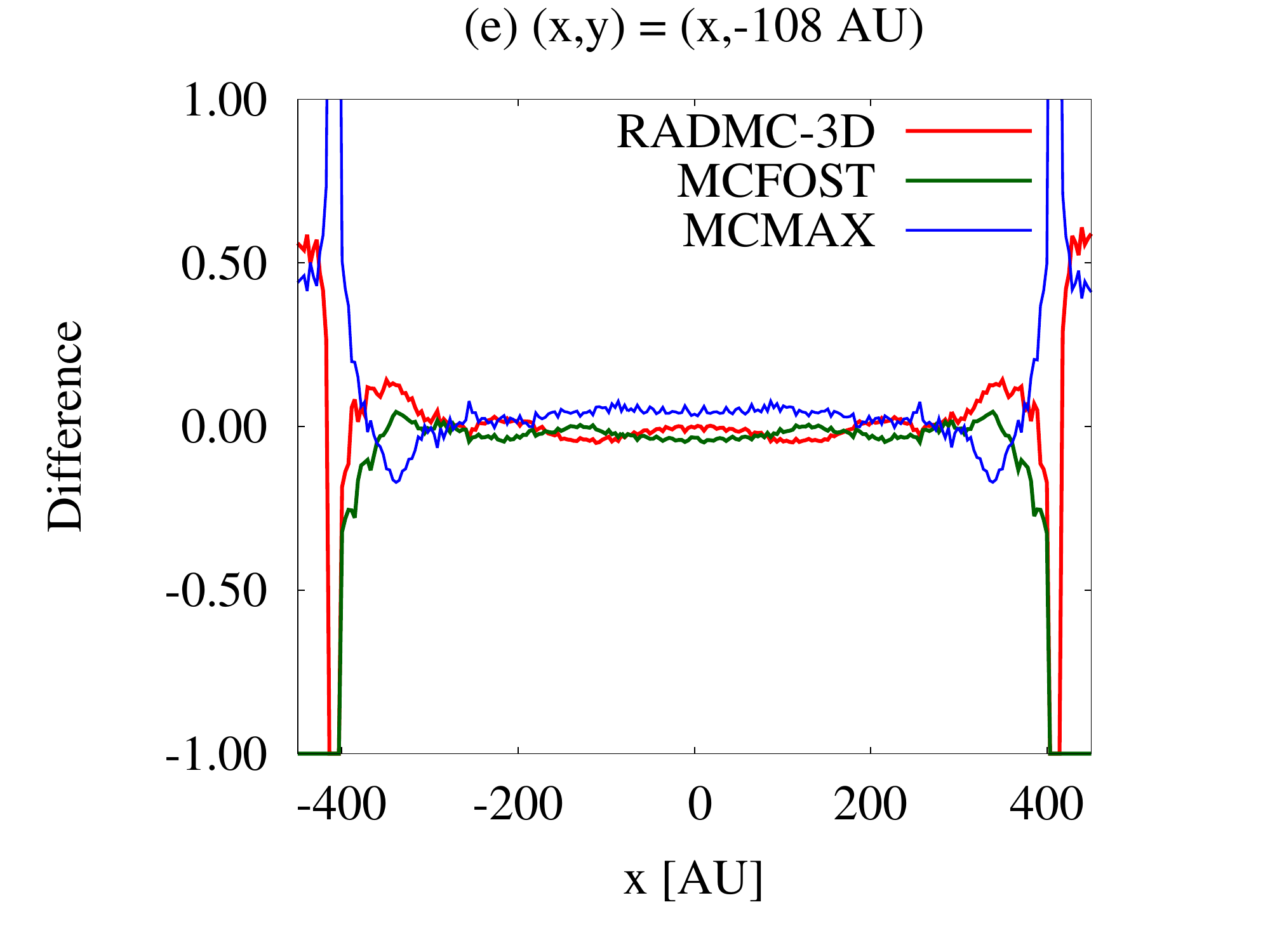}
  }
 \subfigure{
    \includegraphics[width=50mm]{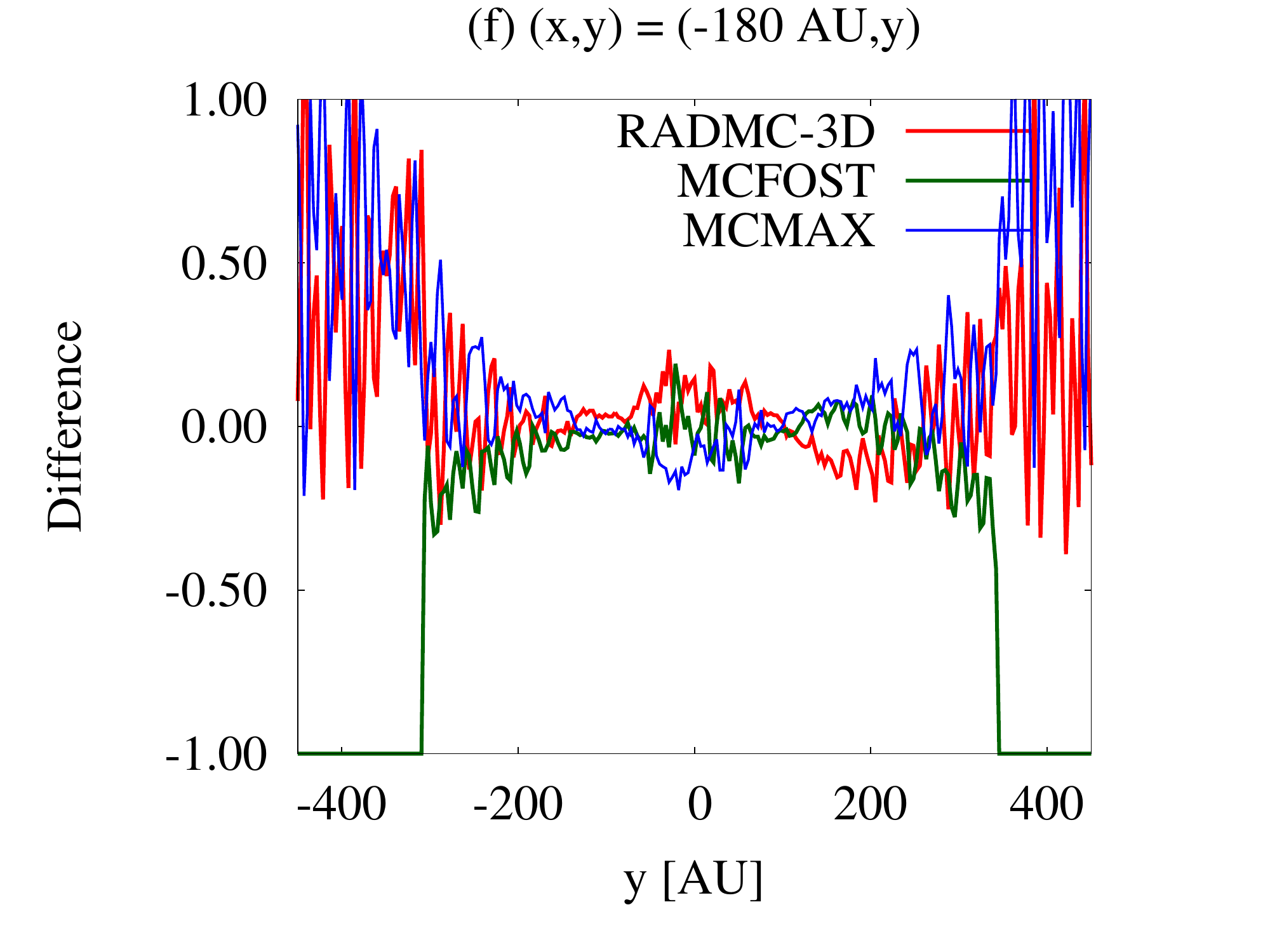}
  }
 \caption{
 The intensity along the three cuts of $(x,y)=(x,108 {\rm~AU})$, $(x,-108 {\rm~AU})$, and $(-180 {\rm~AU},y)$ in the case of $i=87^\circ$.
 The lower panels show the difference against the average.
 Due to the low signal-to-noise ratio, we take the average without the results of TORUS.
 }
 \label{fig:int_i87_cuts}
\end{figure*}

\begin{figure*}[htbp]
\centering
 \subfigure{
    \includegraphics[width=75mm]{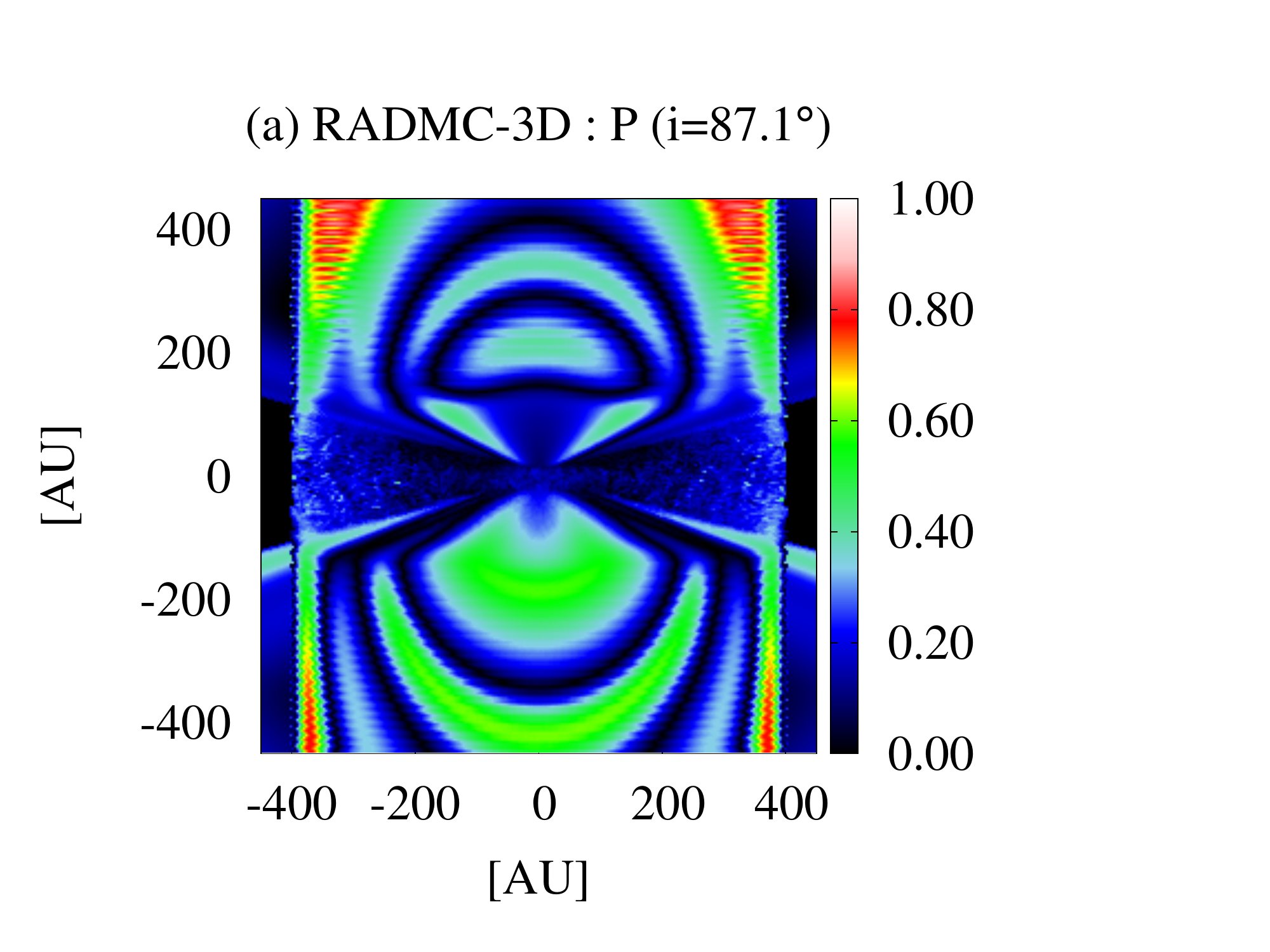}
  }
\subfigure{
  \includegraphics[width=75mm]{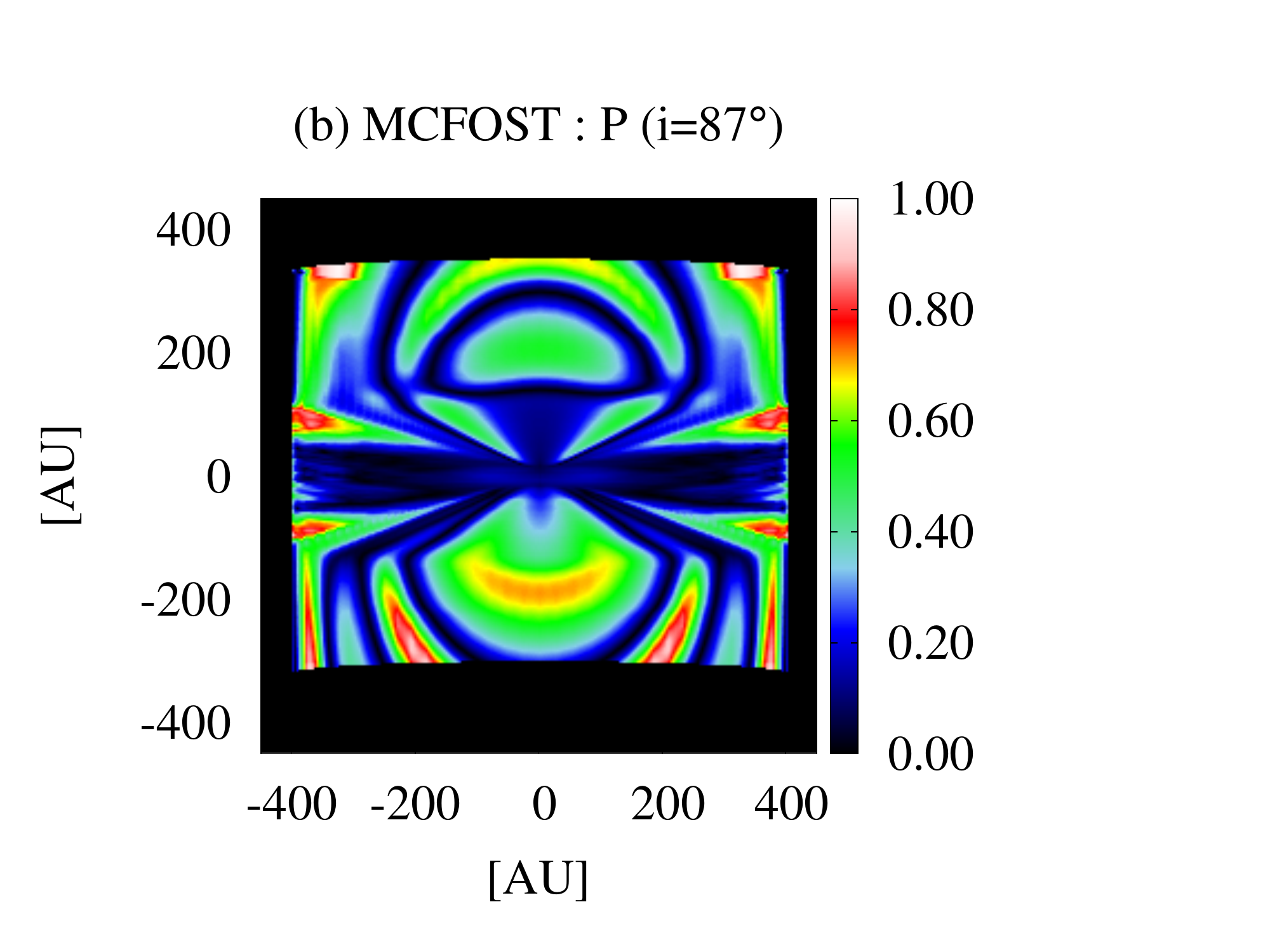}
  }
 \subfigure{
    \includegraphics[width=75mm]{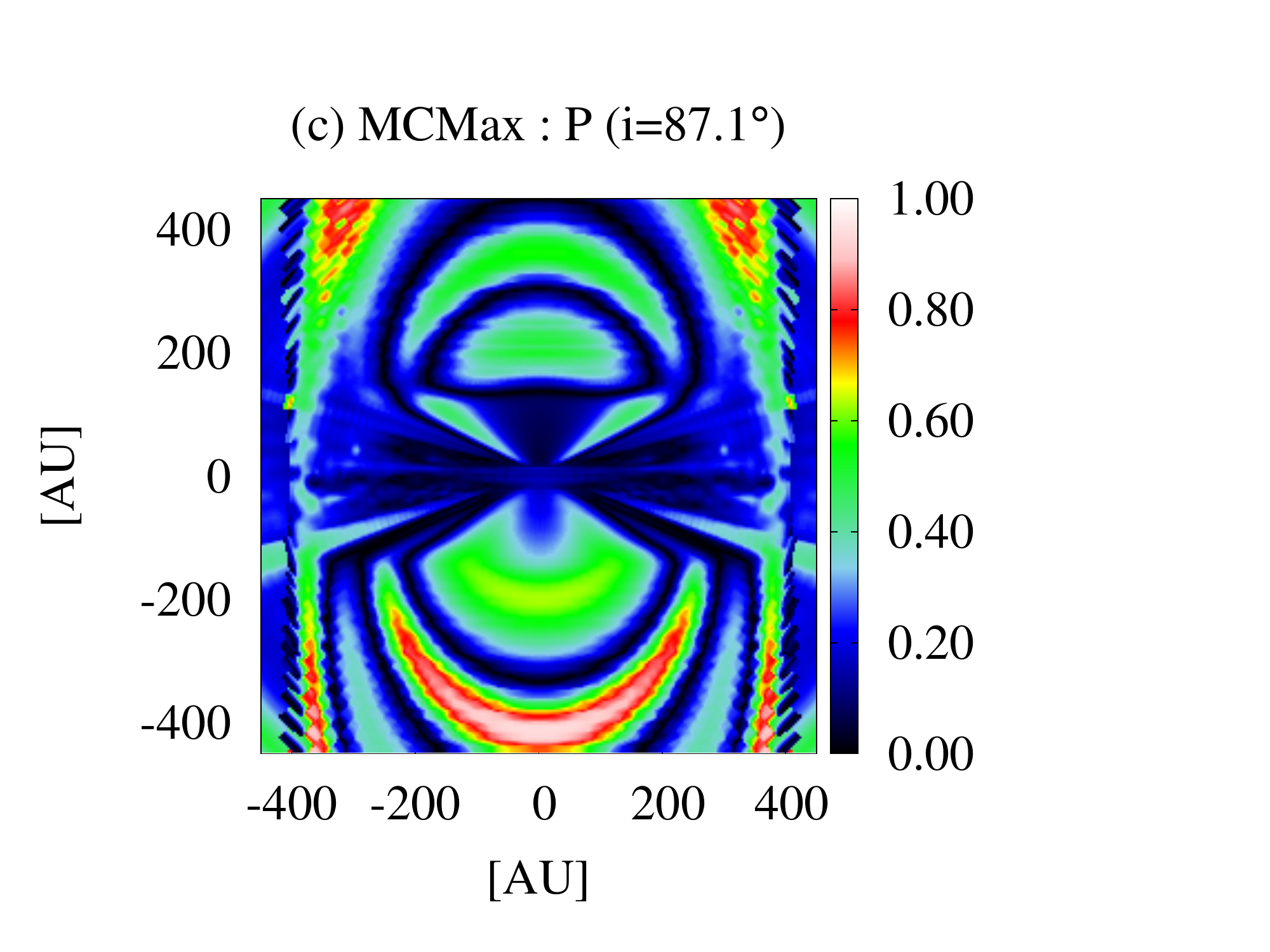}
  }
 \subfigure{
    \includegraphics[width=75mm]{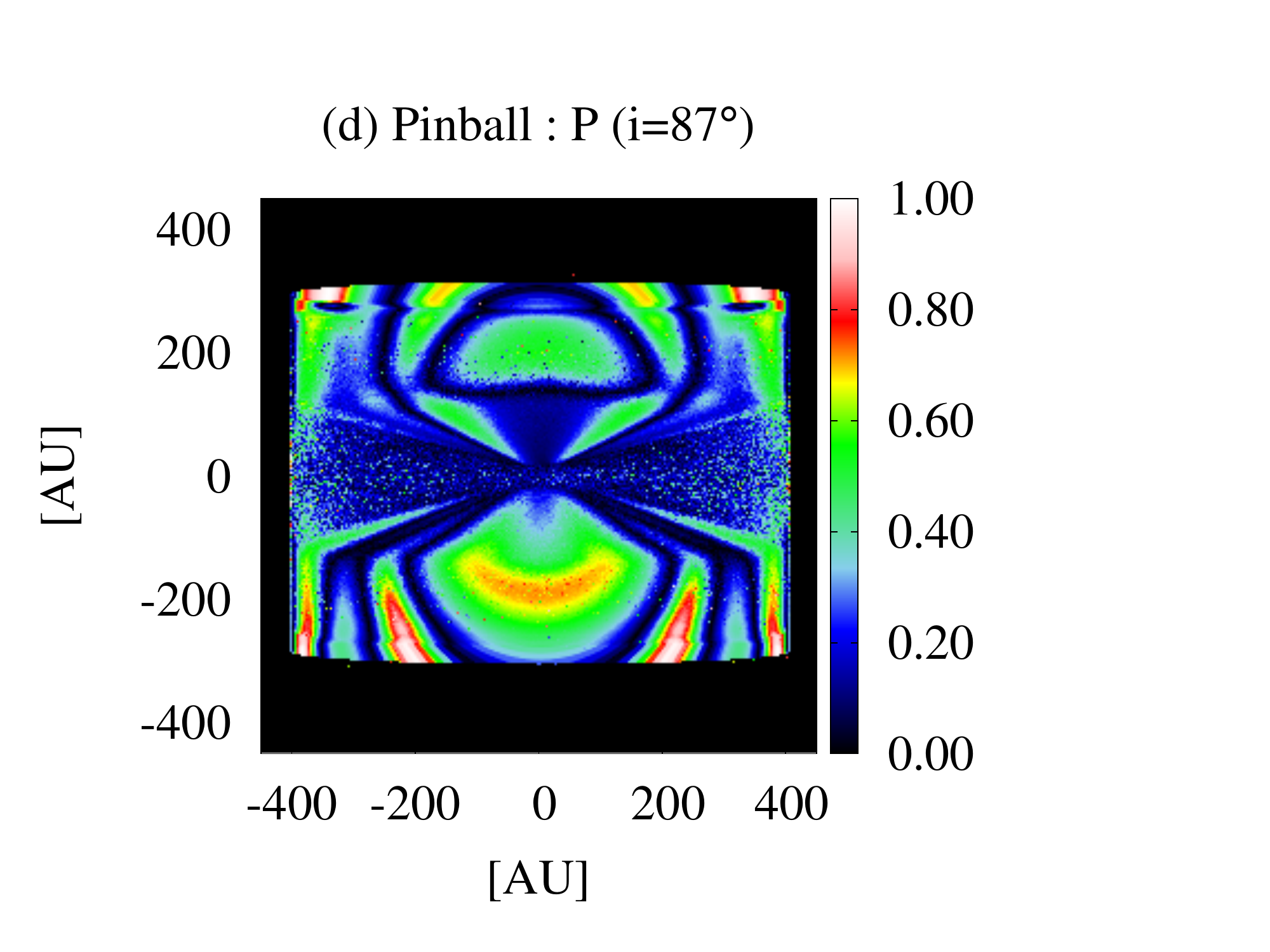}
  }
 \caption{
Polarization maps of other models: RADMC-3D, MCFOST, MCMAX, and Pinball.
The inclination is $87.1^\circ$.
 }
 \label{fig:pol_i87}
\end{figure*}

\begin{figure*}[htbp]
\centering
\subfigure{
    \includegraphics[width=50mm]{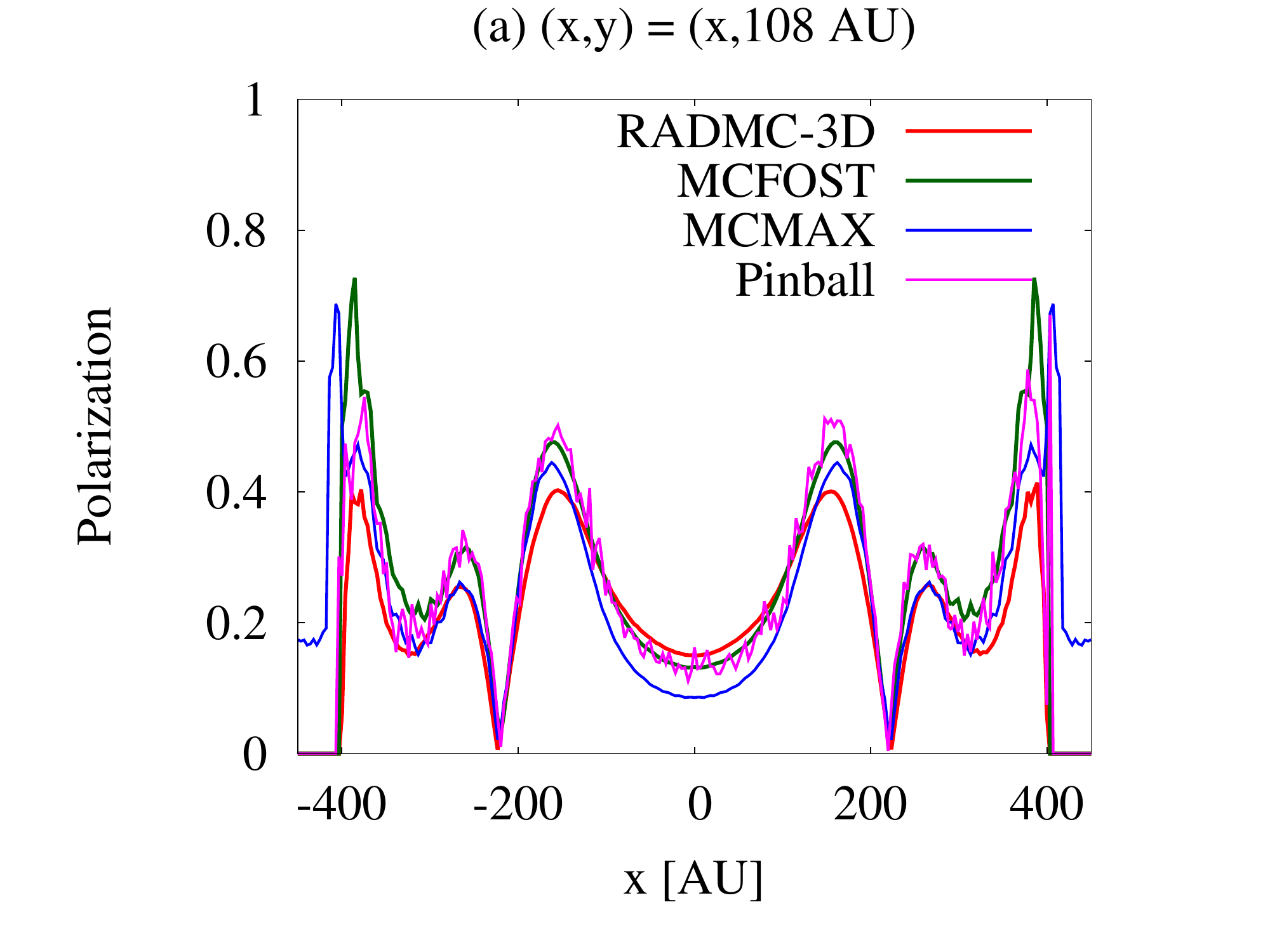}
  }
 \subfigure{
    \includegraphics[width=50mm]{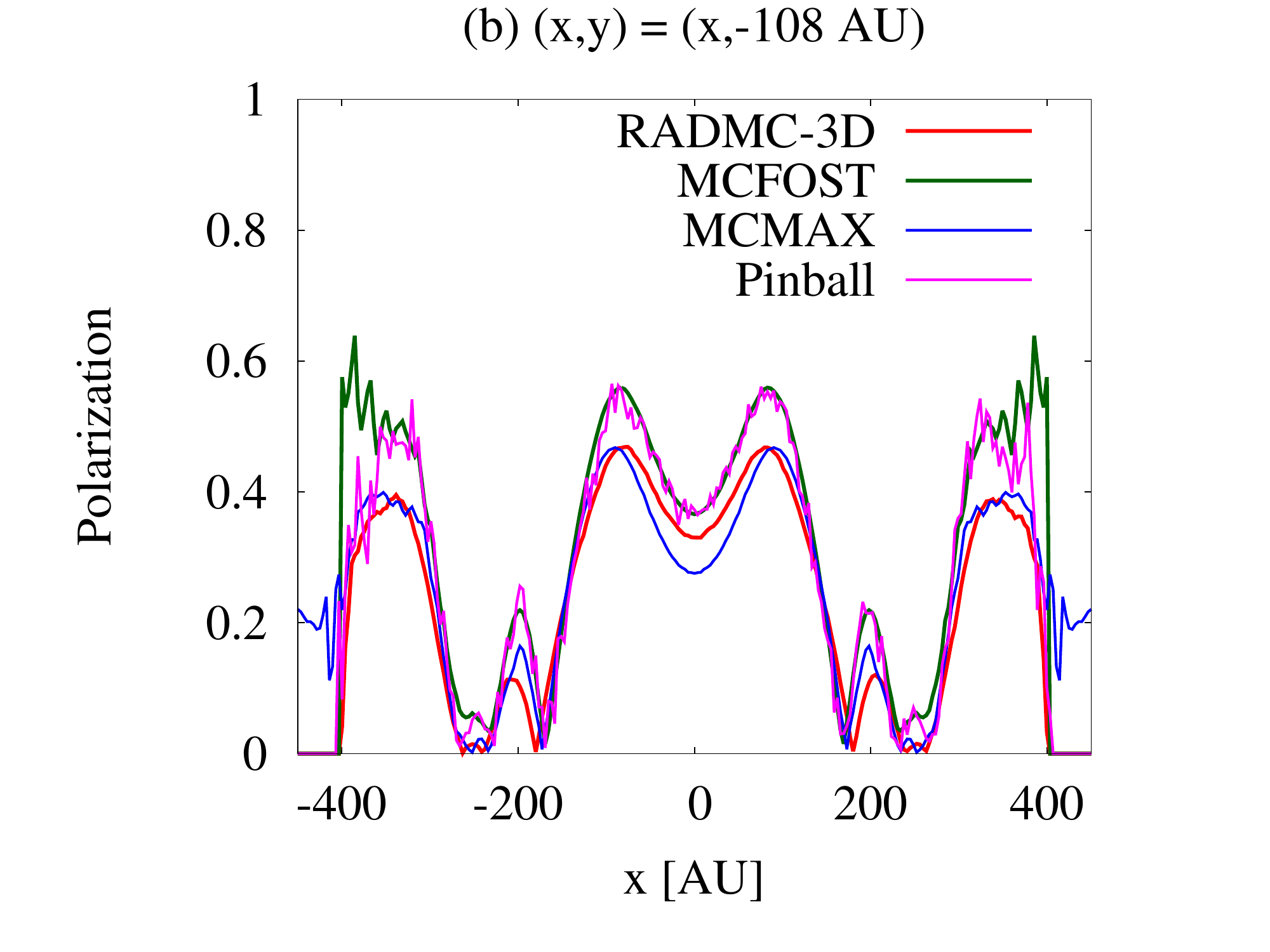}
  }
 \subfigure{
    \includegraphics[width=50mm]{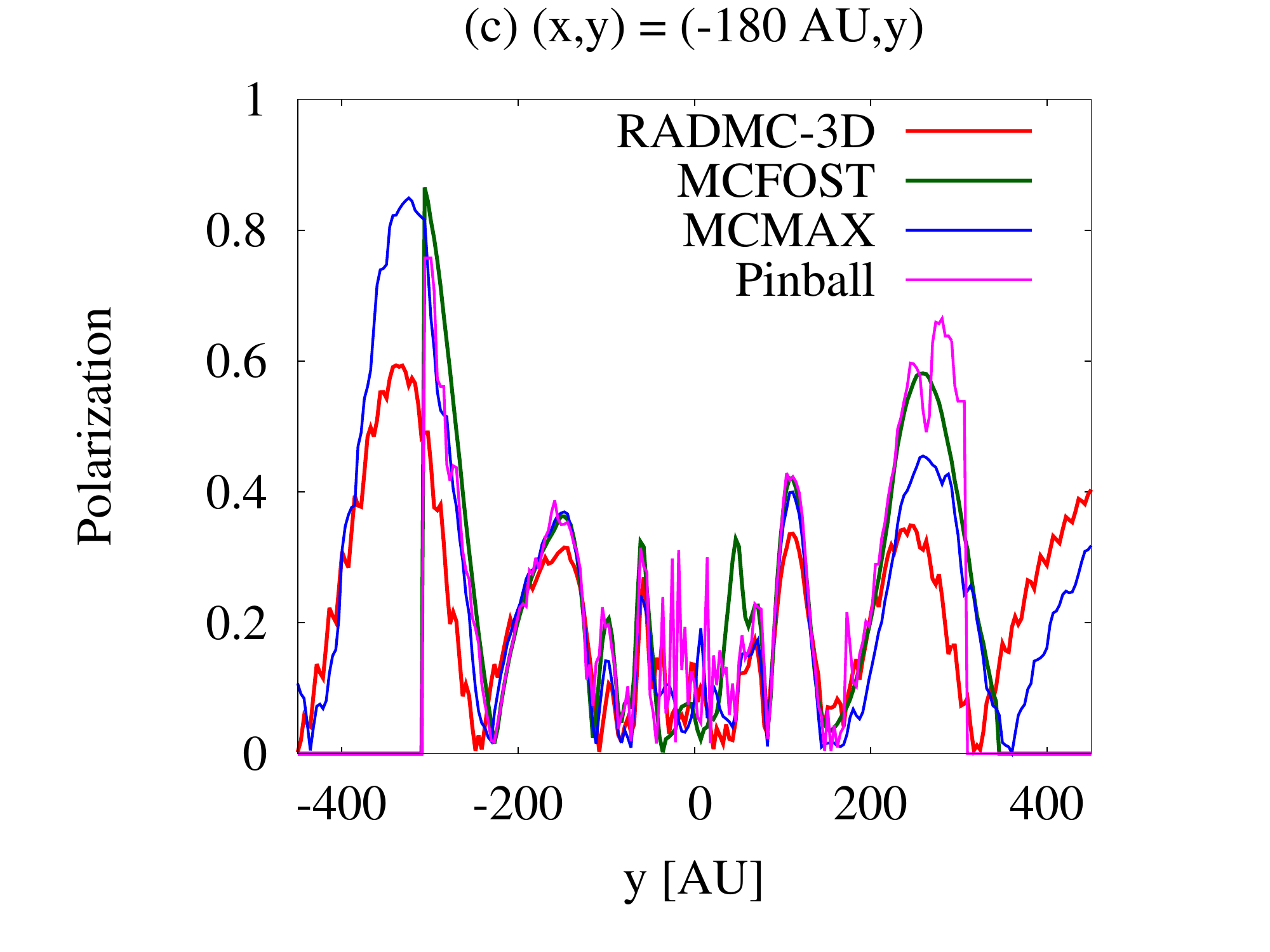}
  }
  \subfigure{
    \includegraphics[width=50mm]{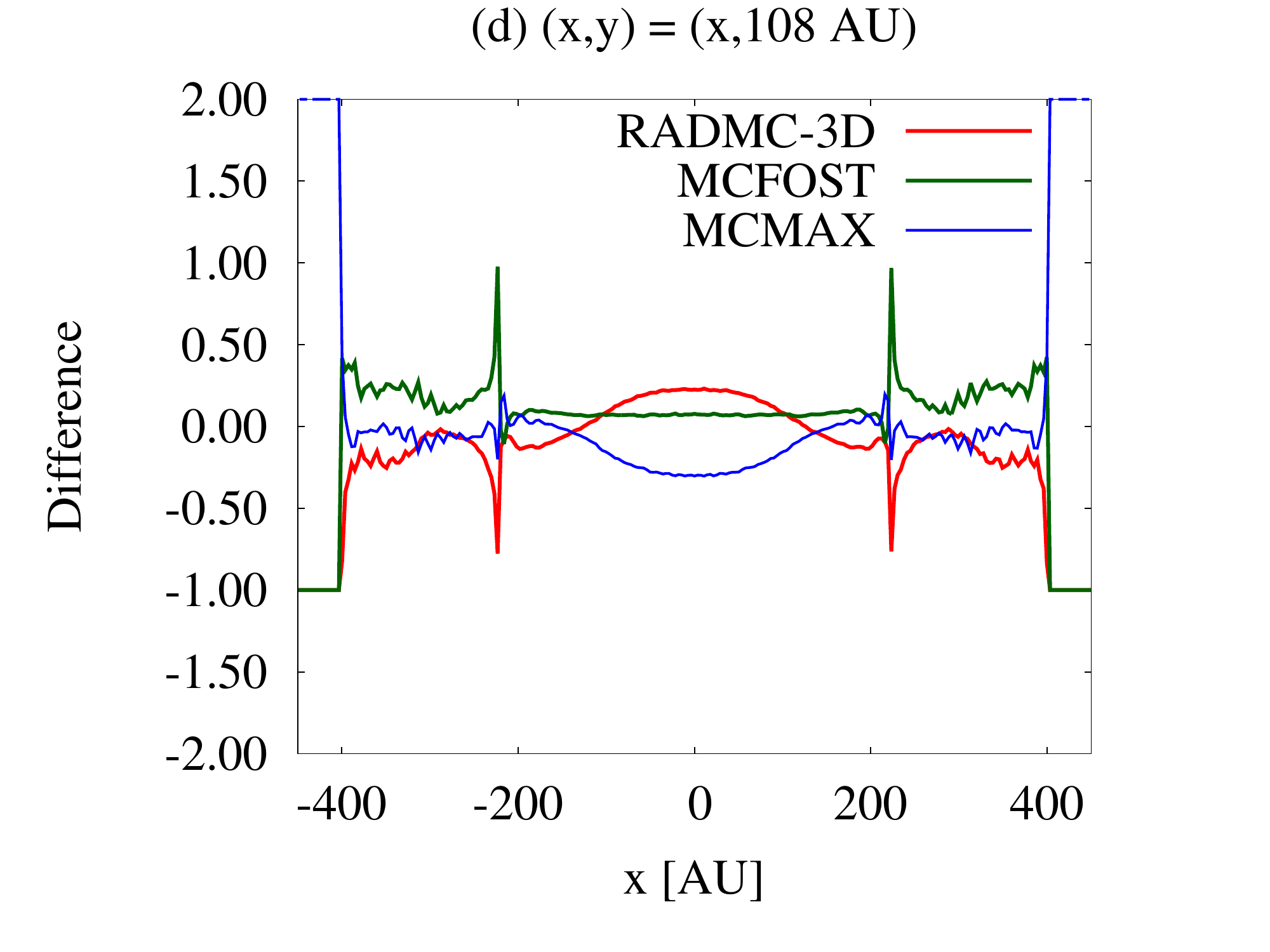}
  }
 \subfigure{
    \includegraphics[width=50mm]{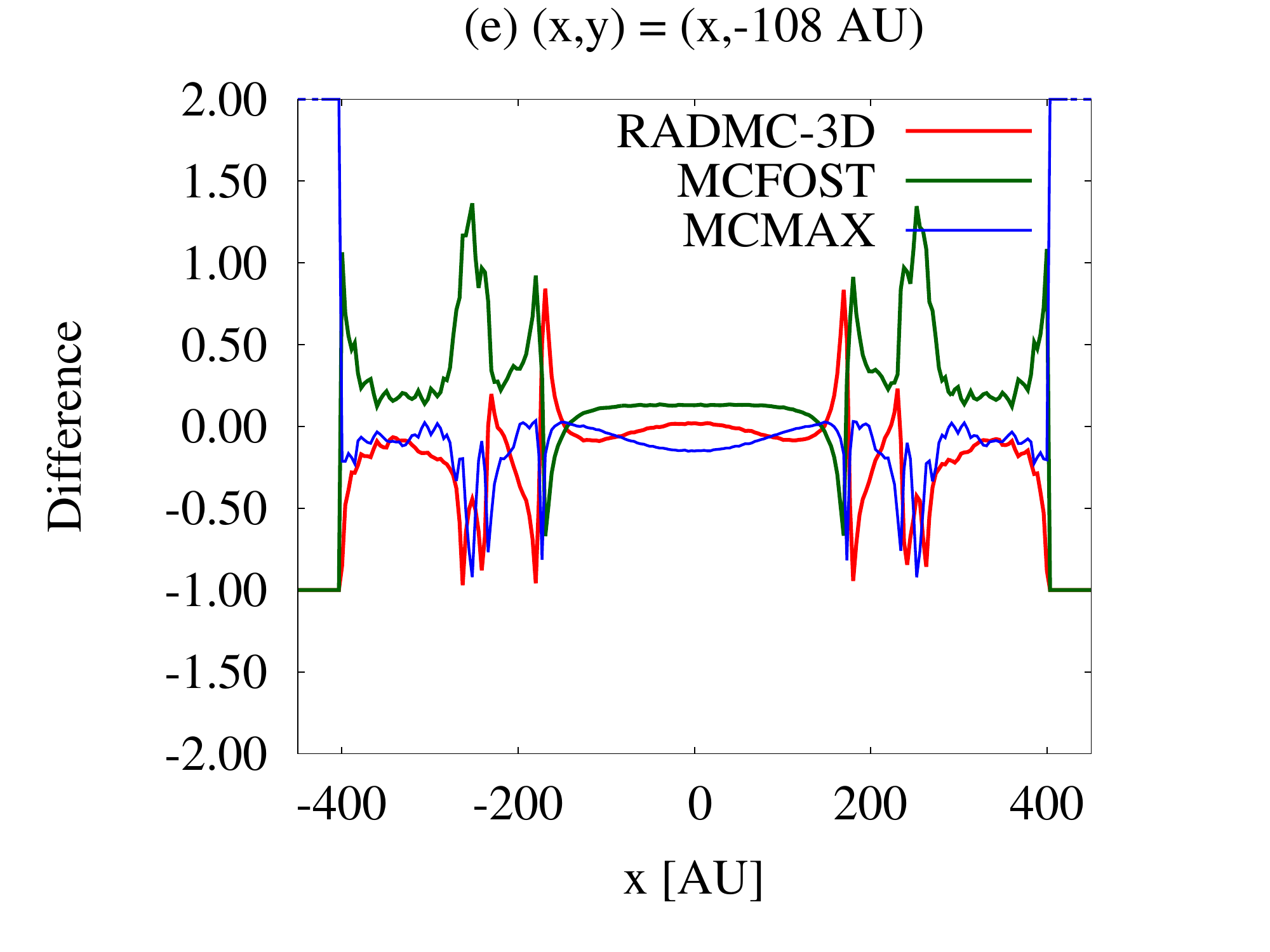}
  }
 \subfigure{
    \includegraphics[width=50mm]{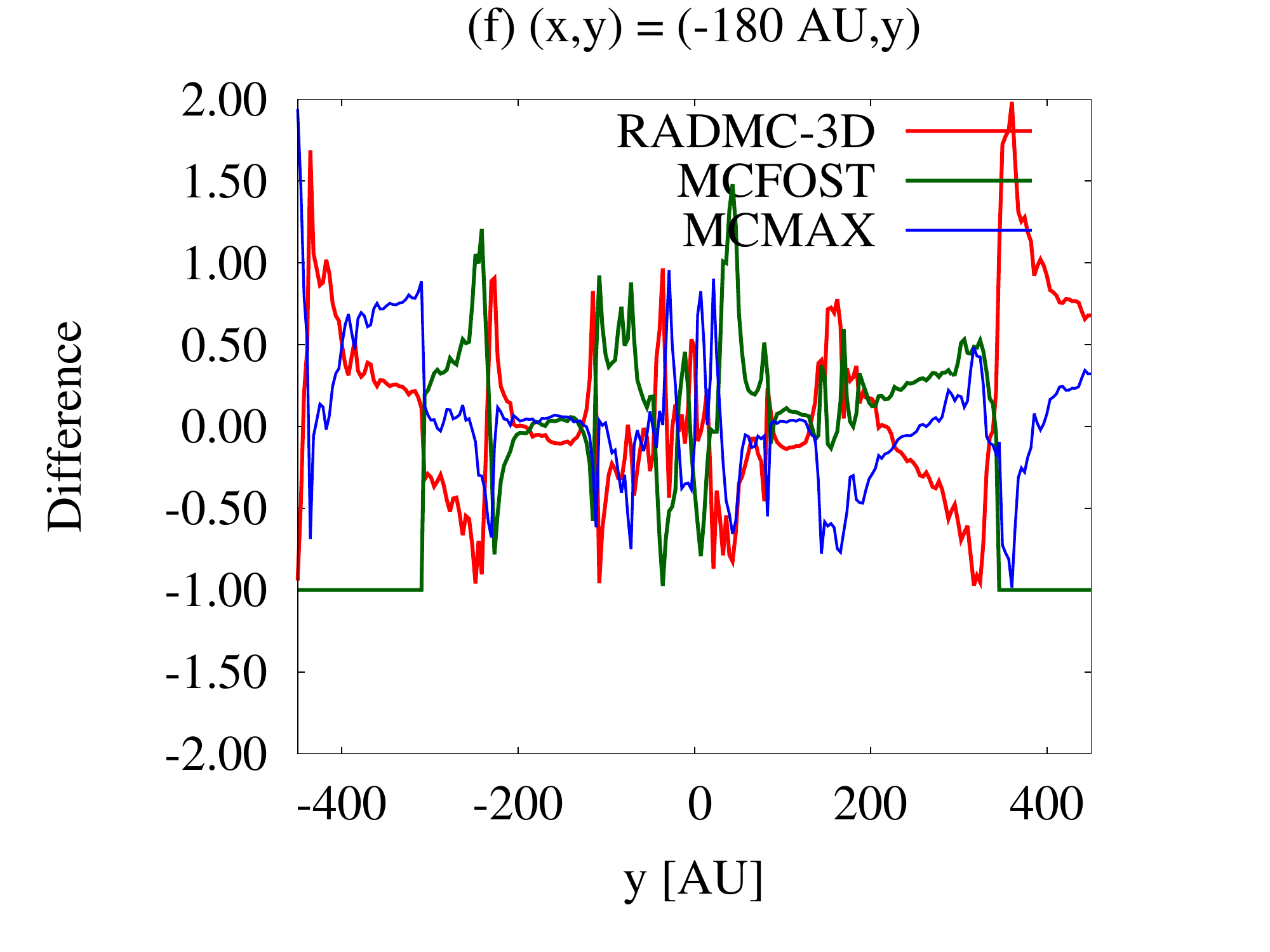}
  }
 \caption{
 The polarization degree along the three cuts of $(x,y)=(x,108 {\rm~AU})$, $(x,-108 {\rm~AU})$, and $(-180 {\rm~AU},y)$ in the case of $i=87.1^\circ$.
 The lower panels show the difference against the average.
 Due to the low signal-to-noise ratio, we take the average without the results of Pinball.
 }
 \label{fig:pol_i87_cuts}
\end{figure*}

Figures \ref{fig:int_i87}, \ref{fig:int_i87_cuts}, \ref{fig:pol_i87}, and \ref{fig:pol_i87_cuts} show the results of the same benchmark test shown in Figures \ref{fig:int_i70}, \ref{fig:int_i70_cuts}, \ref{fig:pol_i70}, and \ref{fig:pol_i70_cuts} in the case of the inclination angle of $i=87.1^\circ$.
Note that the three lines for the cuts are taken at $(x,y)=(x,108 {\rm~AU})$, $(x,-108 {\rm~AU})$, and $(-180 {\rm~AU},y)$.
The results of this case also show the similar results of the previous discussion.

The relative standard deviation in intensity and polarization is summarized in Tables \ref{table:deviation_int} and \ref{table:deviation_pol}.
The relative standard deviation is calculated by dividing the standard deviation by the average value.
To avoid the low signal-to-noise ratio regions, the relative standard deviation is calculated in the range of $-360 {\rm~AU}<x<360 {\rm~AU}$ for the cases of the horizontal lines and $-180 {\rm~AU}<y<180 {\rm~AU}$ for the vertical lines.
The relative difference is up to around 10 \% in intensity.
On the other hand, the relative difference is up to 30 \% in polarization degree.

From this results, we conclude the benchmark test as follows.
The calculations of RADMC-3D provides the intensity which is quite similar to the other results. 
The error is statistical and the relative standard deviation is within $\sim 10 \%$ if the number of photons is high enough.
In terms of the polarization degree, the results of the different codes are qualitatively similar and the relative standard deviation is up to $\sim 30 \%$.
If the polarization degree is low, however, the difference is large, which is due to the combinations of the low degree of polarization, a low number of photons at distant region from the star and, the optically thick disk.
Therefore, it requires more photons to simulate if we investigate the region where the low degree of polarization.
Note that the we confirm that we perform the simulations with enough number of photons in this paper.

\begin{deluxetable}{ccccccc}
\tabletypesize{\scriptsize}
\tablecaption{The relative standard deviation in intensity}
\tablewidth{0pt}
\tablehead{
 & \multicolumn{3}{c}{$i=69.5^\circ$} & \multicolumn{3}{c}{$i=87.1^\circ$}\\
codes & $(x,180 {\rm~AU})$ & $(x,-180 {\rm~AU})$ & $(-108 {\rm~AU},y)$ & $(x,108 {\rm~AU})$ & $(x,-108 {\rm~AU})$ & $(-180 {\rm~AU},y)$
}
\startdata
RADMC-3D  & 3.7 \%& 9.3 \% & 10.1 \% & 3.1 \% &  4.2 \%  & 10.9 \%\\
MCMAX        & 4.0 \% &  3.4 \%  &  5.2 \%  & 5.9 \% &  1.9 \%  & 9.4 \% \\ 
MCFOST     & 4.4 \%&  6.8 \%  & 10.0 \% & 4.0 \%&  4.1 \% & 6.1 \%
\enddata
\label{table:deviation_int}
\end{deluxetable}

\begin{deluxetable}{ccccccc}
\tabletypesize{\scriptsize}
\tablecaption{The relative standard deviation in polarization}
\tablewidth{0pt}
\tablehead{
 & \multicolumn{3}{c}{$i=69.5^\circ$} & \multicolumn{3}{c}{$i=87.1^\circ$}\\
codes & $(x,180 {\rm~AU})$ & $(x,-180 {\rm~AU})$ & $(-108 {\rm~AU},y)$ & $(x,108 {\rm~AU})$ & $(x,-108 {\rm~AU})$ & $(-180 {\rm~AU},y)$
}
\startdata
RADMC-3D & 11.3 \% & 26.0 \%  & 23.9 \%  & 13.0 \% & 11.6 \% & 22.8 \% \\
MCMAX        & 10.2 \% & 12.5 \% & 17.6 \%  & 10.6 \% & 10.3 \% & 16.9 \%\\ 
MCFOST     & 12.0 \%  & 18.1 \% & 27.7 \%  & 13.2 \% & 18.1 \% & 29.3 \%
\enddata
\label{table:deviation_pol}
\end{deluxetable}

\clearpage

\end{document}